\documentclass{siamonline1116}

\usepackage{lipsum}
\usepackage{amsfonts}
\usepackage{graphicx}
\usepackage{epstopdf}
\usepackage{algorithmic}
\usepackage{caption}
\ifpdf
  \DeclareGraphicsExtensions{.eps,.pdf,.png,.jpg}
\else
  \DeclareGraphicsExtensions{.eps}
\fi

\numberwithin{theorem}{section}

\newcommand{\TheTitle}{Sparse Bayesian Imaging of Solar Flares} 
\newcommand{\TheAuthors}{ Federica Sciacchitano, Silvio Lugaro and Alberto Sorrentino}

\headers{\TheTitle}{\TheAuthors}

\author{
  Federica Sciacchitano\thanks{Dipartimento di Matematica, Universit\`{a} di Genova, via Dodecaneso 35, 16146 Genova, Italy (\email{silvio.lugaro@gmail.com,sciacchitano@dima.unige.it}).}
  \and
  Silvio Lugaro\footnotemark[1] \and
  Alberto Sorrentino\thanks{Dipartimento di Matematica, Universit\`{a} di Genova, via Dodecaneso 35, 16146 Genova, Italy and CNR - SPIN Genova,via Dodecaneso 33 16146 Genova, Italy (\email{sorrentino@dima.unige.it}).}
}

\usepackage{amsopn}

\usepackage{mathtools}
\newcommand{\dd}{{\text{d}}}
\usepackage{gensymb}
\usepackage{multirow}

\newcommand{\lapprox} {\, \lower3pt\hbox{$\sim$}\llap{\raise2pt\hbox{$<$}}\,}

\usepackage{xcolor}
\usepackage{relsize}

\usepackage{verbatim}
\ifpdf
\hypersetup{
  pdftitle={\TheTitle},
  pdfauthor={\TheAuthors}
}
\fi

\title{Sparse Bayesian Imaging of Solar Flares}

\begin{document}

\maketitle

\begin{abstract}
	We consider imaging of solar flares from NASA RHESSI data as a parametric imaging problem, where flares are represented as a finite collection of geometric shapes.
	We set up a Bayesian model in which the number of objects forming the image is a priori unknown, as well as their shapes.
	We use a Sequential Monte Carlo algorithm to explore the corresponding posterior distribution.
	We apply the method to synthetic and experimental data, largely known in the RHESSI community.
	The method reconstructs improved images of solar flares, with the additional advantage of providing uncertainty quantification of the estimated parameters.
\end{abstract}

\begin{keywords}
Sparse imaging, Bayesian inference, Sequential Monte Carlo, Astronomical imaging, Solar flares.
\end{keywords}

\begin{AMS}
65R32, 62F15, 65C05, 85A04
\end{AMS}

\section{Introduction}\label{sec:intro}
In the last couple of decades, sparsity has emerged as a key concept in a number of imaging applications including geophysics \cite{gholami2010regularization}, magnetic resonance \cite{lustig2007sparse}, photoacoustics \cite{frikel2018efficient}, image restoration problems \cite{elad2006image}, and also astronomical imaging \cite{duval2018solar}. The fundamental intuition underlying this revolution is that the information content of natural images is small compared to the number of pixels we use to represent them.
As a consequence images can be compressed substantially by representing them on a proper basis, in which they turn out to have few non--zero coefficients, i.e. they are sparse; crucially, such compression typically entails little loss of information. 


In many imaging problems, sparsity would be ideally obtained by introducing a regularization term with the $\ell^0$--norm of the image, represented on a suitable basis \cite{bruckstein2009sparse}. However, such regularization term is non--convex and leads to a problem of combinatorial complexity, for which two computational strategies have been proposed: greedy methods (for instance matching pursuit) \cite{tropp2004greed} and the relaxation into a convex term \cite{tropp2006just}. Since the relaxation to $\ell^p$--norm with $0<p<1$ leads again to a non-convex minimization problem, the most common approximation is the $\ell^1$--norm, which represents a good trade--off between sparsity--promotion and computational tractability. In the last decade this approach has been widely investigated, also thanks to the development of new efficient algorithms for convex optimization \cite{yin2008bregman,beck2009fast}. 

In this study we present an alternative approach to sparse imaging, based on three ingredients: (i) the parametrization of the image in terms of a small set of objects, each one described by a small number of parameters; (ii) a Bayesian model in which sparsity is encouraged through the choice of a Poisson prior distribution on the number of objects with a small mean; (iii) a Monte Carlo algorithm which is able to sample the resulting complex posterior distribution. 


We apply the proposed approach to the problem of reconstructing images from data recorded by the NASA RHESSI satellite \cite{hurford2002rhessi}. RHESSI \cite{lin2002reuven}, launched in 2002, records X--rays from the Sun, with the main objective of improving our understanding of the mechanisms underlying the generation of solar flares. After more than 16 years of activity, the X--ray observations ended on April 11 2018, due to communication problems, and RHESSI was decommissioned on August 16 2018\footnote{\url{https://hesperia.gsfc.nasa.gov/rhessi3/}}. During its mission, RHESSI covered more than a complete 11--year solar cycle, recording over 120,000 X--ray events. A successor of RHESSI, STIX  (Spectrometer/Telescope for Imaging X-rays), is expected to be launched in 2020\footnote{\url{https://solar-orbiter.cnes.fr/en/SOLO/GP_stix.htm}}. 

Due to the specific hardware configuration, RHESSI imaging can be synthetically described as an inversion of the Fourier transform, with undersampled data. In the past fifteen years RHESSI data have been converted to images using several different methods: from old--fashioned, heuristic approaches such as CLEAN \cite{hogbom1974aperture}, MEM \cite{cornwell1985simple}, PIXON \cite{pina1993bayesian} and Forward Fitting \cite{aschwanden2003reconstruction}, to more mathematically sound regularization methods such as UV--smooth \cite{massone2009hard}, Space--D \cite{bonettini2010nonnegative}, Expectation--Maximization \cite{benvenuto2013expectation},  and Semi--blind deconvolution \cite{bonettini2013new}. Only recently, sparse imaging has been proposed through compressed sensing approaches in a few studies \cite{duval2018solar, felix2017compressed}. Specifically, the former (VIS\_WV) proposes to use a finite isotropic wavelet transform combined with FISTA \cite{beck2009fast}; the second one (VIS\_CS) uses a Coordinate Descent with Active Set heuristic \cite{friedman2010regularization} and a parametrized image space. 

These recent approaches are providing improved images to the solar physics community, particularly for low signal--to--noise ratios. However, one of the goals of RHESSI imaging is to make quantitative inference on the physical objects appearing in the images. For instance, one of the variables of interest is the total flux of X--rays coming from a flare; in order to estimate this quantity from an image, it is required to integrate over the area occupied by the flare; however, evaluating what part of the image actually contains the flare can be a difficult task, and may make results subjective. 

In our method, making inference on the individual objects of an image is made easy by the presence of an underlying Bayesian parametric model.
Our approach has some similarities with the work of \cite{felix2017compressed}, inasmuch as both methods assume that the image can be described by a combination of very simple geometric shapes. However, our method approximates the whole posterior distribution of a Bayesian model, while the method described in \cite{felix2017compressed} is an optimization technique that only provides the global minimum of a regularization functional; as a consequence, our approach provides additional information in terms of uncertainty quantification of the parameters of the estimated image.

We set up a statistical model in the Bayesian framework, in which the number of objects in the image is itself a random variable. Technically, this is done by a so--called variable dimension model, i.e. a collection of spaces of differing dimensions. 
Since the posterior distribution is often complex and possibly multi--modal, deterministic optimization algorithms would have difficulties in finding its maximum; here we resort to sampling, rather than optimization. While this approach has an inherently high computational cost, it provides additional benefits such as uncertainty quantification and possibly multi--mode detection. 

We sample the posterior distribution using an adaptive Sequential Monte Carlo (SMC) sampler \cite{del2006sequential}. SMC samplers are effective algorithms for sampling complex distributions of interest; the main idea behind SMC samplers is to construct an auxiliary sequence of distributions such that the first distribution is easy to sample, the last distribution is the target distribution, and the sequence is ``smooth''. In the present case, we draw a random sample from the prior distribution and then let these sample points evolve according to properly selected Markov kernels, until they reach the posterior distribution. In less technical terms, SMC samplers start with a set of candidate solutions, and proceed in parallel with a stochastic local optimization -- done through these Markov kernels. The key to the efficiency of the proposed algorithm is highly dependent on the construction of effective sets of transition kernels, to explore the space. 
Here we adopt three kind of moves: individual--parameter--updating moves, such as modifying the abscissa of the location of an object; birth and death moves, where the number of objects is increased or decreased by addition or deletion; split and merge moves, where a single object is split into two, or two objects are merged into one.



The outline of the paper is the following. In  \cref{sec:astro} we provide the necessary background on RHESSI imaging. In \cref{sec:bay} we present the Bayesian model and the adaptive sequential Monte Carlo method (ASMC) used to approximate the posterior distribution. In \cref{sec:experiments} we show the results obtained by our method on a set of synthetic and real data, and compare them to those obtained by commonly used techniques. Finally we draw our conclusions.

\section{Parametric models for RHESSI imaging}
\label{sec:astro}
The Reuven Ramaty High Energy Solar Spectroscopic Imager (RHESSI) is a hard X--ray imaging instrument which was launched by NASA on February 5, 2002. The goal of this NASA mission is to study solar flares and other energetic solar phenomena \cite{lin2002reuven}.  
RHESSI observes X-ray emissions from the Sun through a set of nine coaligned pairs of rotating modulation collimators, and the transmitted radiation is recorded on a set of cooled high-purity germanium detectors \cite{hurford2002rhessi}. The raw data provided by RHESSI are nine count profiles of the detected radiation as a function of time, modulated by the grid pairs. By combining rotation and modulation it is possible to estimate the \textit{visibilities}, which are image Fourier components at specific spatial frequencies. 
Each detector samples Fourier components on a different circle in the Fourier space, over the course of the $\simeq$4~s spacecraft rotation period. More formally, let $F: \mathbb{R}^2 \to \mathbb{R}$ be the spatial photon distribution to be recovered from the observed visibilities $V: \mathbb{R}^2 \to \mathbb{C}$, which are defined as:
\begin{equation}
V(u, v)= \int_{\mathbb{R}^2} F(x,y) e^{2 \pi i(ux+vy)} \dd x \dd y ~~~;
\label{eq:vis}
\end{equation}
for more details we refer to  \cite{bonettini2013new} and \cite{hurford2002rhessi}. 

\begin{figure}[htbp]
  \centering
\includegraphics[height=4cm]{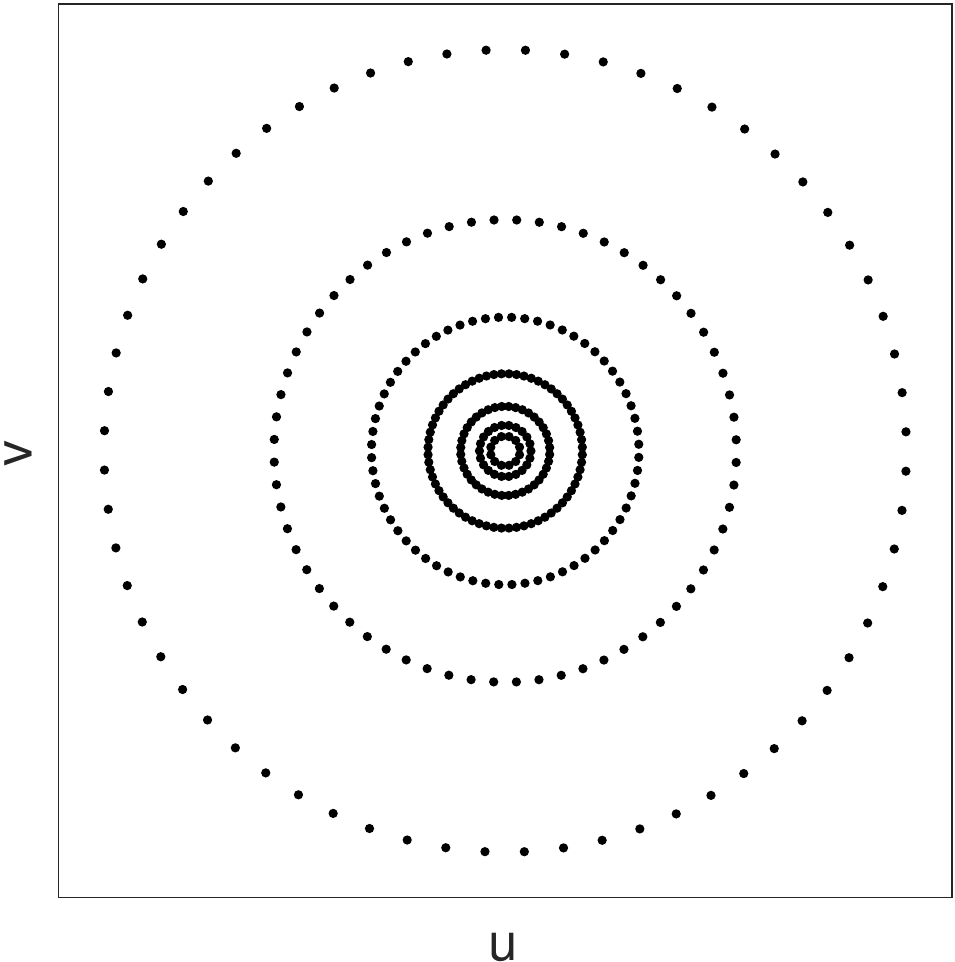}\hspace{0.5cm}
 \includegraphics[height=4cm]{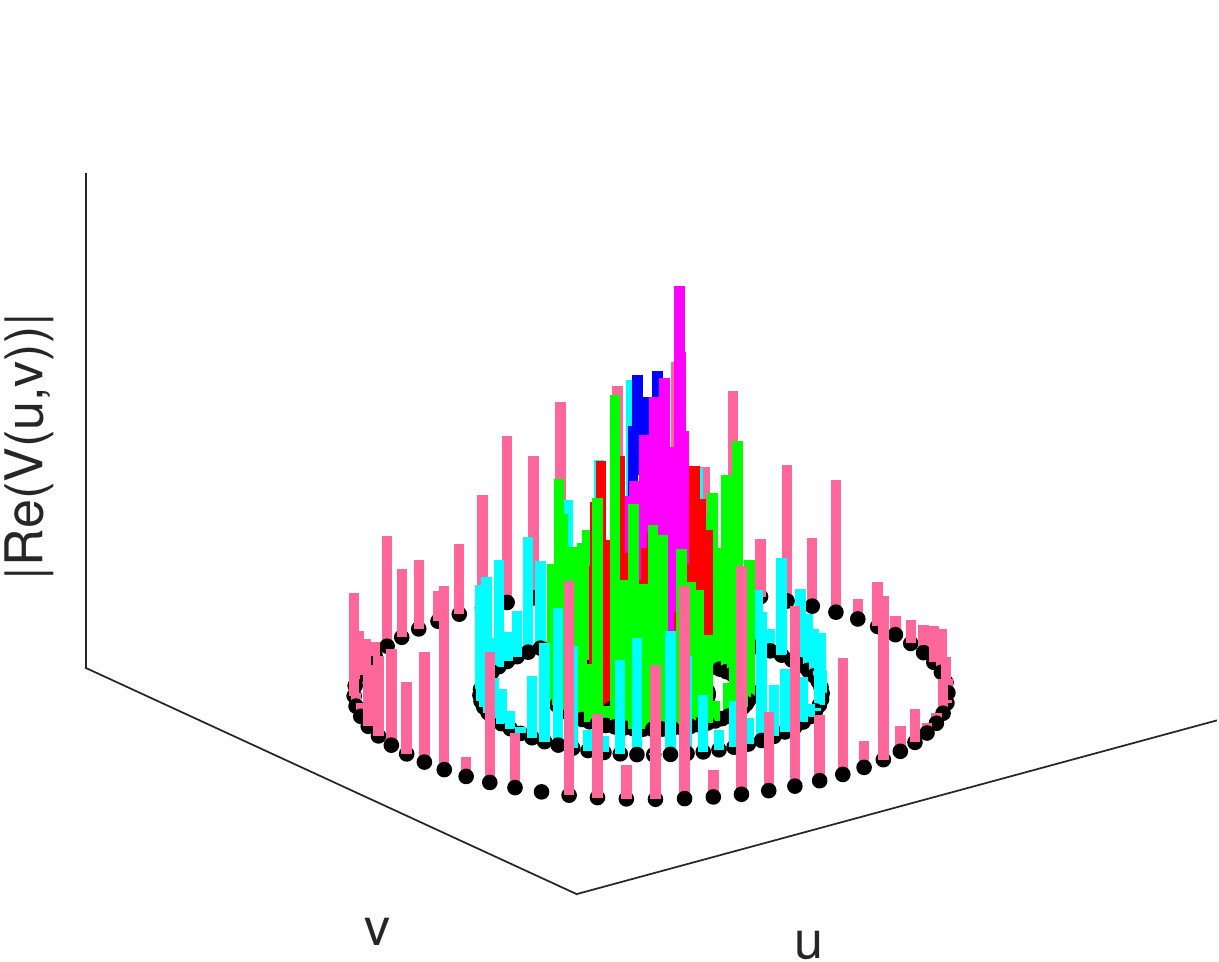}\hspace{0.5cm}
 \includegraphics[height=4cm]{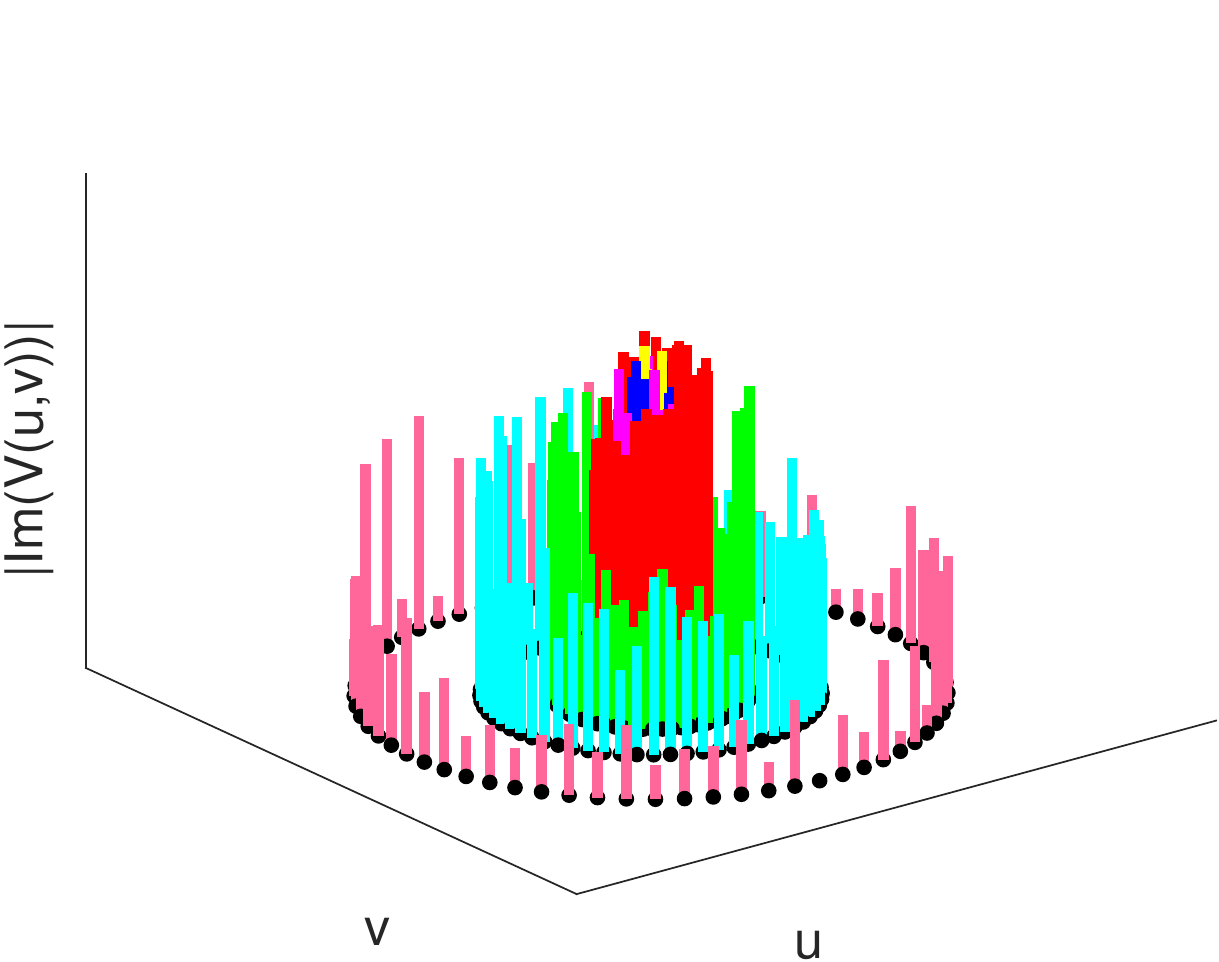}
  \caption{Left: sampling of the spatial frequency $(u, v)$ plane by seven detectors on RHESSI.  Absolute value of amplitude (middle) and phase (right) of the visibilities for a particular event. For graphical reason, we do not show the detectors 1 and 2.}
  \label{fig:vis}
\end{figure}

In the left panel of \cref{fig:vis} we include the sampled $(u,v)$-plane, where each of the sampled circles correponds to a single detector (here, only detectors from 3 to 9 are shown) and each dot of a circle represents a sampled $(u,v)$ frequency at which a visibility is measured. In the middle and in the right panel of \cref{fig:vis} an example of the absolute value of the amplitude and the phase of some visibilities is shown.

Equation \cref{eq:vis} defines a linear inverse problem, and indeed, in the last decade most of the methodological efforts have been devoted to devising novel linear inverse methods. However, early works such as \cite{kosugi1995yohkoh} and \cite{sato1998yohkoh} have shown that with a little simplification, which is acceptable given the limited spatial resolution provided by RHESSI, a solar flare can show up in an image under a very limited range of shapes: the lower parts of the flare, called footpoints, can be circular or elliptical depending on the viewing angle; the upper part of the flare may either appear as an elliptical source or as an arc, a sort of bent ellipsoid, often referred to as \textit{loop}.

\begin{figure}
\hspace{-3.0cm}
\begin{center}
\includegraphics[height=8cm]{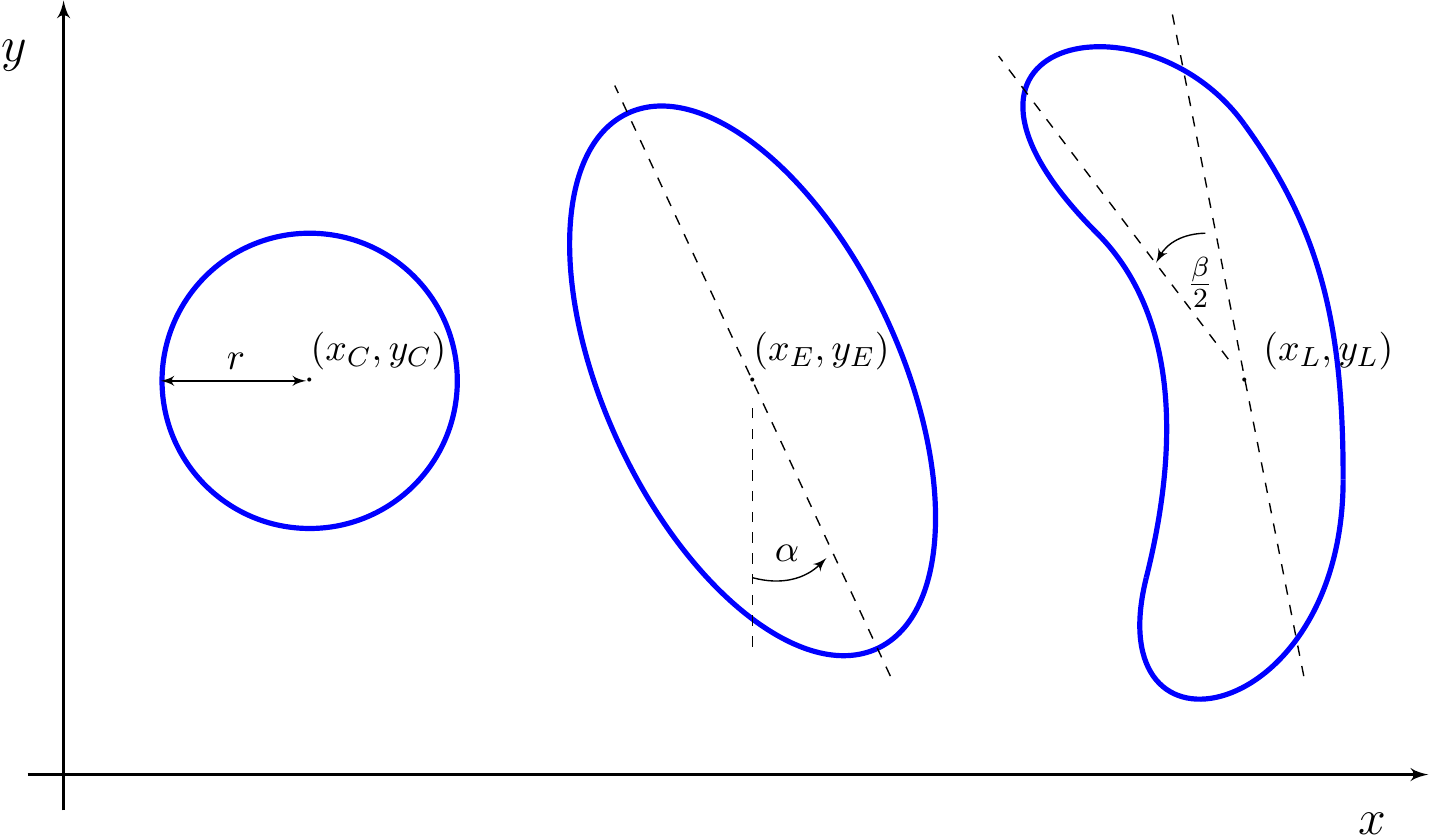}
\end{center}
\caption{The three geometrical objects: circular source, elliptical source, and loop.}
\label{Fig:shapes}
\end{figure}

An image containing one of these geometrical shapes can be parameterized easily; for instance, a circular source $C$ centered at $(x_C, y_C)$, of radius $r_C$ and total photon flux $\phi_C$ can be properly represented by a 2-dimensional Gaussian distribution defined by 
\begin{equation}
\label{eq:circle}
F_C(x,y)= \phi_C \exp\biggl[ - \frac{(x-x_C)^2+(y-y_C)^2}{2r_C^2} \biggr].
\end{equation}
An elliptical source $E$ oriented along the $x,y$ axes can be easily derived from \cref{eq:circle}, by adding the eccentricity parameter $\varepsilon_E$
\begin{equation}
\label{eq:ellipse}
F_E(x,y)= \phi_E \exp\biggl[ - \frac{(x-x_E)^2}{2 r_E^2}- \frac{(y-y_E)^2}{2 r_E^2}(\varepsilon_E+1)^2 \biggr].
\end{equation}
One can obtain a generic elliptical shape by adding a rotation angle $\alpha$; and finally, a loop source is obtained by adding a loop angle $\beta$. In the following, we will be calling these distributions circle, ellipse and loop; we refer to \cref{Fig:shapes} for a pictorial representation of these geometric objets, and to \cite{aschwanden2003reconstruction} for the explicit definition of these type of sources.

The key point is that it is straightforward to compute the visibilities associated to these simple objects. For instance, by computing the 2D-Fourier transformation for the circular Gaussian source, we obtain the following

\begin{equation}
V_C(u, v) = \phi_C \exp\biggl[ 2 \pi i(ux_C+vy_C) - \frac{\pi^2 r_C^2}{4\log 2}(u^2+v^2) \biggr] ~~~,
\end{equation}

and similar results, although more complicated, hold for ellipses and loops.

Eventually, we can build an imaging model in which the final image is composed of an arbitrary -- but small -- number of elementary objects 
\begin{equation}
F(x,y)= \sum_{s=1}^{N_S} F_{T_s}(x_{T_s},y_{T_s},r_{T_s},\phi_{T_s}, \dots) ~~~,
\label{eq:F}
\end{equation}
where $N_S$ is the number of sources in the image,  
and the corresponding visibilities will be the sum of the visibilities of the single objects
$$V(u,v) = \sum_{s=1}^{N_S} V\bigl(F_{T_s}(x_{T_s},y_{T_s},r_{T_s},\phi_{T_s}, \dots)\bigr) ~~~.$$

It therefore makes sense to re--cast the imaging problem \cref{eq:vis} as a parametric problem, in which one aims at reconstructing the parameters of one or more circles, ellipses, or loops. 
In \cite{aschwanden2003reconstruction} this approach was pursued using a non--linear optimization algorithm; however, the number and type of shapes has to be defined a priori within a very small set of pre--defined configurations (one circle, one ellipse, one loop, two circles). 

We also notice that Eq. \cref{eq:F} corresponds to a representation of the image on the basis of circles, ellipses and loops, in which the flux $\phi$ of each source can be seen as the coefficient of the corresponding basis element.

\section{Bayesian sparse imaging with Sequential Monte Carlo}
\label{sec:bay}
We cast the problem of reconstructing images of solar flares from the visibilities measured by RHESSI as a Bayesian inference problem. After formalization of the parameterized image space, we describe our choice in terms of prior distribution, embodying the sparsity penalty term, and the sequential Monte Carlo algorithm that we use to obtain samples of the posterior distribution of interest.

\subsection{The image space}

In our model an image is a collection of pre--defined shapes, or objects. The number of objects is not known a priori, nor are the object types; therefore our aim is to make inference on the following unknown:
\begin{equation}
	F=({N_S},T_{1:{N_S}}, \theta_{1:{N_S}})
	\label{source_model}
\end{equation}
where $N_S=0, \dots, N_{\max}$ represents the number of sources, $T_{1:N_S}=(T_1, \dots, T_{N_S})$ the source types and $\theta_{1:{N_S}}=(\theta_1, \dots, \theta_{N_S})$ the source parameters for each source.  

We call $\mathcal{T}$ the set of source types $\mathcal{T} = \{C, E, L \}$; as explained in the previous section, different source types are characterized by different numbers of parameters: circles are defined by four parameters, ellipses by six parameters and loops by seven parameters. As a consequence, and for convenience, we define the single--source parameter space as the union of these fixed--dimensional spaces

\begin{equation}
	\Theta = \mathbb{R}^4 \cup \mathbb{R}^5 \cup \mathbb{R}^7
\end{equation}

The most complex source, the loop, is defined by seven parameters, i.e. $\theta_s=(x_s, y_s, r_s, \phi_s, \alpha_s, \varepsilon_s, \beta_s)$, where $(x_s,y_s) \in \mathbb{R}^2 $ indicates the position of the source, $r_s \in \mathbb{R}_{> 0}$ the full width at half maximum (FWHM), $\phi_s \in \mathbb{R}_{> 0}$ the flux, $\alpha_s \in [0,360\degree]$ the rotation angle, $\varepsilon_s \in [0,1]$ the eccentricity and $\beta_s \in [-180\degree,180\degree]$ the curvature. The circle only has the first four parameters, the ellipse the first six.

Since the number of sources is not known {\it a priori}, we need to consider a variable-dimension model, i.e., the state space $\mathcal{F}$ of the sources is defined as follows:

\begin{equation}
\label{eq:space}
\mathcal{F} \coloneqq \displaystyle\bigcup_{s=1}^{N_S} \{s\}\times \mathcal{T}^s \times \Theta^s \,\,\, ,
\end{equation}
where $\Theta^s$ is the Cartesian product of $\Theta$, $s$ times (and analogously for $\mathcal{T}^s$).

\subsection{The statistical model}\label{subsec:prior}

In a Bayesian setting, we aim at characterizing the posterior distribution 

\begin{equation}
\pi(f|\nu)\propto \pi(f) \pi(\nu|f), 
\end{equation} 
where: $\nu$ and $f$ indicate the measured visibilities and the image parameters, respectively; $\pi(f)$ is the prior distribution, which represents the information that is known before any measurements; and $\pi(\nu|f)$ is the likelihood function.  \\

\subsubsection{Prior} 
Our choice of the prior probability density  $\pi(f)$ reflects basic expectations that are based on available
knowledge about the solar flare geometries.
In our prior distribution we consider the number of sources, the
source types and the parameter values of each source and we assume that they are all independent. Thus, the prior distribution $\pi$ can be written as follows
\begin{equation}
\label{eq:pr}
\pi(f)= \pi(N_S, T_{1:N_S}, \theta_{1:N_S})= \rho_1(N_S) \, \prod_{s=1}^{N_S} \rho_2(T_s) \, \rho_3(\theta_s) \,\,\, ,
\end{equation}

where $\rho_j$, with $j \in \{1,2,3\}$, indicates the distribution of the number of sources, the shape of the sources, and the parameters, respectively. 

We assume that the number of sources in an image is  Poisson distributed with mean $\lambda>0$, 
\begin{equation}
\label{eq:pr_num}
\rho_1(N_S)=\frac{e^{-\lambda}\lambda^{N_S}}{N_S!}.
\end{equation} 
In the simulations below we will be using $\lambda \in [1, 4]$, and limit the maximum number of sources to 5, for computational reasons; in \cref{sec:experiments} we study the influence of the parameter $\lambda$  in the reconstruction.

We adopt the following distribution for the type of the sources
\begin{equation}
\label{eq:pr_sh}
\rho_2(T_s)=
\begin{cases}
p_C \quad &\text{if } T_s \text{ is a circle} \\
p_E \quad &\text{if } T_s \text{ is an ellipse} \\
p_L \quad &\text{if } T_s \text{ is a loop},
\end{cases}
\end{equation}
where $p_C,p_E,p_L \in [0,1]$. In general the morphology of solar flare hard X-ray sources is dependent on the photon energy observed:
 images at lower energies $\lapprox 15$~keV generally are dominated by extended loop-like (or elliptical) structures, while images at higher photon energies are more likely to include compact sources (possibly exclusively so), see \cite{xu2008rhessi} and \cite{brown1972directivity}. Thus, in our simulations, for low energy levels we set $(p_C,p_E,p_L)=\big(\frac{1}{4}, \frac{1}{4}, \frac{1}{2}\big)$, while for high energy levels we use $(p_C,p_E,p_L)=\big(\frac{1}{2}, \frac{1}{4}, \frac{1}{4}\big)$. However, we observed that moderate changes to these values do not have an impact on the result.
 
Given the $s$-th source, the associated source parameters have uniform distributions 
\begin{equation}
\label{eq:pr_par}
\begin{aligned}
(x_s,y_s) &\sim \mathcal{U}\bigl([ x_m-FOV/2,x_m+FOV/2], [y_m-FOV/2,y_m+FOV/2]\bigr) \\
r_s &\sim \mathcal{U}\biggl(\Bigl[R_{\min},R_{\max}\Bigr]\biggr) \\
\phi_s &\sim \mathcal{U}\bigl([0,V_{\max}]\bigr) \\
\varepsilon_s &\sim \mathcal{U}\bigl([0,1]\bigr) \\
\alpha_s &\sim \mathcal{U}\bigl([0,360\degree]\bigr) \\
\beta_s &\sim \mathcal{U}\bigl([-180\degree,180\degree]\bigr) ,
\end{aligned}
\end{equation}
where $FOV$ is the \textit{field of view},  $(x_m, y_m)$ is the center of the image, $R_{\min}$ and $R_{\max}$ are a lower and an upper bound for the source FWHM, and $V_{\max}=\max(\text{Re}(V))$ represents  the maximum of the real  component of the visibilities; in the simulations below we will be using for convenience $R_{\min} = 0$ and $R_{\max} = FOV/3$. 

\subsubsection{Likelihood} We make the standard assumption that the visibilities are affected by additive Gaussian noise; although this may not be an accurate model, particularly for low photon counts, it is widely used in literature for lack of a better one. Therefore,  the likelihood is given by

\begin{equation}
\pi(\nu|f) \propto \exp  \Biggl(- \mathlarger{\mathlarger{‎‎\sum}}_j \frac{\bigl(\nu-V(f)\bigr)_j^2}{2 \sigma_j^2} \Biggr)
\label{eq:likelihood}
\end{equation}

where $V(f)$ indicates the forward operator, which maps the image parameters $f$ onto its exact visibilities, and  $\sigma_j$ the noise standard deviation. Since uncertainty on the measured visibilities comes mainly from the statistical error on the X--ray counts, plus a known systematic error, we can assume knowledge of the standard deviations $\{\sigma_j\}_j$ \cite{aschwanden2003reconstruction}. In practice, these values are regularly provided together with the visibilities themselves.

\subsubsection{Parallel with regularization theory}

We now provide an informal parallel between the Bayesian model just described and the more well-known regularization approach to imaging.

The regularization approach consists in minimizing a functional with the following form

\[
\|\nu - V (f )\|_2^2 + \lambda R(f) 
\]
where the first term represents the data fitting term (in this case with the $\ell^2$--norm), $\lambda$ is the regularization parameter and $R(f)$ is the regularization term, which is responsible of promoting sparsity when sparse solutions are seeked.

It is well known that this functional can be interpreted, up to a normalizing constant, as the negative logarithm of a posterior distribution, where the likelihood is of the form \cref{eq:likelihood}; we now discuss what sort of regularization term is implied by our choice of the prior distribution.

\begin{figure}
\begin{center}
	\includegraphics[width=5cm]{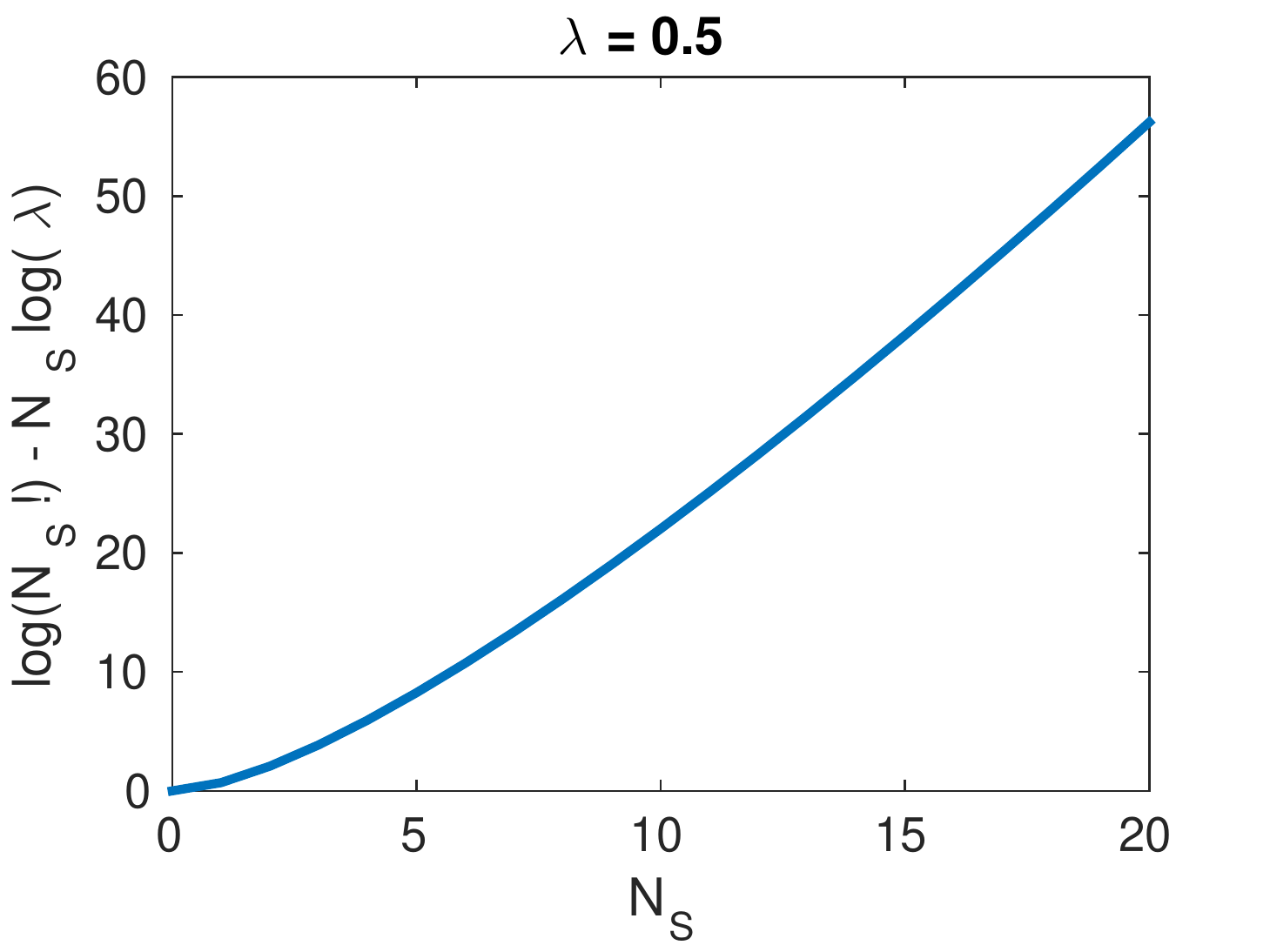}
	\includegraphics[width=5cm]{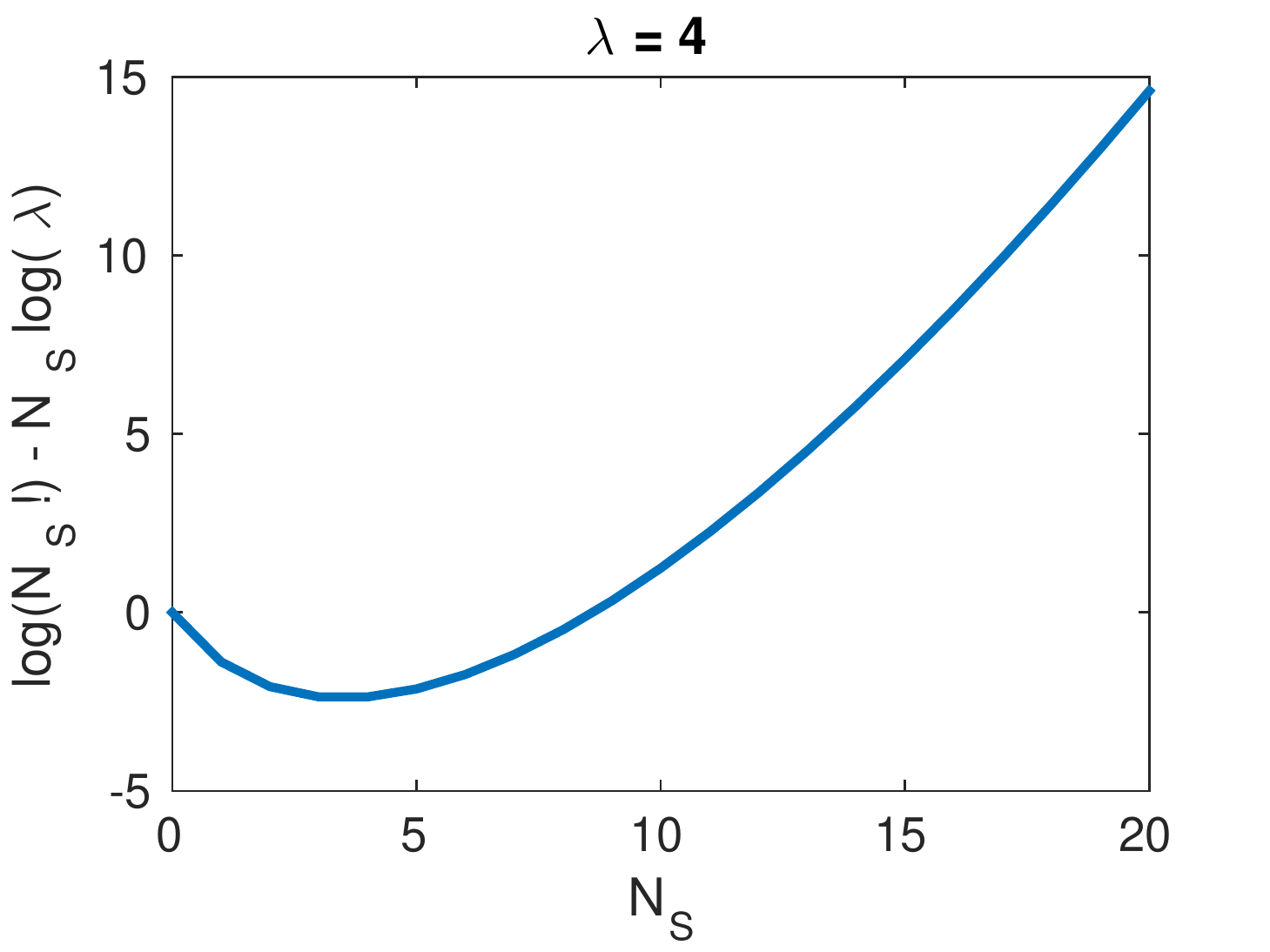}
\end{center}
\caption{The regularization term corresponding to our Poisson prior, for two different values of $\lambda$.}
\label{fig:reg_lambda}
\end{figure}

We first observe that all parameters, except for the number of sources, are assigned a uniform prior distribution, which corresponds to a specific choice of the set of basis elements but with no explicit regularization term. As far as the number of sources is concerned, the Poisson distribution with mean $\lambda$ corresponds to the following regularization term

\begin{eqnarray} 
R(f) & = & - \log \bigg(\frac{e^{- \lambda} \lambda^{N_S}}{N_S!}\bigg) = \lambda + \log(N_S!) - N_S \log(\lambda) \nonumber \\
 &\simeq &\log(N_S!) - N_S \log(\lambda) ~~~.
 \end{eqnarray}
 
which, for small values of $\lambda$ (less than $4$, the maximum value used below) and reasonable values of $N_S$, grows almost linearly with $N_S > \lambda$ (see \cref{fig:reg_lambda}). Therefore, our regularization term is counting the number of sources, i.e. the number of non--zero components in our basis expansion, exceeding the expected value $\lambda$; in this sense, for $\lambda<1$ it correponds to an $\ell^0$--regularization term. We also remark, however, that the proposed approach is not an optimization approach, but rather a Monte Carlo sampling approach, that aims at characterizing the whole posterior distribution rather than just finding its mode; in this way, our method provides more rich information about the unknown that is not typically provided in regularization.

\subsection{Sequential Monte Carlo samplers}\label{sec:ASMC}

\subsubsection{Monte Carlo sampling}
In order to approximate the posterior distribution we employ a sequential Monte Carlo method \cite{del2006sequential}. The general goal of Monte Carlo sampling is to obtain a set of $N_P$ sample points $f^{(p)}$, also called \textit{particles}, with corresponding weights $w^{(p)}$, such that the following approximation holds for a general enough class of functions $h(\cdot)$:
\begin{equation}
	\sum_{p=1}^{N_P} w^{(p)} h(f^{(p)}) \simeq \int h(f) \pi(f) df ~~~;
	\label{eq:MC}
\end{equation}
the relevance of \cref{eq:MC} is manifest, since all the moments of the distribution can be obtained by simple choices of $h(f)$, such as $h(f) = f$ for the mean; in addition, \cref{eq:MC} can be interpreted as providing an approximation of the distribution $\pi(f)$:
\begin{equation}
	\sum_{p=1}^{N_P} w^{(p)} \delta(f- f^{(p)}) \simeq \pi(f) ~~~,
\end{equation}
being $\delta(\cdot - \cdot)$ the Dirac delta function.
Ideally, one would like to have uniformly weighted samples, as small--weight samples provide little contribution to the sum; therefore a standard measure of the quality of the sample set $W = \{ w^{(p)}\}_{p=1,...,N_P}$ is given by the so--called Effective Sample Size, defined as

$$\text{ESS}(W) \coloneqq \Biggl( \displaystyle \sum_{p=1}^{N_P} \Bigl(w^{(p)} \Bigr)^2\Biggr)^{-1} ~~~;$$
the ESS ranges between $1$, for bad choices of the sample points, and $N_P$, the optimal value attained when all samples carry the same probability mass.

\subsubsection{Sequential Monte Carlo samplers}\label{subsec:SMCsamplers}
\textit{Sequential} Monte Carlo samplers produce such weighted sample set through a smooth iterative procedure, and are particularly useful when the target distribution is complex and/or highly concentrated in few areas of the state--space. The derivation of SMC samplers is relatively complex and will not be reported here; instead, we provide here the minimal amount of details that should allow reproduction of the results.

The key point is to consider a sequence of distributions $\{\pi_i\}_{i=1}^{N_I}$, with $N_I$ the number of iterations, such that $\pi_1(f)=\pi(f)$ is the prior distribution and  $\pi_{N_I}(f)=\pi(f|\nu)$ is the posterior distribution; a natural choice is given by 
\begin{equation}
\label{eq:pi}
\pi_i(f) \propto \pi(\nu|f)^{\gamma_i}\pi(f),
\end{equation}
where $\gamma_i \in [0,1]$ for $i=1,\dots,N_I$, $\gamma_{i+1}>\gamma_i$, $\gamma_1=0$ and $\gamma_{N_I}=1$. Our SMC sampler starts by drawing an i.i.d. sample set $\{f^{(p)}_1\}_{p=1}^{N_P}$ from $\pi_1(f)$, and then produces, at iteration $i$, an approximation of $\pi_i(f)$ with the sample points $f^{(p)}_i$ and the corresponding weights $w^{(p)}_i$ through the following procedure: 
\begin{itemize} 
	\item[i] the sample $f^{(p)}_i$ is drawn from the $\pi_{i}$--invariant Markov kernel $K_i(f | f^{(p)}_{i-1})$; technically, these Markov kernels are built as a composition of vanilla Metropolis--Hastings \cite{chib1995understanding} and reversible jump \cite{green1995reversible} moves; in practice, the particle at the next iteration is obtained by perturbing the current particle in such a way that the new state is neither too different nor too similar to the current one; the choice of these kernels is crucial in determining the effectiveness of the procedure, and will be discussed more in detail in the next subsections;
	\item[ii] the weights are given by 
	\begin{equation}
	w^{(p)}_i \propto w^{(p)}_{i-1} \/ \frac{\pi_{i}\bigl(f^{(p)}_{i-1}\bigr)}{\pi_{i-1}\bigl(f^{(p)}_{i-1}\bigr)}~~~, 
	\label{eq:w}
	\end{equation}	
	i.e. the right side is computed and then weights get normalized; in order to obtain good performances, it is important that the sequence is smooth, i.e. that $\pi_i \simeq \pi_{i-1}$. We control such smoothness by increasing adaptively the exponent: we first attempt $\gamma_{i+1}= \gamma_i+\delta_{\max}$ and compute the ratio 
	$$\frac{\text{ESS}\bigl(W_{i+1}\bigr)}{\text{ESS}\bigl(W_{i}\bigr)}~~~; $$ 
	if the value of this ratio is in the interval $[0.90, 0.99]$ we confirm the choice $\gamma_{i+1}$, otherwise we proceed with $\gamma_{i+1} = \gamma_i + \delta$ by bisection on $\delta$ until the ESS ratio falls in the prescribed interval. In the simulations below we use $\delta_{\max} = 0.1$. Notice that the particle weights \textit{do not depend} on the current particle, but only on the previous particle; this makes this adaptive calculation very cheap.
	\item[iii] finally, we notice that the Effective Sample Size naturally diminishes as the iterations proceed: when its value gets below $N_P/2$, we perform a resampling step \cite{douc2005comparison}, where the particles with high weights are copied multiple times and the ones with low weights are discarded. At the end of the resampling, all the new particles are assigned equal weights. 

\end{itemize} 

Importantly, since in our model the likelihood is Gaussian, every distribution of the sequence \cref{eq:pi} can be interpreted as the posterior distribution corresponding to a different (decreasing) scaling factor of the noise (co)variance. Although in our application the noise on the visibilities is assumed to be known, in the simulations below we exploit this fact to investigate what happens when such estimate is not correct.






\subsubsection{Transition kernels}\label{sec:trans}

As discussed in the previous subsection, the generation of new samples from the existing ones is the key point of the algorithm, as it allows exploration of the parameter space. Due to the relative complexity of the state space, which is a variable--dimension model with an unknown number of objects of unknown type, we had to construct a relatively complex transition kernel. Specifically, the transition kernel has the following abstract form:
\begin{equation}
K_i(f_{i}|f_{i-1})= K_i^{\text{u}}(f_i| f''') K_i^{\text{sm}}(f'''| f'') K_i^{\text{c}}(f''| f') K_i^{\text{bd}}(f'|f_{i-1} )~~~.
\label{multiple_transition}
\end{equation}
Namely, the whole transition kernel is given by the composition of several $\pi_i$--invariant kernels implementing different types of move, specifically: $K_i^{\text{bd}}(\cdot, \cdot)$ implements a birth/death move, $K_i^{\text{c}}(\cdot, \cdot)$ implements a change move, $K_i^{\text{sm}}(\cdot, \cdot)$ implements a split/merge move and $K_i^{\text{u}}(\cdot, \cdot)$ implements an update move.

In practice, all these moves are attempted in sequence, starting from the current state $f^{(p)}_{i-1}$ and ending up in the next state $f^{(p)}_{i}$, through several intermediate states $f'$,\dots,$f'''$. In order for each move to be a draw from a $\pi_i$--invariant kernel,  we use the classical Metropolis--Hastings accept/reject, where proposal distributions and acceptance probabilities are defined according to the theory of reversible jumps \cite{green1995reversible}. The general mechanism for moving from $f'$ to $f''$ is as follows. Given the particle $f'$, a new particle $f^*$ is proposed from a proposal distribution $q(\cdot|f')$; notice that this can also be viewed as the random variable $F^*$ being a deterministic function of the current state and of the random quantity $u$: $F^* = g(f',u)$. The proposed particle is then either accepted, with probability

\begin{equation}
\label{eq:acc_rate}
\alpha(f^*,f')=\min \Biggl\{ 1, \frac{\pi_i(f^*)  q(f'|f^*)}{\pi_i(f') q(f^* | f')} \biggl| \frac{\partial g}{\partial(f',u)} \biggr|\Biggr\} ~~~,
\end{equation}
or rejected: in case of acceptance, we set $f'' = f^*$; otherwise, we set $f'' = f'$. We now explain more in detail how these kernels work, by providing the explicit form of the birth/death case, which is the most complex one. \\

\paragraph{Birth and death moves}
Birth and death moves consist in adding/deleting one source to/from the image. We realize this by using the following proposal
\begin{equation}
\label{eq:proposal}
\begin{aligned}
q(f^*|f') & = & \frac{2}{3} \delta(N_S^*, N_S'+1) \; \rho_2(T_{N_S'+1}) \rho_3(\theta^{N_S'+1}) \prod_{s=1}^{N_S'} \delta(\theta'_{s}-\theta^*_{s}) + \\ 
& & \frac{1}{3} \delta(N_S^*, N_S'-1)\; \frac{1}{N_S'}\sum_{t=1}^{N_S'} \prod_{s=1}^{N_S'-1} \delta (\theta'_{a(s,t)}-\theta^*_{s}),
\end{aligned}
\end{equation}
where $\delta(\cdot, \cdot)$ represents the Kronecker delta function.
In the first row a birth move is proposed with probability $2/3$, and the new ($N'+1$)--th source type and the new source parameters are drawn from their (uniform) prior distributions, collectively denoted here as $\rho_3(\theta^{N_S'+1})$; the other sources remain identical, as represented here by the Dirac delta functions. In the second row a death move is attempted with probability $1/3$, and one of the existing sources is removed at random. This entails a re--organization of the indices: assume, for instance, that the starting configuration $f'$ has three sources, and source 2 disappears; then source 2 in $f^*$ will evolve from source 3 of $f'$: formally, $a(t,s)$ is the new index of source $s$ when the $t$--th source has disappeared, $a(t,s) = s$ for $s<t$, $a(t,s)=s-1$ for $s>t$.\\

\paragraph{The change move}
Change moves are attempted  at each iteration, and consist in proposing the transformation of a circle into an ellipse (and vice versa) and an ellipse into a loop (and vice versa). We do not consider the transition from a circle to a loop (and vice versa) because we retain only ``smooth'' transformations. From the mathematical perspective, these moves resemble birth/death moves, but they foresee the appearance/disappearance of just one or two parameters, rather than of a whole source; like in the birth/death move, the new parameters are drawn from the prior.\\

\paragraph{The merge and split moves}
The merge and split moves are attempted at each iteration if the current state has at least one or two ellipses, respectively. The merge move consists of combining two close ellipses to create a new one. The proposal here is deterministic: given two ellipses, the proposed source is defined by the combination of their parameters, i.e. the position and the rotation angle are the mean of the corresponding parameters of the two ellipses, while the flux, FHWM and the eccentricity are the sum. 
The split move generates two ellipses from a single one. This move is semi-deterministic: the centers of the new ellipses are drawn along the major axis of the old ellipse, at a random distance; the flux is split equally, and the major axis is a random quantity around half the major axis of the old ellipse.\\

\paragraph{The update move}
 The last move is the classical parameter update move of the Metropolis - Hastings algorithm, and is  proposed at each iteration. Each parameter is updated independently, i.e. a separate move is attempted for each parameter.
We highlight that even if the prior distribution of the rotation angle  and the loop angle  are uniform on a bounded set, during the update move we let these parameters evolve freely without any constraints, in order to exploit their periodic nature. \\

Finally, let us observe that ellipses with small eccentricity are not practically different from circles, and loops with small loop angle are not practically different from ellipses. Therefore, at the end of each iteration the source type $T_s$ of each source is updated based on the parameter values, according to the following scheme:
\begin{equation*}\label{eq:class} T_s = 
\begin{cases}
 \text{circle:} &\text{ if }\varepsilon_s\leq 0.1 \\
 \text{ellipse:} &\text{ if } 0.1 \leq \varepsilon_s < 0.4, |\beta_s| \leq 45\degree\\
 &  \text{ and } \varepsilon_s \geq 0.4  \text{ and } |\beta_s| \leq 10\degree \\
  \text{loop:} & \text{otherwise.}
\end{cases}
\end{equation*}

\begin{algorithm}[!ht]
\caption{Adaptive sequential Monte Carlo algorithm (ASMC)}
\label{alg:ASCMC}
\begin{algorithmic}
\STATE{\textit{Initialization of the sample:}}
\FOR{$p=1,\dots, N_P$}
 \STATE{draw $f_1^{(p)}$ from $\pi$}
 \STATE{set $w_1^{(p)} = \frac{1}{N_P}$ }
\ENDFOR
\STATE{Set $i=1$ and $\gamma_1=0$}
\WHILE{$\gamma_i \leq 1$}
\STATE{Increase $i$}
\STATE{\textit{Possible resampling step:}}
\IF{$\text{ESS}(W_{i})\leq {N_S}/2$}
\STATE{Apply the systematic resampling}
\ENDIF
\STATE{\textit{MCMC sampling:}}
\FOR{ $p=1,\dots, N_P$}
\STATE{propose birth/death/split/merge/change moves, then accept/reject}
\STATE{for each parameter, update the value, then accept/reject}
\STATE{compute the weights $	w^{(p)}_i \propto w^{(p)}_{i-1} \/ \frac{\pi_{i}\bigl(f^{(p)}_{i-1}\bigr)}{\pi_{i-1}\bigl(f^{(p)}_{i-1}\bigr)}~~~, 
$}
\ENDFOR 
\STATE{\textit{Normalize weights and compute the effective sample size:}}
\STATE{ compute the $\text{ESS}(W_{i+1})$}
\STATE{\textit{Adaptive determination of the next exponent:}}
\WHILE{$\text{ESS}(W_{i+1})/\text{ESS}(W_{i}) \notin [0.9,0.99]$ }
\STATE modify $\delta$ and set $\gamma_{i+1} = \gamma_i + \delta$
 \STATE recompute weights $w_i^{(p)}$ and $\text{ESS}(W_{i+1})$
\ENDWHILE
 \STATE{Type re--assignment}
\ENDWHILE
\RETURN $\{f_i^{(p)},w_i^{(p)} \}_{i=1,...,N_I}^{p=1,...,N_P }$
\end{algorithmic}
\end{algorithm}

\subsection{Point estimates}
At every iteration we can compute the point estimates for each parameter. 
The number of sources at the $i$--th iteration $(\hat{N}_S)_i$ is given by the mode of the marginal distribution of the number of sources in the sample, i.e.
$$\hat{\pi}_i(N_S=k|\nu)=\displaystyle \sum_{p=1}^{N_P} w_i^{(p)} \delta\biggl(k,(N_S^{(p)})_i\biggr),$$
where  $\bigl(N_S^{(p)}\bigr)_i$ is the number of sources in the $p$-th particle at the iteration $i$.

In the same way, the estimated locations are obtained by finding the local modes of the approximated posterior distribution, conditioned on the estimated number of sources, i.e.
$$\hat{\pi}(c|\nu, (\hat N_S)_i)=\displaystyle \sum_{p=1}^{N_P} w_i^{(p)} \delta\biggl((N_S^{(p)})_i, (\hat{N}_S)_i\biggr) \Biggl( \sum_{s=1}^{\bigl(N_S^{(p)}\bigr)_i} \delta(c,c_{is}^{(p)}) \Biggr),$$
where $c=(x, y)$ and $c_{is}^{(p)}=\bigl(x_{i,s}^{(p)}, y_{i,s}^{(p)}\bigr)$ indicate the center of the $s$-th source in the $p$-th particle at the iteration $i$.

The source types are determined as the modes of the type distributions, conditioned on the estimated source locations and the number of sources. The flux, FWHM, and the eccentricity are estimated by the mean values of the conditional distribution conditioned on the sources location, the type and number of sources. The rotation and loop angles are estimated by taking the mode of their conditional posterior distribution, conditioned on the sources location, the type and number of sources. 

In \cref{alg:ASCMC}, we summarize the proposed adaptive sequential Monte Carlo algorithm (ASMC). 



\section{Numerical results}
\label{sec:experiments}
We now demonstrate the effectiveness of the proposed technique in reconstructing images of solar flares. We present first the results obtained by applying the Bayesian method to simulated data which have been already used for testing and validating imaging methods \cite{massone2009hard,hespelink,CSlink,felix2017compressed}. In order to prove that the method is feasible also on real data, we then proceed with the analysis of four experimental data sets.

Henceforth, we will refer to the proposed method with the acronym ASMC (Adaptive Sequential Monte Carlo).
In all the numerical tests below, unless otherwise specified, in the ASMC algorithm we set  $(p_C,p_E, p_L)=\Big(\frac{1}{2}, \frac{1}{4}, \frac{1}{4}\Big)$ and $\lambda=1$.  As we will see later (\cref{fig:lambda} and \cref{fig:lambda_hist}), in most of the cases, especially when the noise level is low, the value of $\lambda$ does not influence the final result.
We systematically compare the results obtained by our Bayesian approach with those obtained by other alternative methods available in IDL SSW (Solar SoftWare, \cite{freeland1998data}): VIS{\_}CLEAN, VIS{\_}FWDFIT, VIS{\_}CS, and VIS{\_}WV.  We used VIS{\_}CLEAN, VIS{\_}CS, and VIS{\_}WV algorithms with the default parameter values\footnote{Some of the default parameters of RHESSI imaging algorithms can be found at \url{https://hesperia.gsfc.nasa.gov/ssw/hessi/doc/params/hsi_params_image.htm}}, because the number of parameters is high and appropriate tuning requires expert knowledge. 
Regarding VIS{\_}FWDFIT \cite{xu2008rhessi} we set either the loop-keyword when a loop is expected or the multi-keyword when more sources have to be reconstructed. 

All of the numerical experiments were carried out on a computer equipped with an Intel Core i5-4570 CPU @ 3.20 GHz x 4 and 8GB memory.
We notice that the computational cost of the ASMC algorithm is substantially larger than that of the competing algorithms: running times in the numerical tests below ranged from few minutes to some hours. The computational cost is mostly due to the likelihood calculations, whose number grows linearly with the number of parameters generating the image (and the number of iterations, and the number of Monte Carlo samples). Simple configurations, such as two circles, can therefore be obtained in much shorter times than complex configurations.

%


\subsection{Simulated data}\label{sect:simul}
\begin{figure}[htbp]
  \centering
\includegraphics[height=3.5cm]{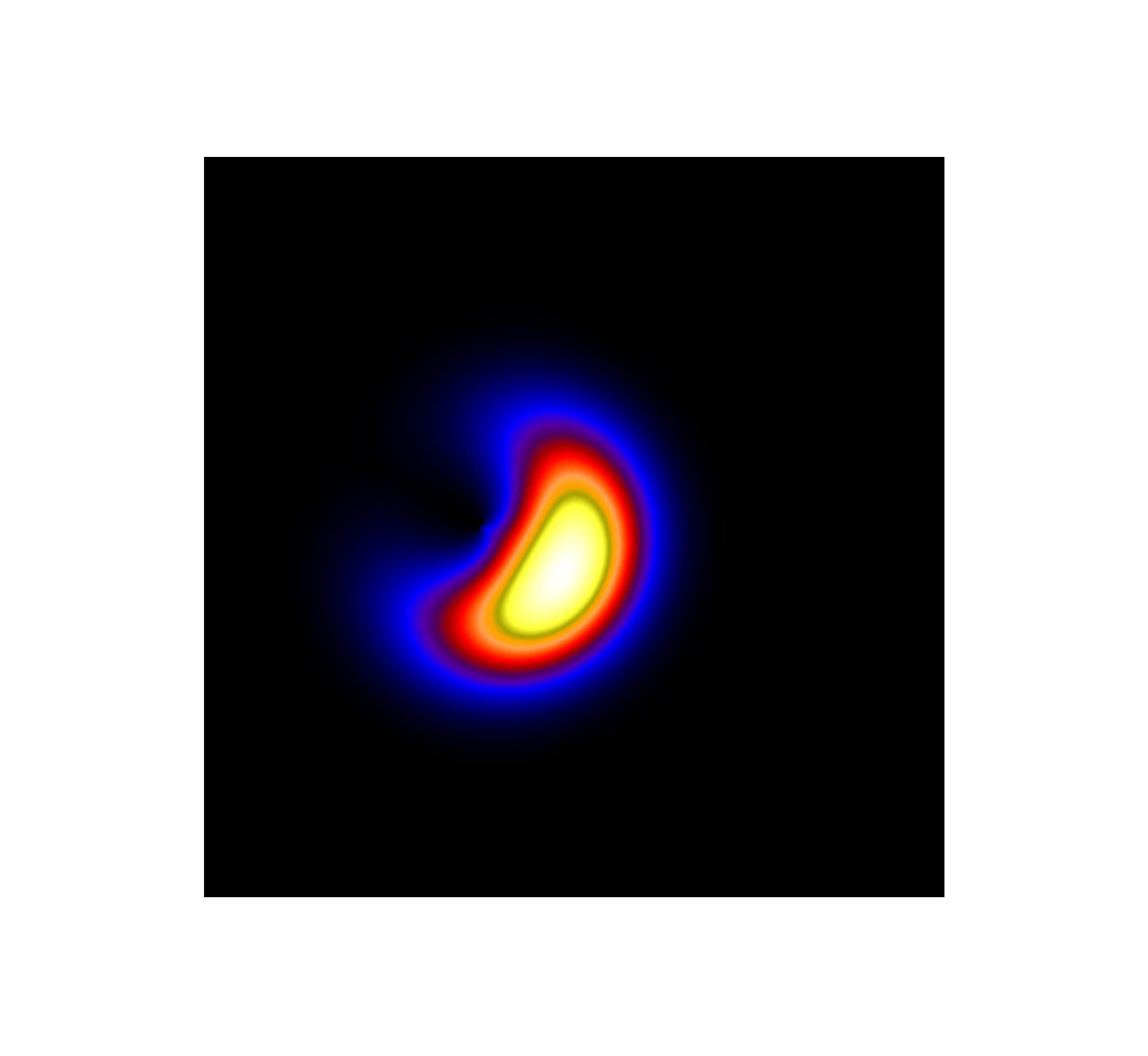}
\includegraphics[height=3.5cm]{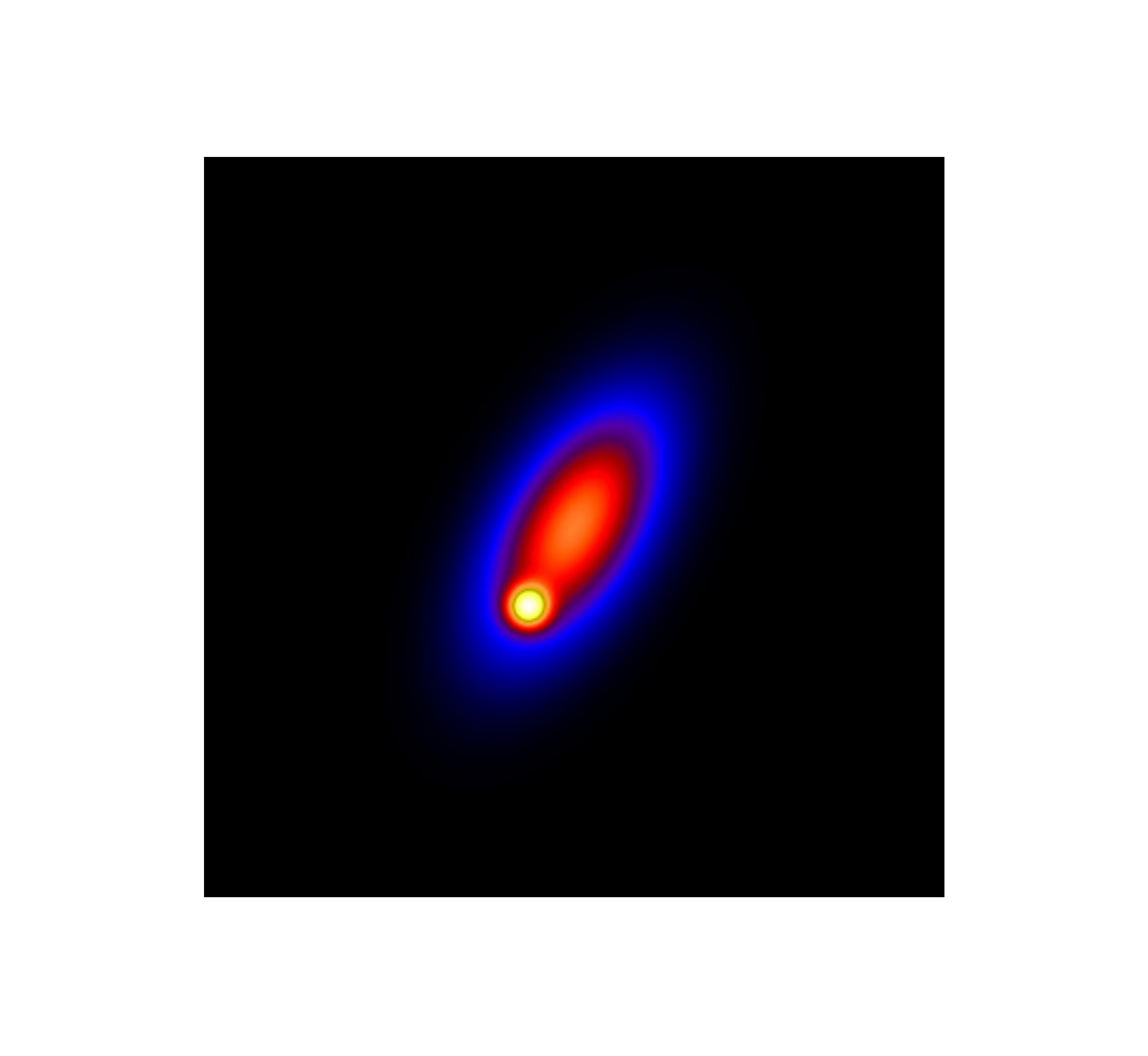}
\includegraphics[height=3.5cm]{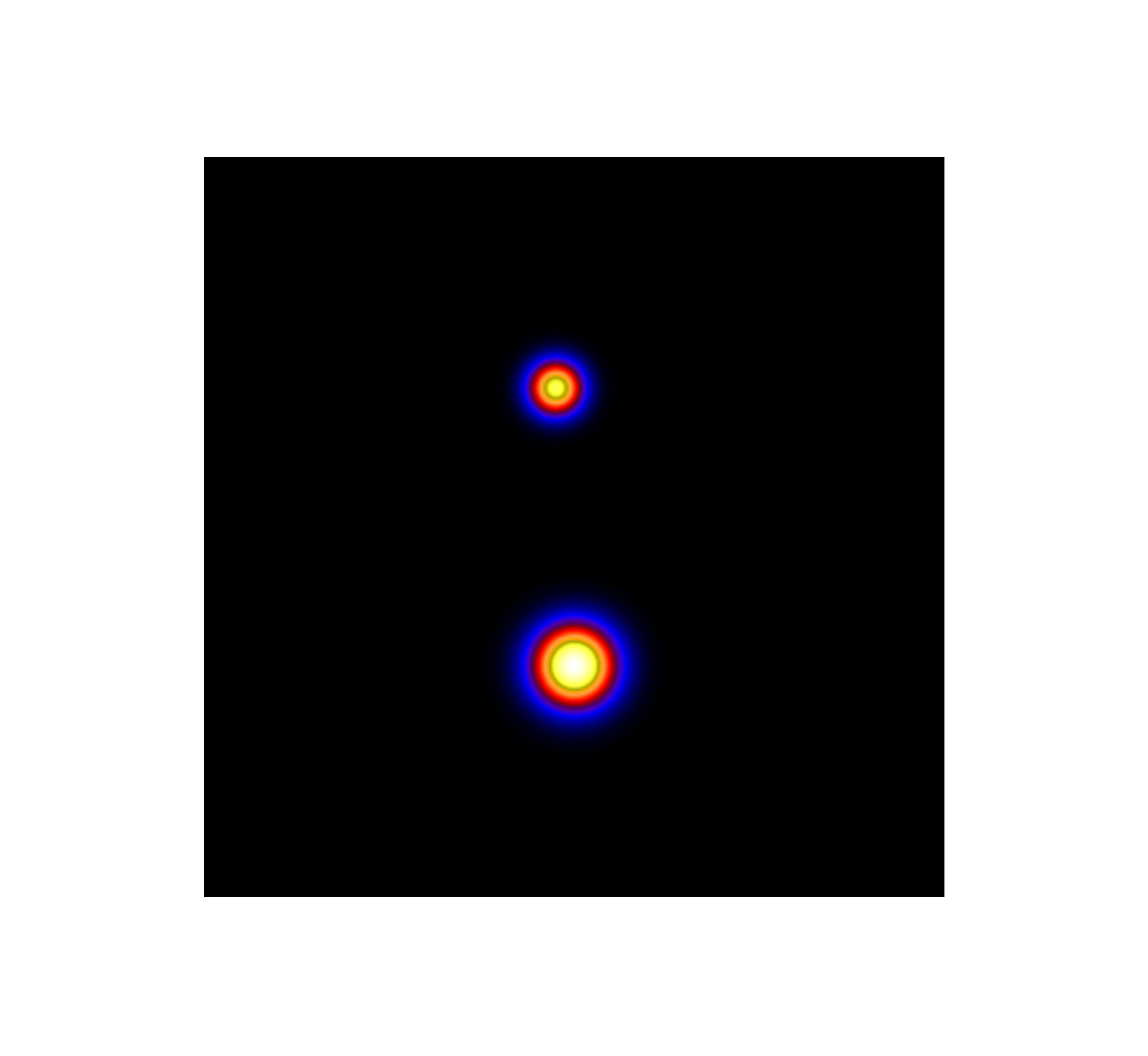}
\includegraphics[height=3.5cm]{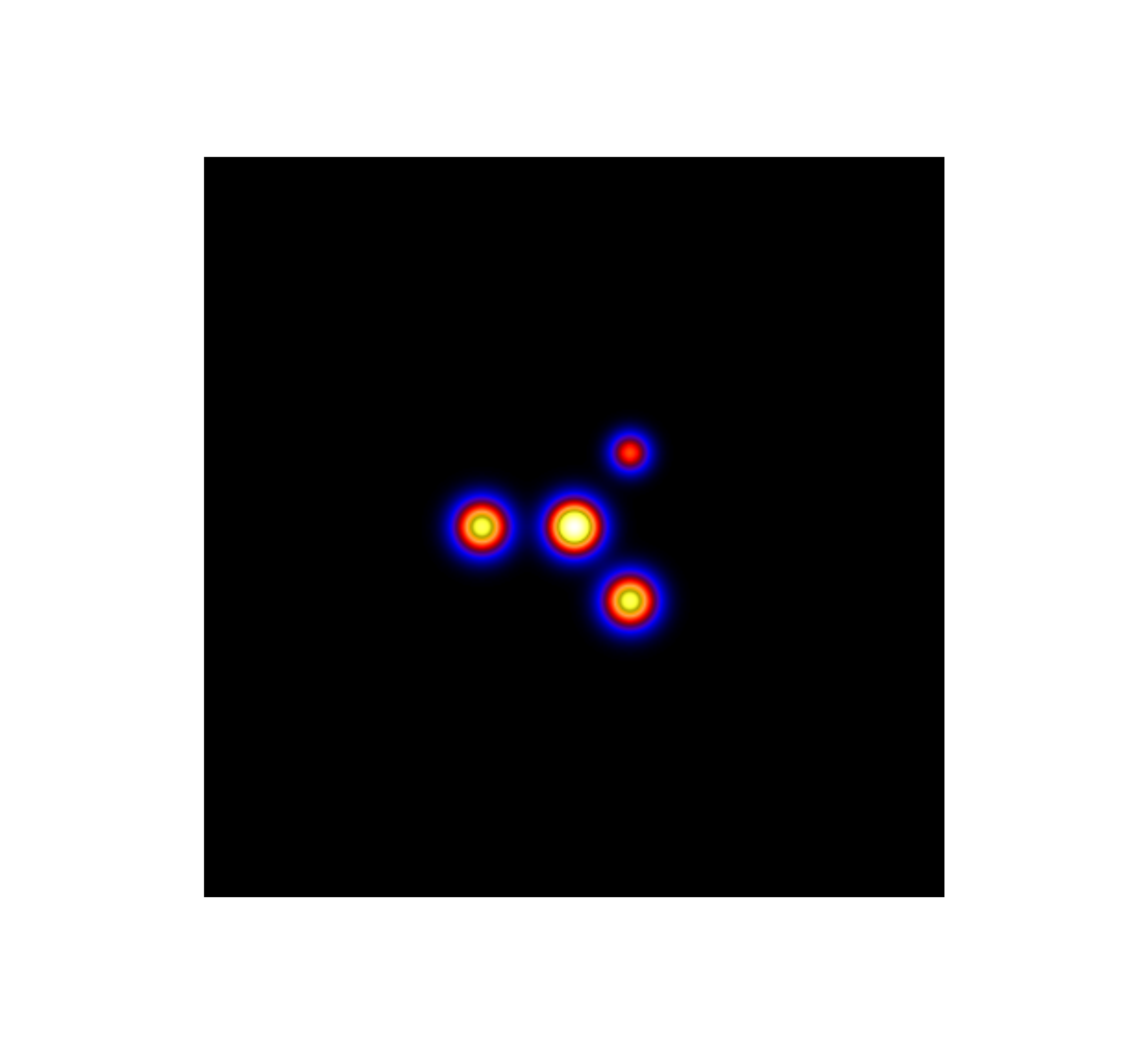}
  \caption{Ground truth images: loop (S1L), ellipse with circle (S1E1C), two circles (S2C), four circles (S4C). }
  \label{fig:original}
\end{figure}
 
In this section we use a well--known simulated data set to test the proposed Bayesian method. The data set comprises several sets of synthetic visibilities, generated from four underlying source configurations, that have been constructed in order to mimick four real solar flares that happened between 2002 and 2005. The four configurations are characterized by different degrees of complexity: one single loop (S1L), one ellipse with a small circular source (S1E1C), two circles (S2C) and four circles (S4C); see  \cref{fig:original} for the true configurations. 
For each flaring configuration, two different synthetic calibrated event lists mimicking a RHESSI acquisition were at our disposal, corresponding to two different levels of statistics (an average of 1000 counts per detector for the low level -- high noise, 100000 for the high level -- low noise). 


\subsection{Reconstructions}

\begin{figure}[h!]
\begin{minipage}{.11\textwidth}
\centering
VIS{\_} CLEAN
\end{minipage}
\begin{minipage}{.88\textwidth}
 \includegraphics[height=3.7cm]{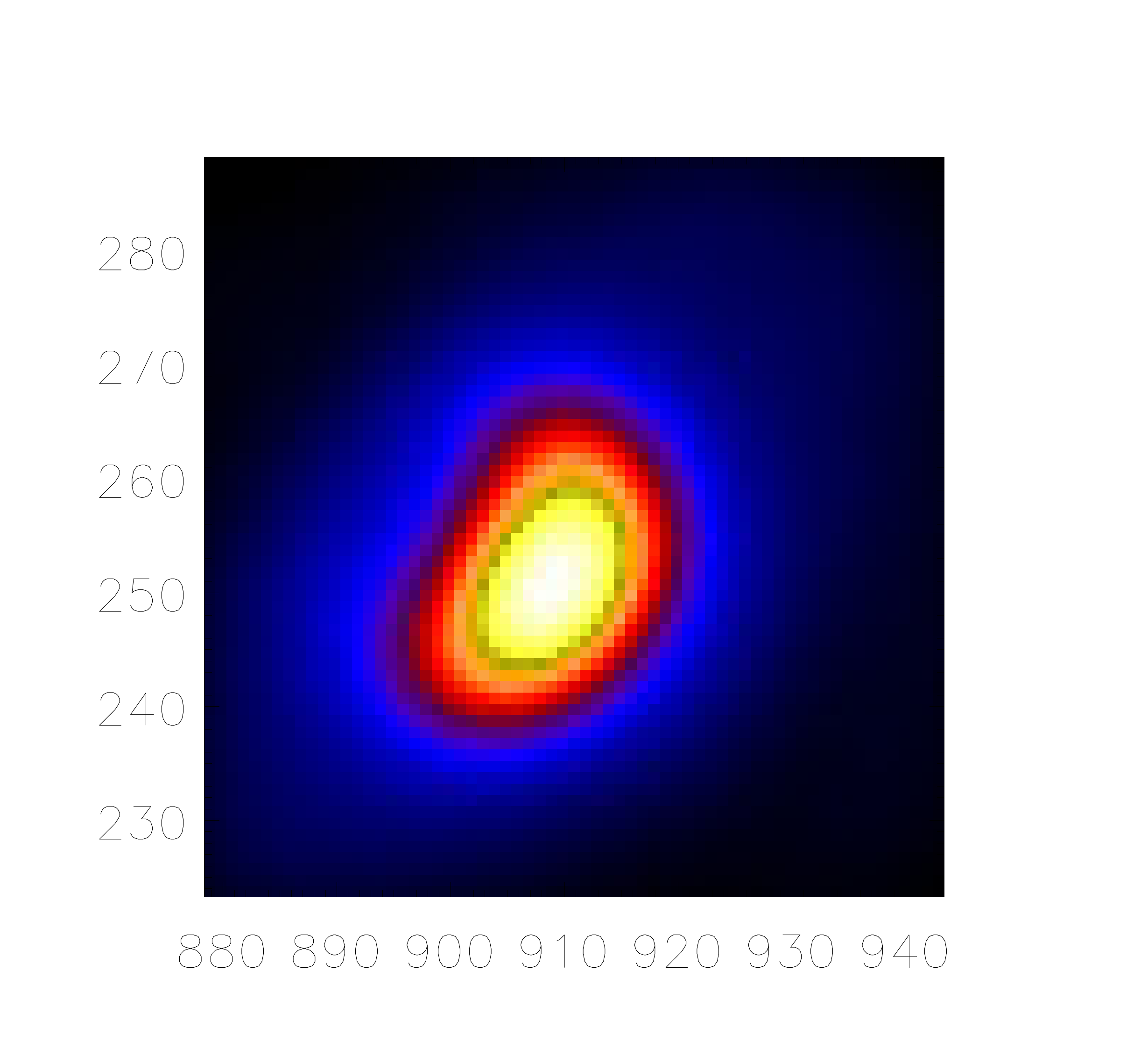}\hspace{-1cm}
    \includegraphics[height=3.7cm]{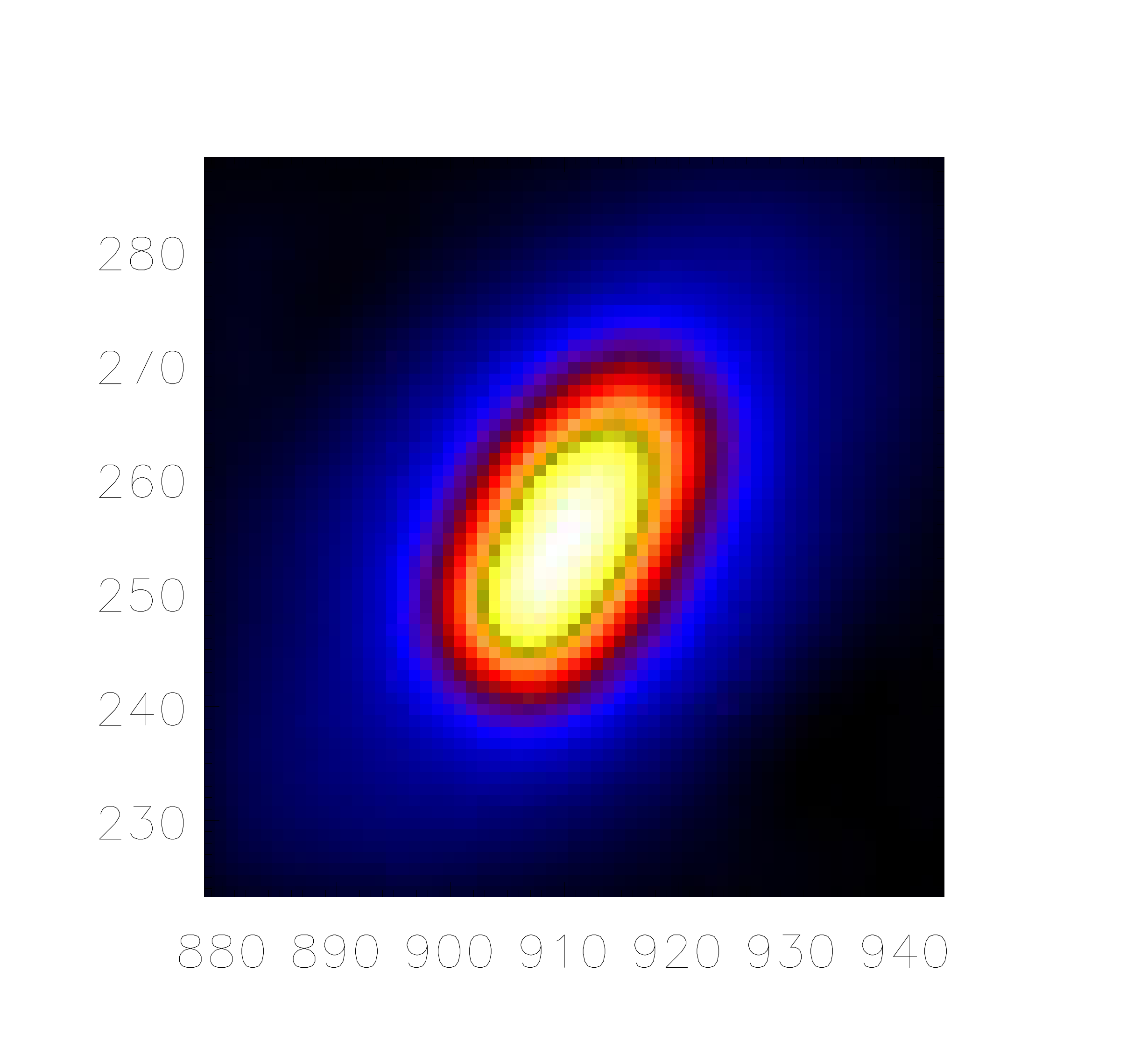}\hspace{-1cm}
    \includegraphics[height=3.7cm]{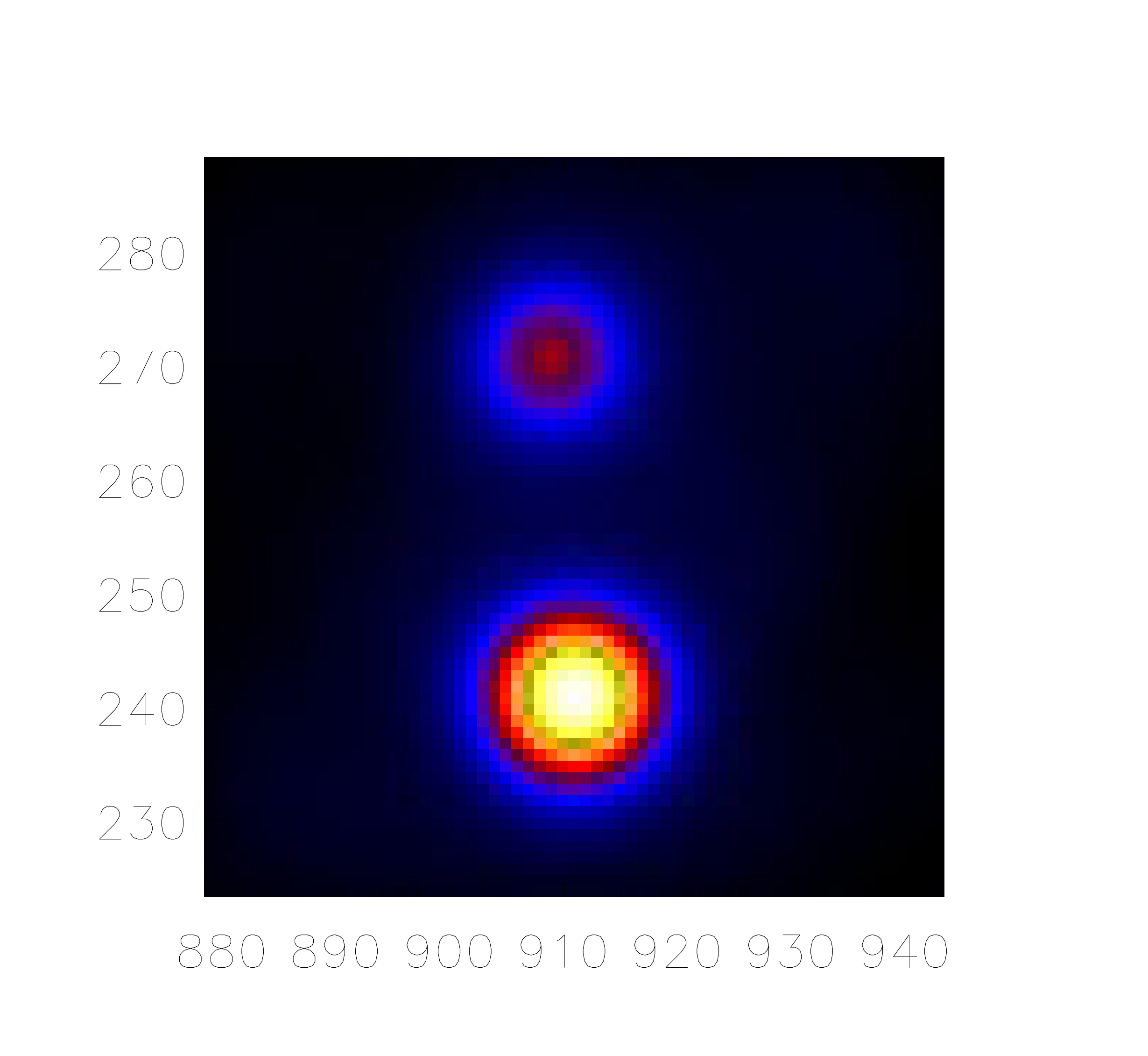}\hspace{-1cm}
    \includegraphics[height=3.7cm]{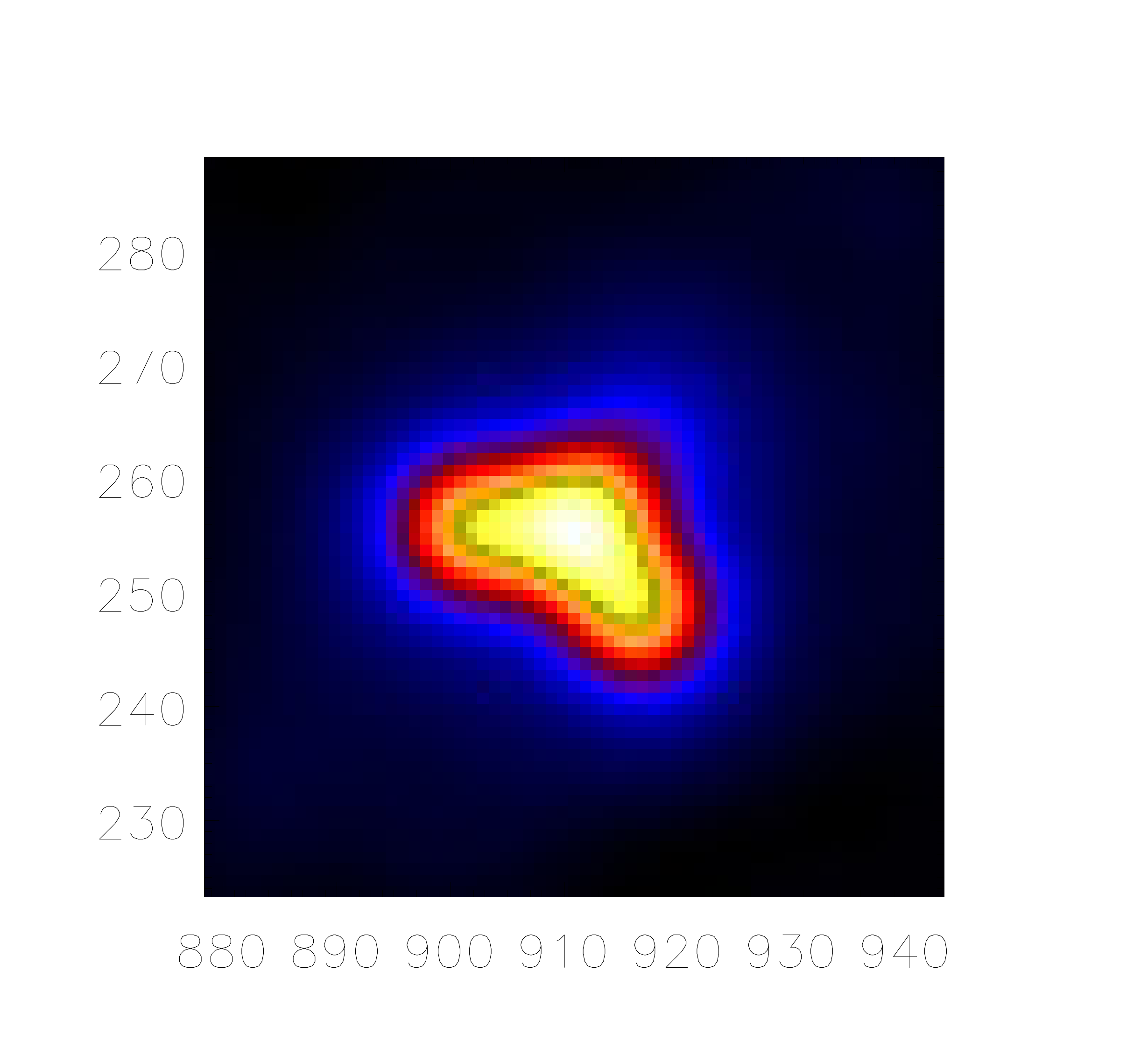}\\ \vspace{-1cm}
    \end{minipage}
    \begin{minipage}{.11\textwidth}
\centering
VIS\texttt{\_} FWDFIT
\end{minipage}
\begin{minipage}{.88\textwidth}
 \includegraphics[height=3.7cm]{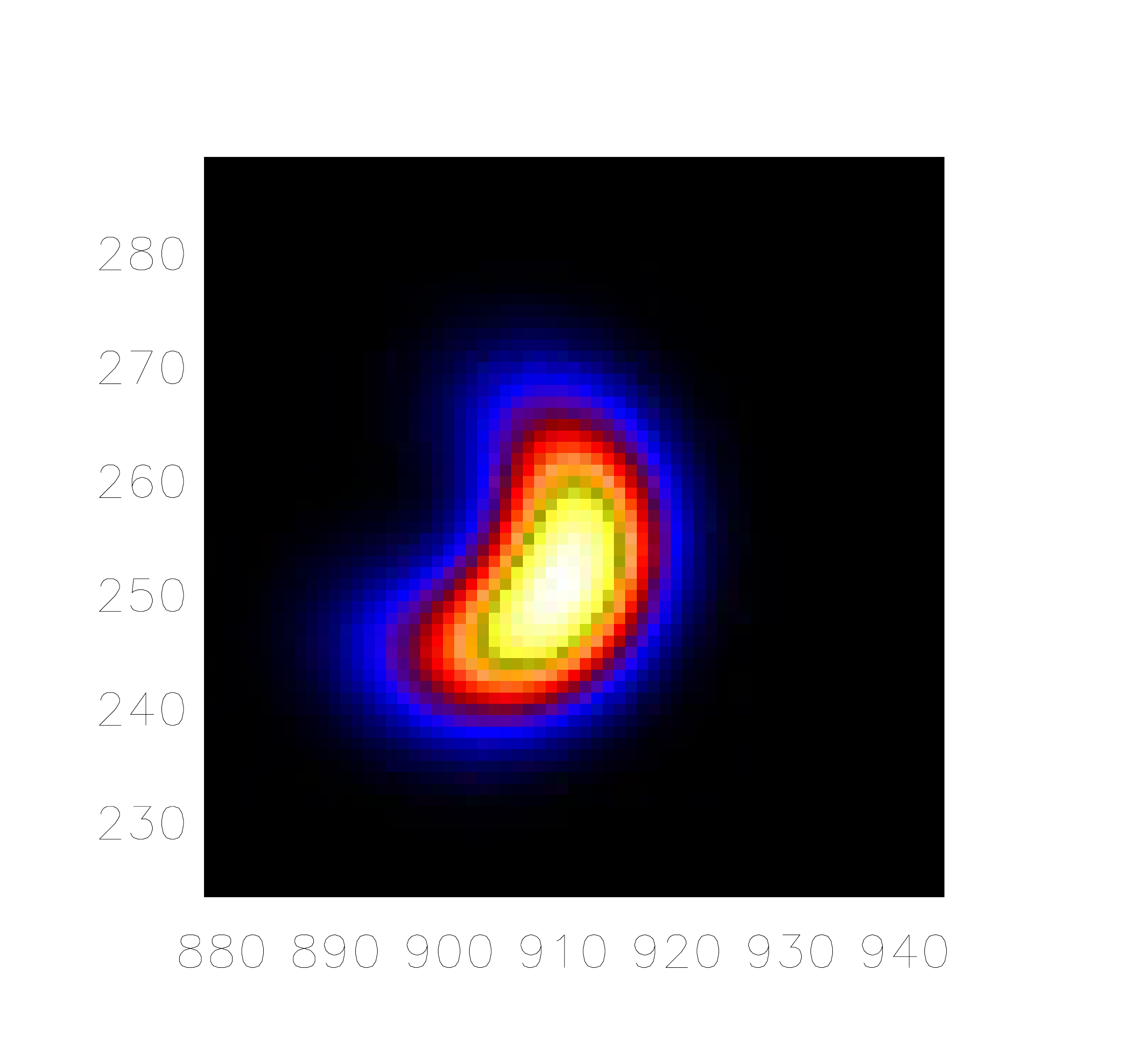}\hspace{-1cm}
    \includegraphics[height=3.7cm]{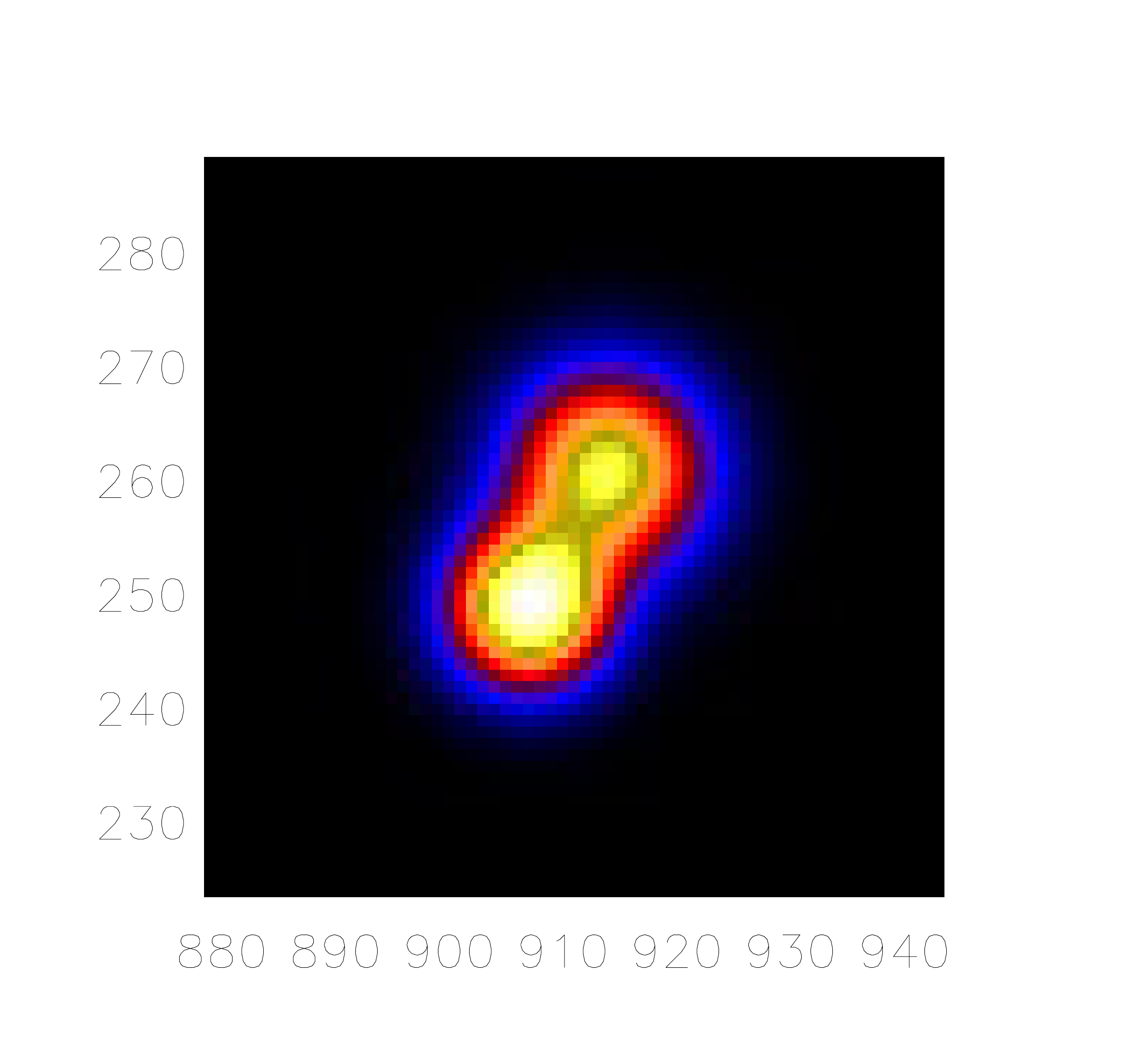}\hspace{-1cm}
    \includegraphics[height=3.7cm]{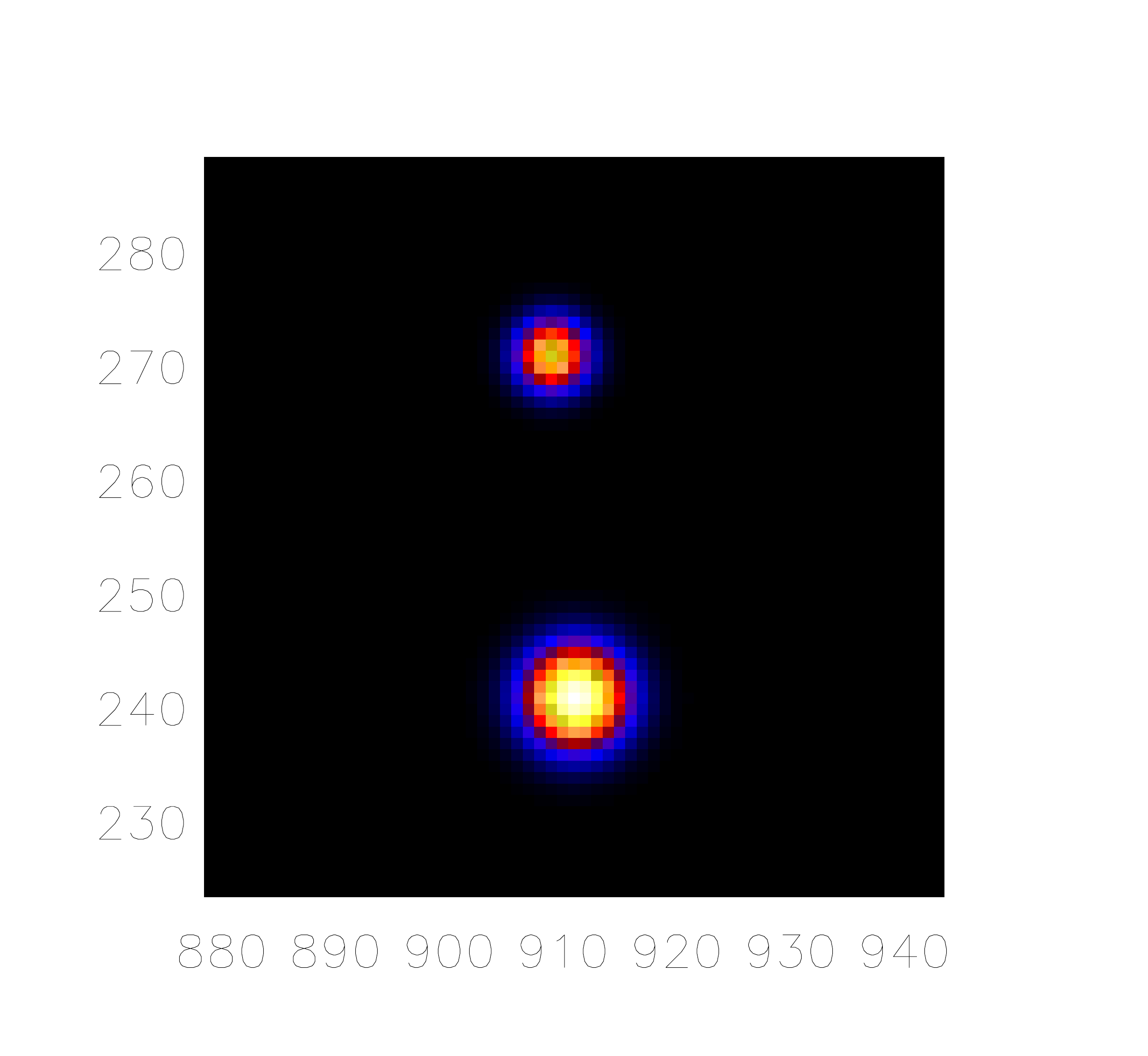}\hspace{-1cm}
    \includegraphics[height=3.7cm]{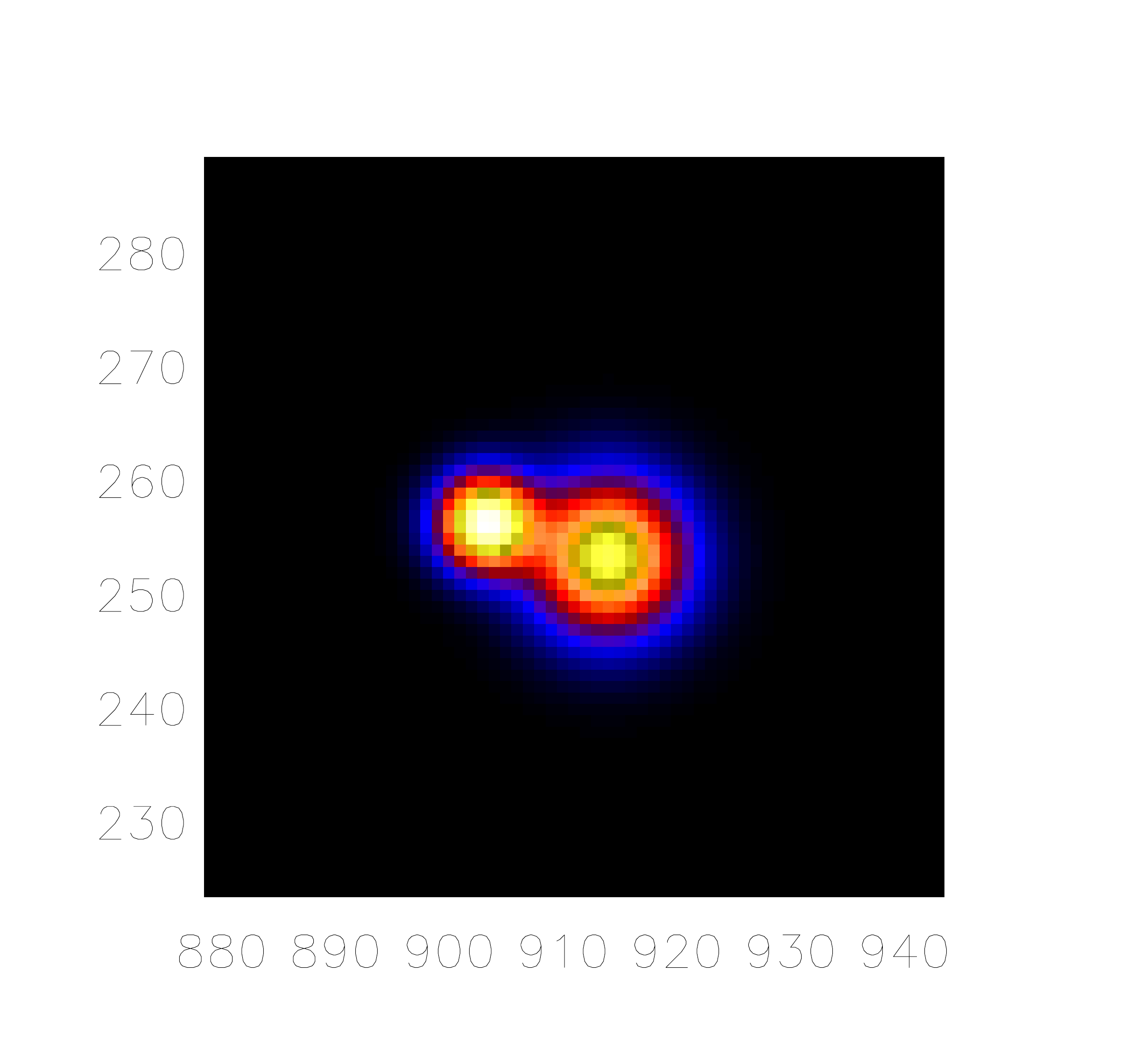}\\ \vspace{-1cm}
    \end{minipage}
        \begin{minipage}{.11\textwidth}
\centering
VIS\texttt{\_}CS
\end{minipage}
\begin{minipage}{.88\textwidth}
 \includegraphics[height=3.7cm]{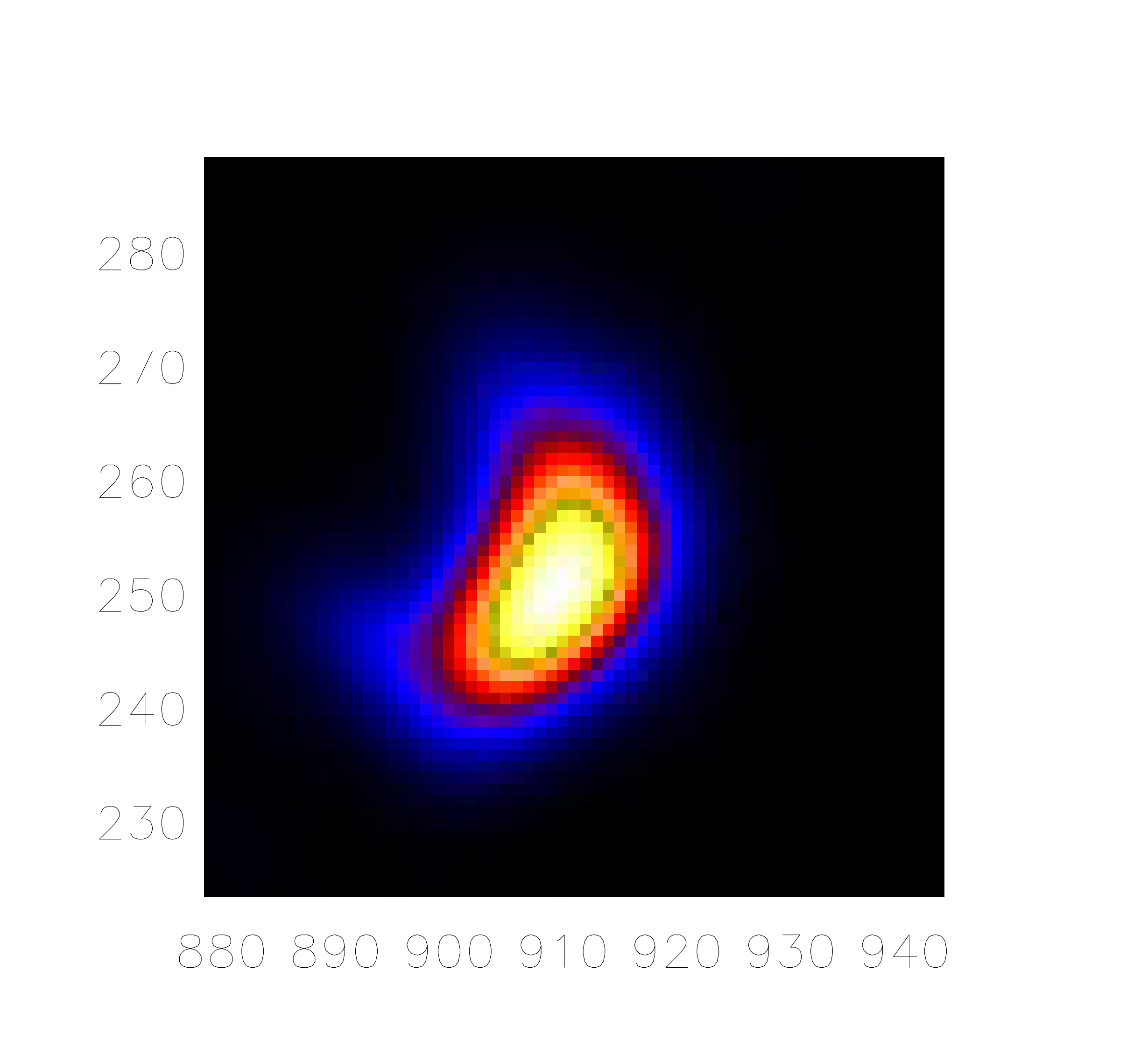}\hspace{-1cm}
    \includegraphics[height=3.7cm]{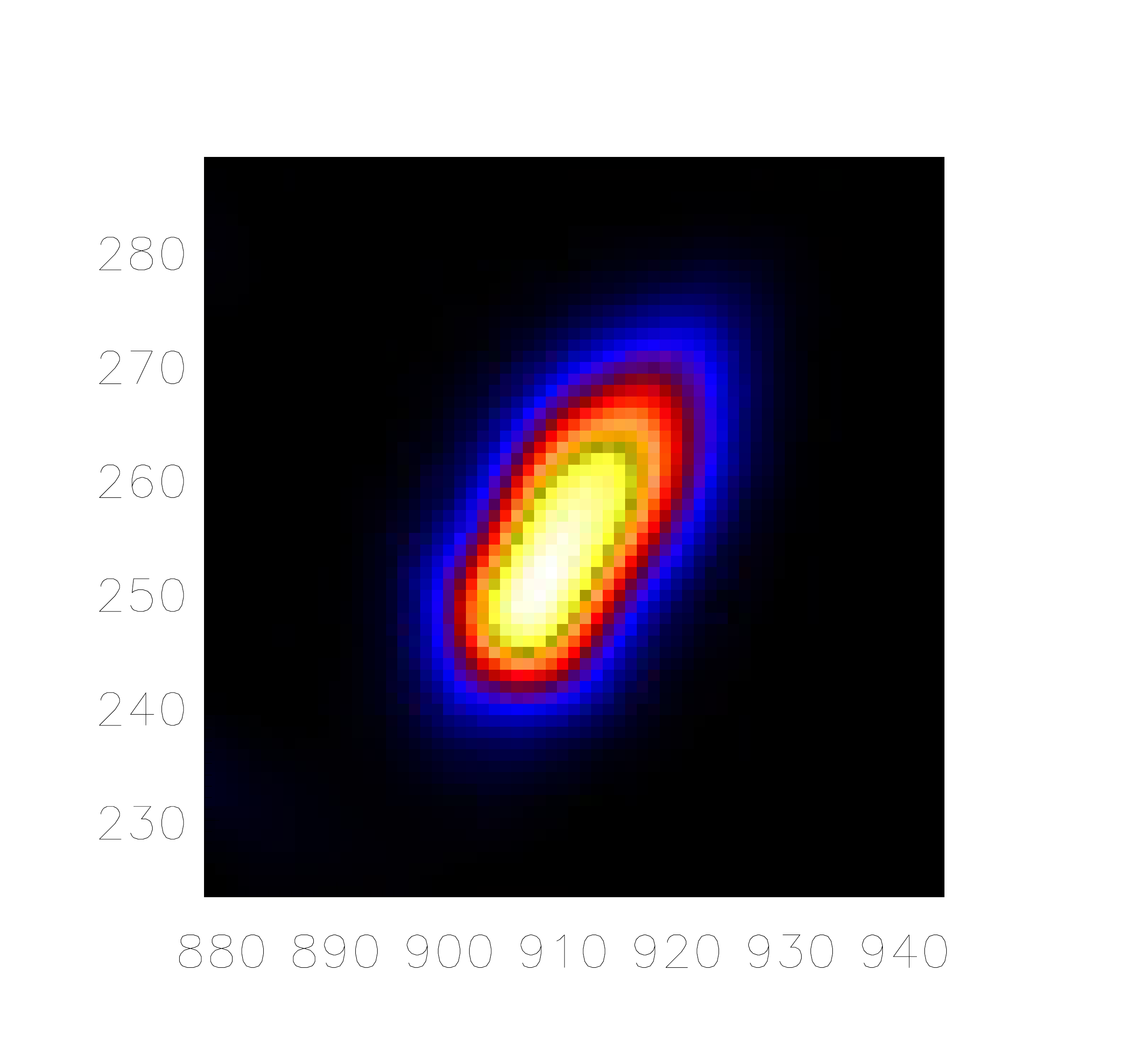}\hspace{-1cm}
    \includegraphics[height=3.7cm]{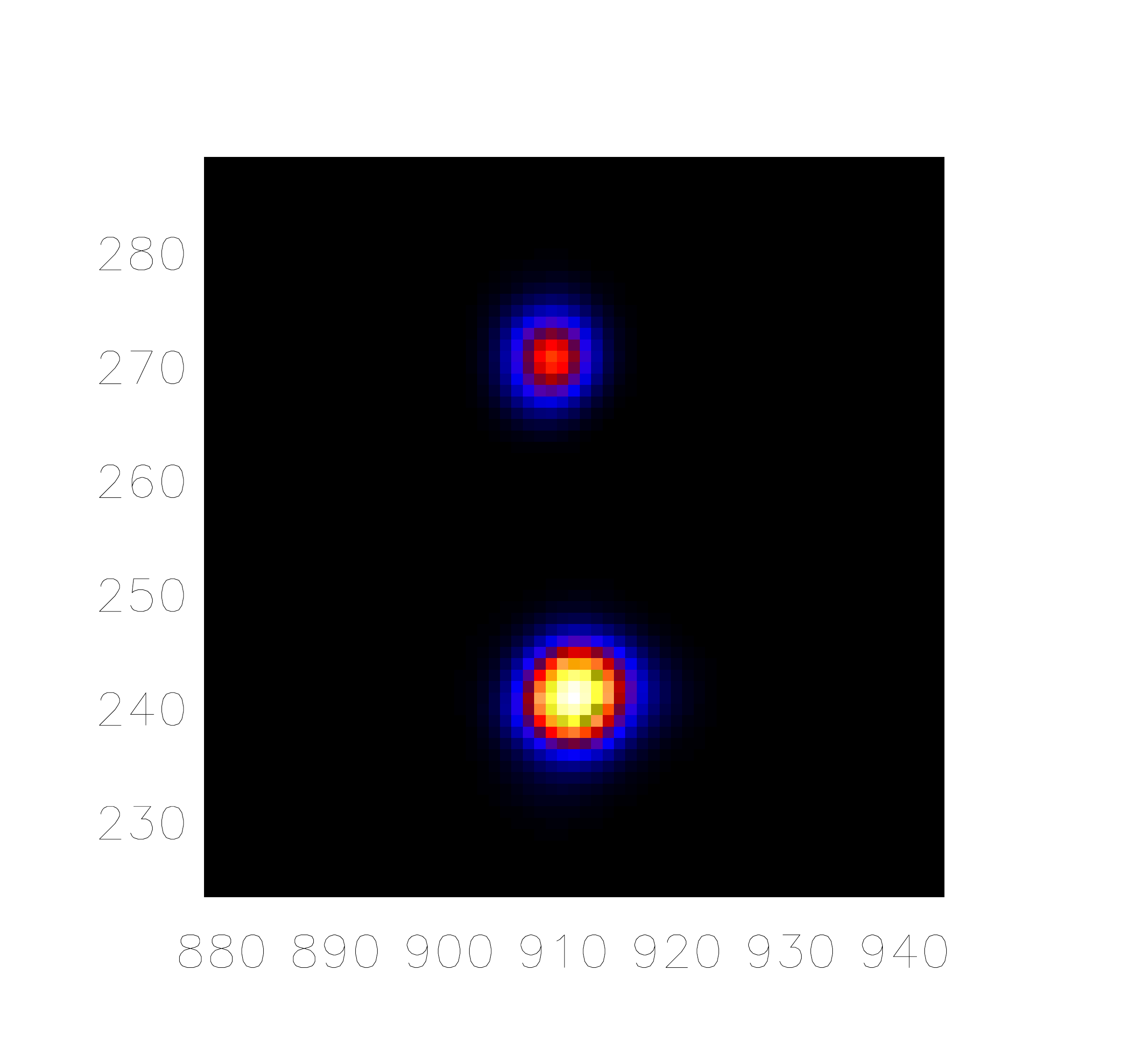}\hspace{-1cm}
    \includegraphics[height=3.7cm]{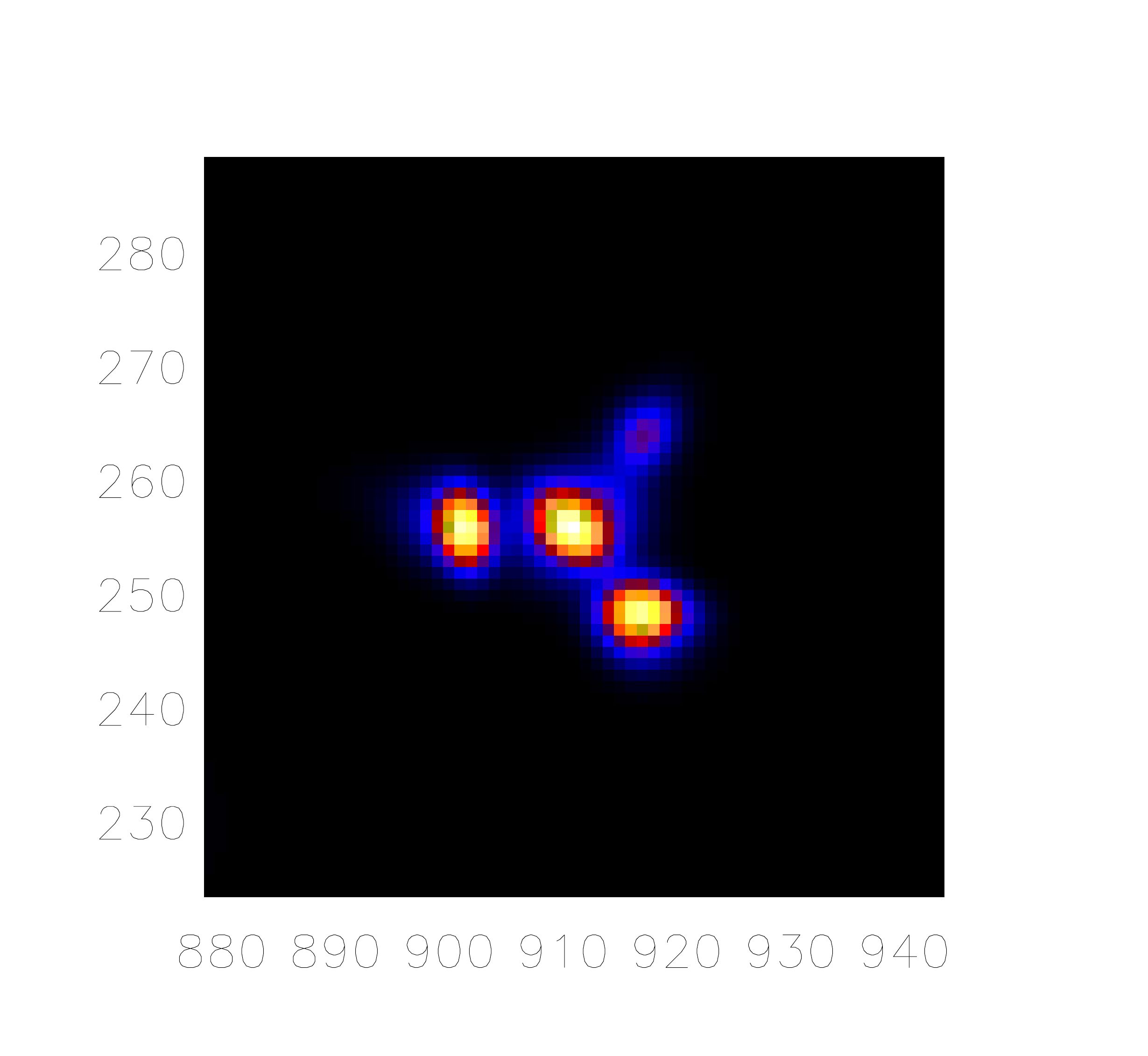}\\ \vspace{-1cm}
    \end{minipage}
        \begin{minipage}{.11\textwidth}
\centering
VIS\texttt{\_}WV 
\end{minipage}
\begin{minipage}{.88\textwidth}
 \includegraphics[height=3.7cm]{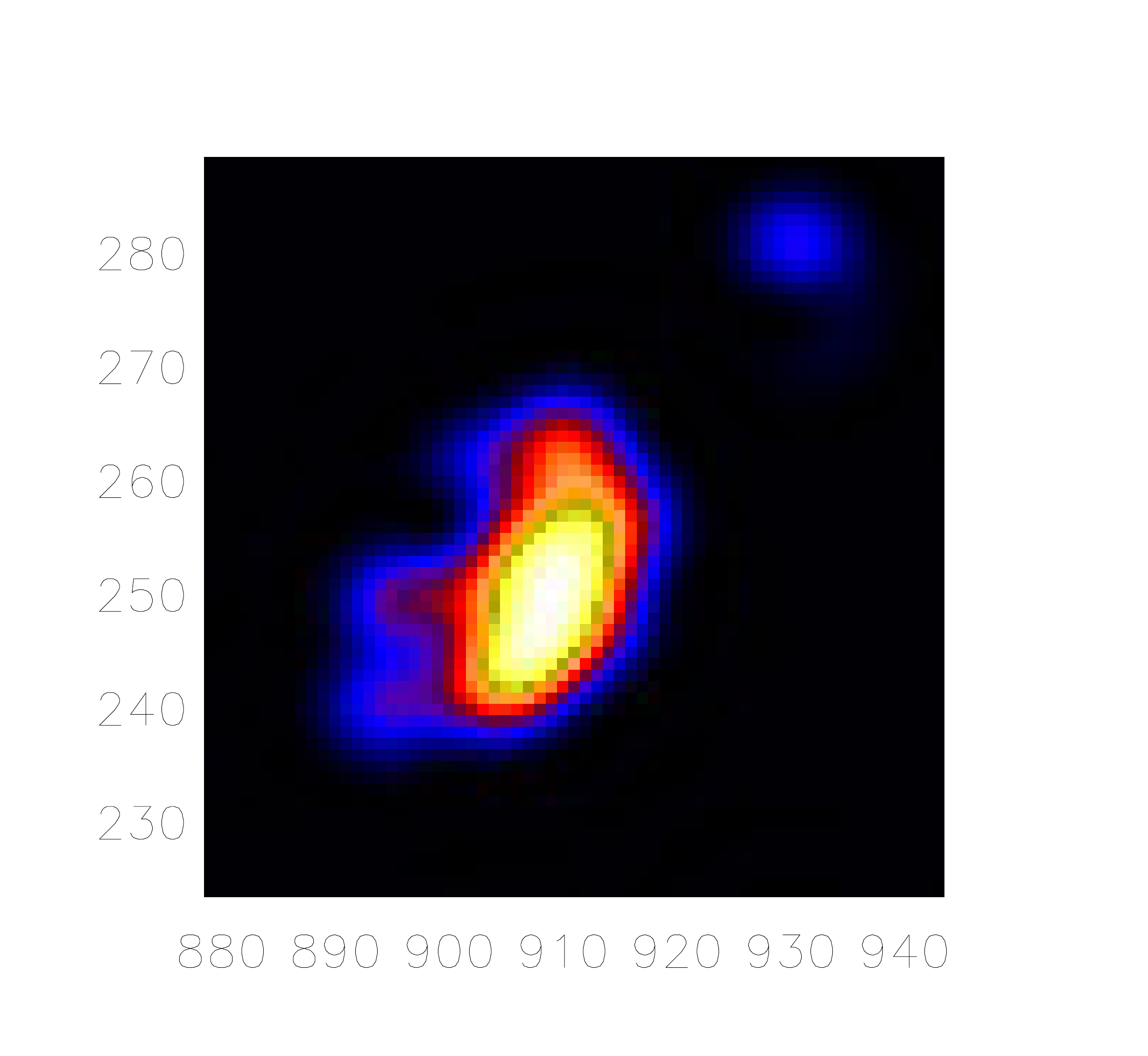}\hspace{-1cm}
    \includegraphics[height=3.7cm]{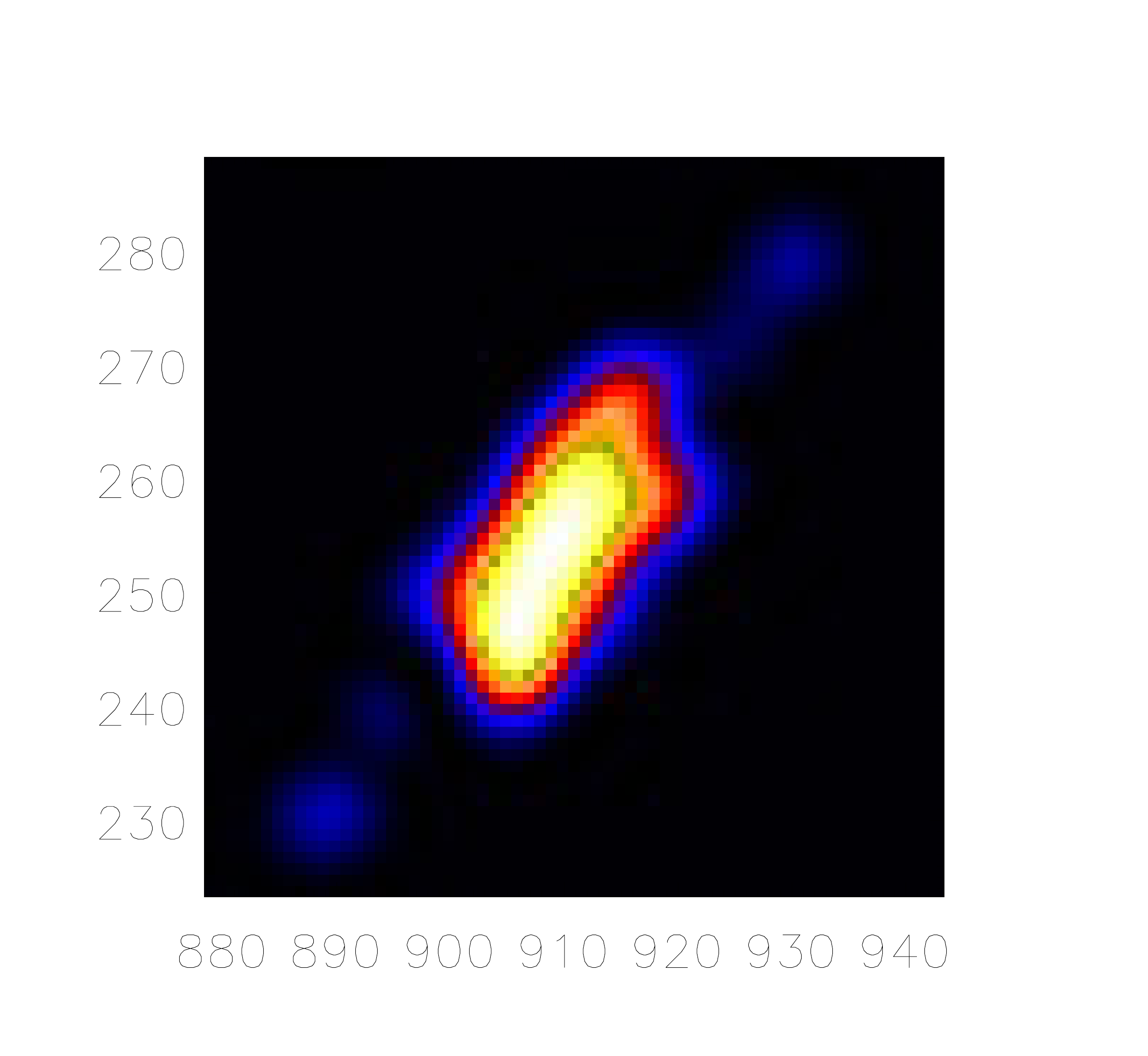}\hspace{-1cm}
    \includegraphics[height=3.7cm]{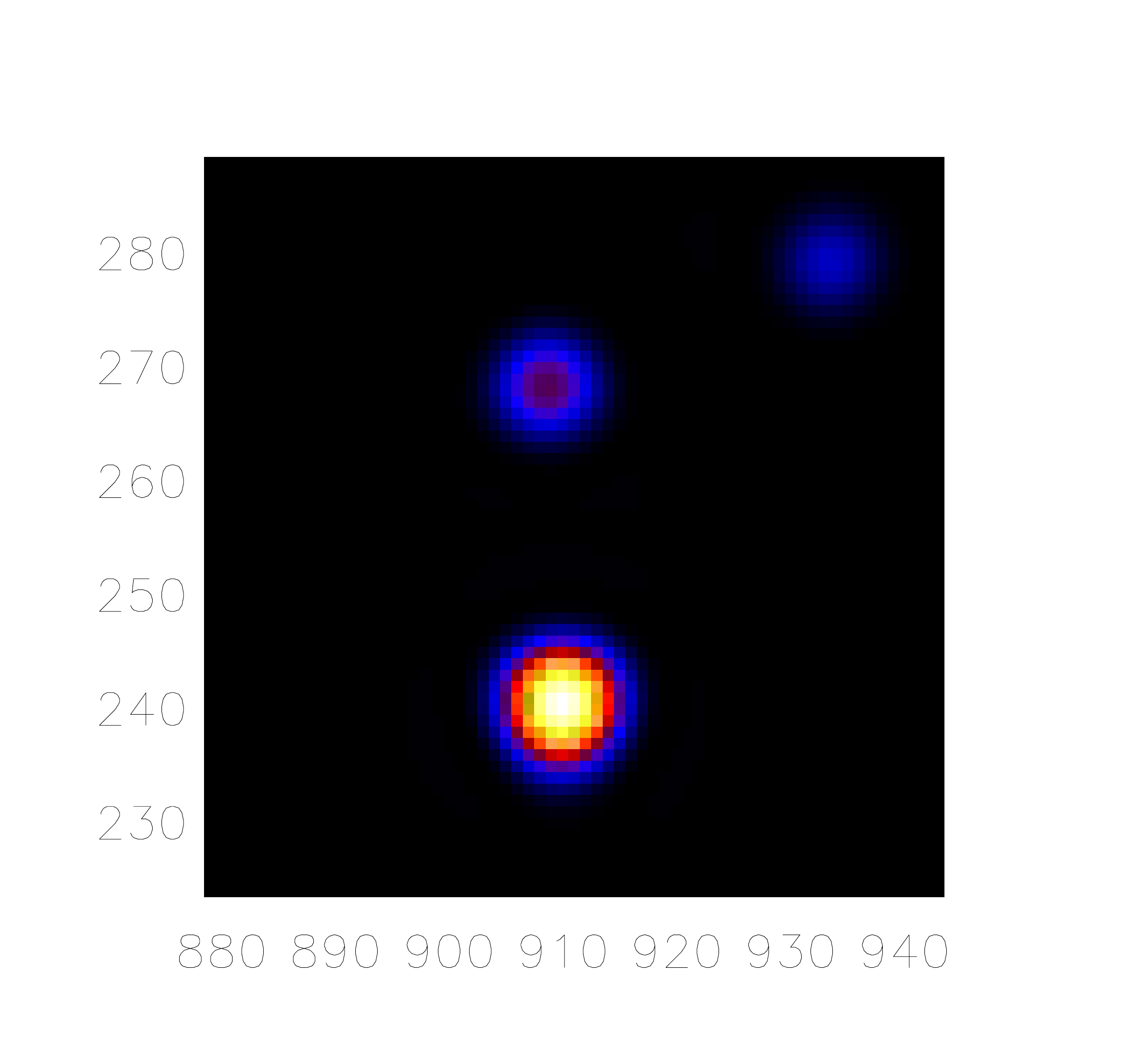}\hspace{-1cm}
    \includegraphics[height=3.7cm]{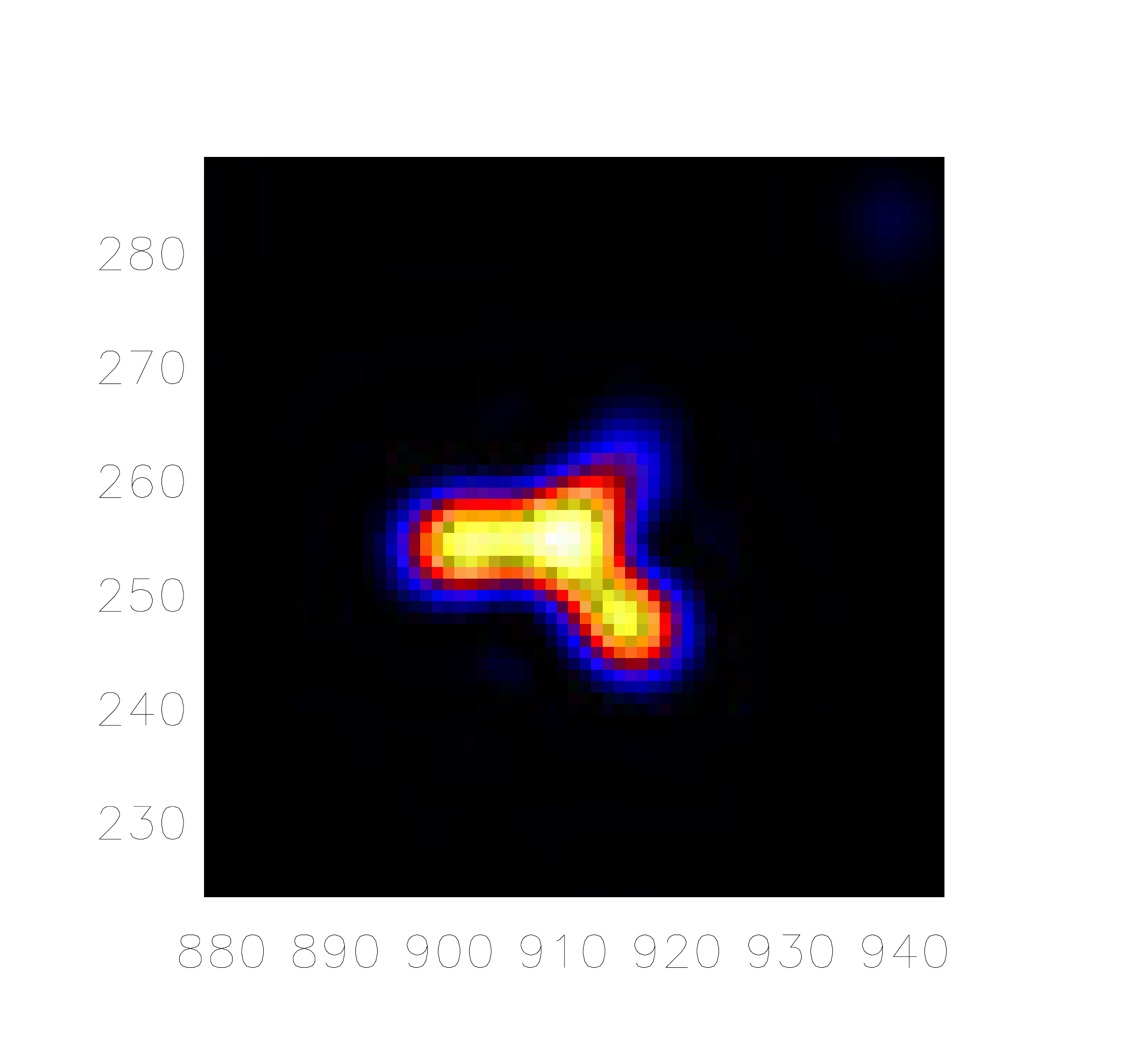}\\ \vspace{-1cm}
    \end{minipage}  
        \begin{minipage}{.11\textwidth}
\centering
ASMC \quad
\end{minipage}
\begin{minipage}{.88\textwidth}
 \includegraphics[height=3.7cm]{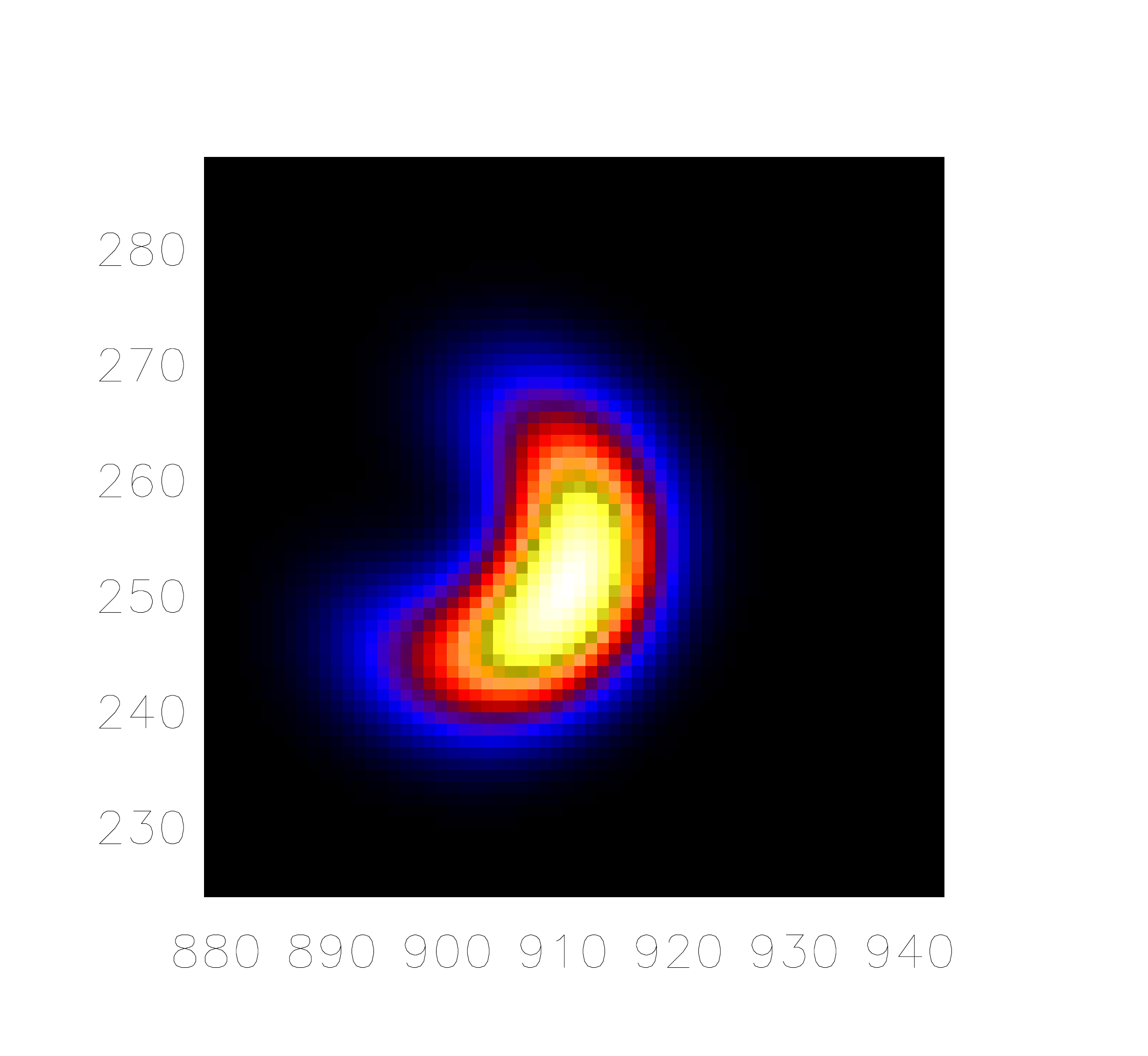}\hspace{-1cm}
    \includegraphics[height=3.7cm]{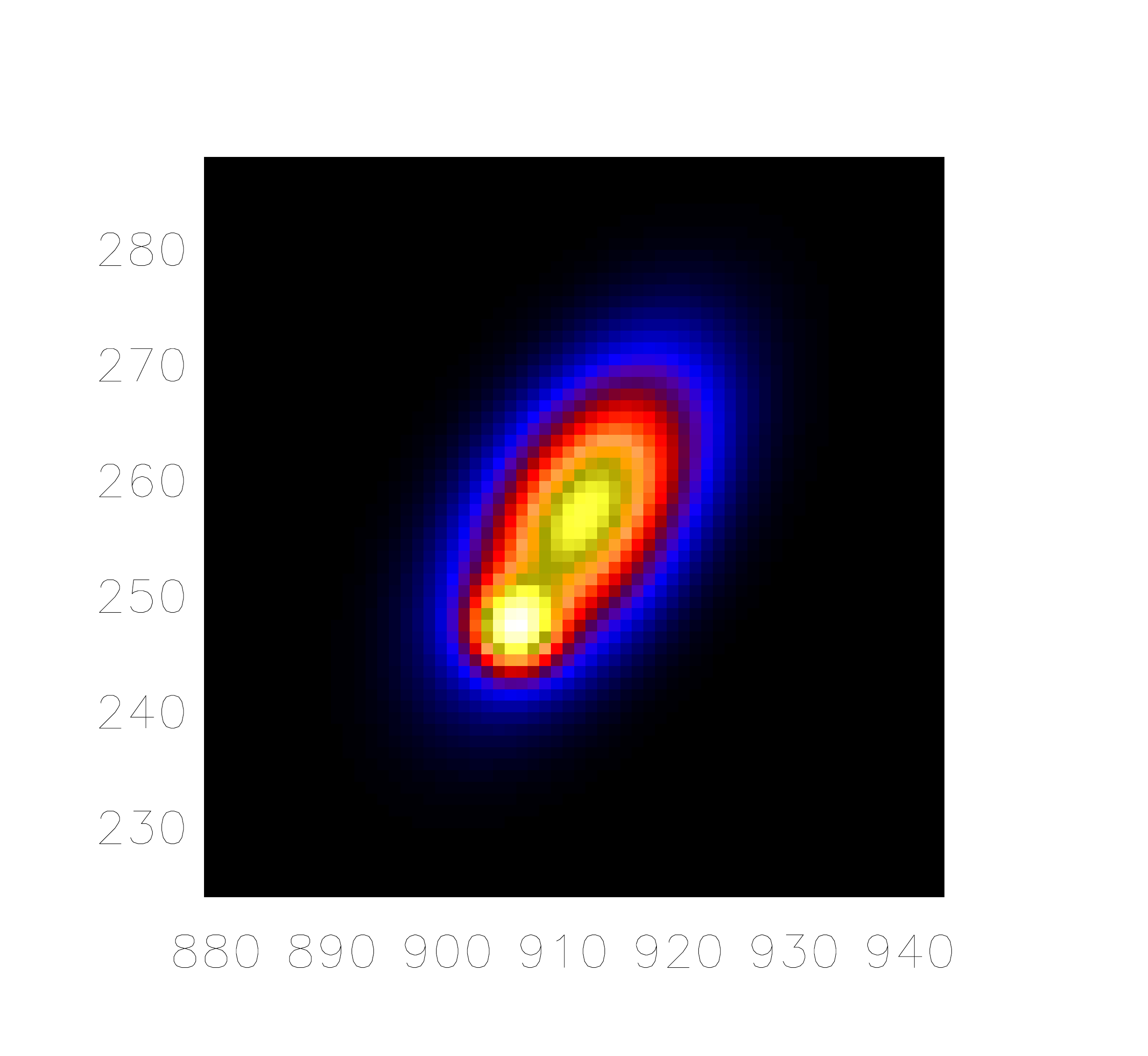}\hspace{-1cm}
    \includegraphics[height=3.7cm]{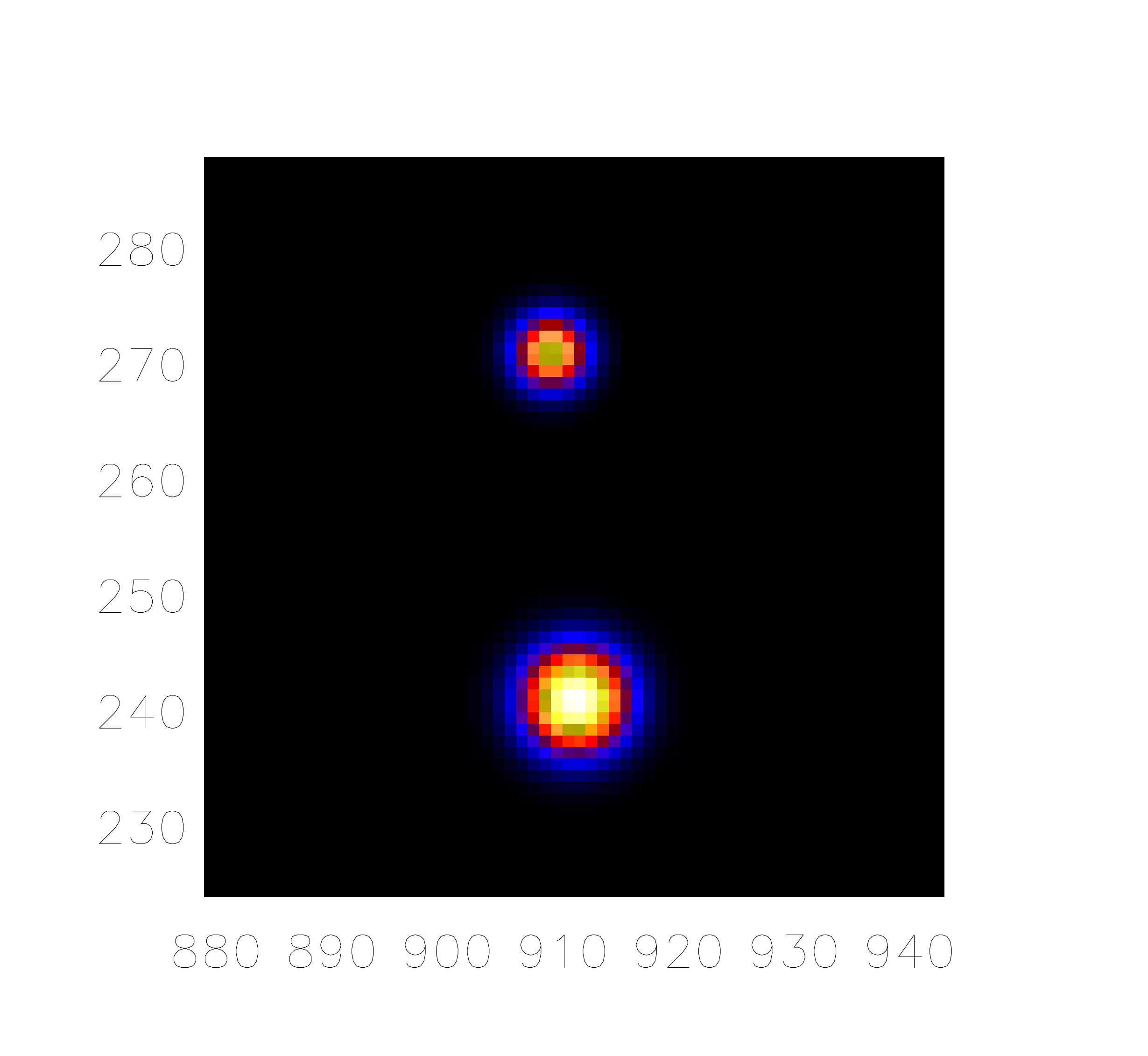}\hspace{-1cm}
    \includegraphics[height=3.7cm]{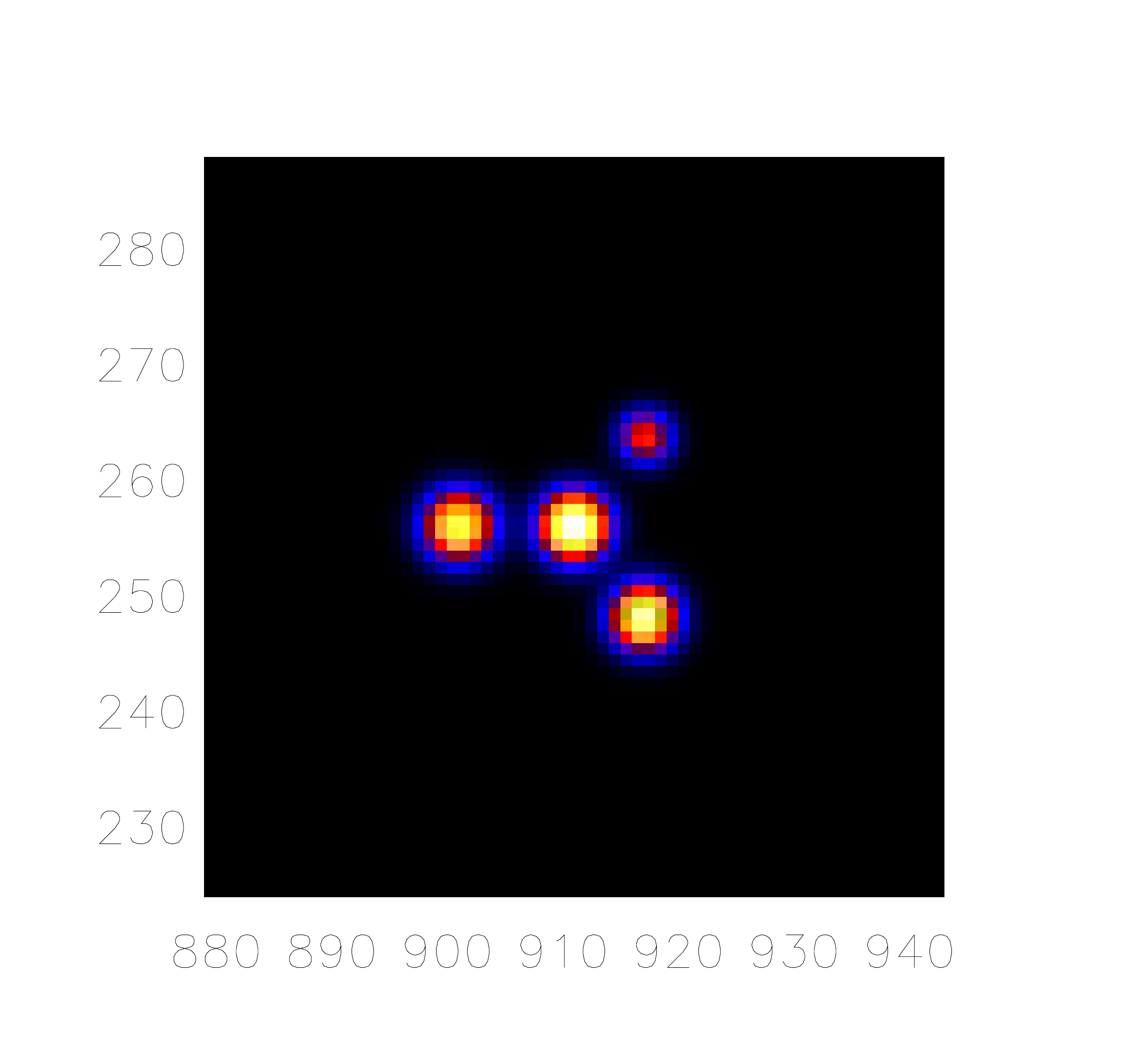}\\ \vspace{-.5cm}
    \end{minipage}
  \caption{Low noise case. Reconstructions obtained by VIS{\_}CLEAN (first row), VIS{\_}FWDFIT (second row), VIS\texttt{\_}CS (third row), VIS\texttt{\_}WV (fourth row) and ASMC (fifth row). True sources in \cref{fig:original}.}
  \label{fig:testfig}
\end{figure}

\begin{figure}[h!]
\begin{minipage}{.11\textwidth}
\centering
VIS{\_} CLEAN
\end{minipage}
\begin{minipage}{.88\textwidth}
 \includegraphics[height=3.7cm]{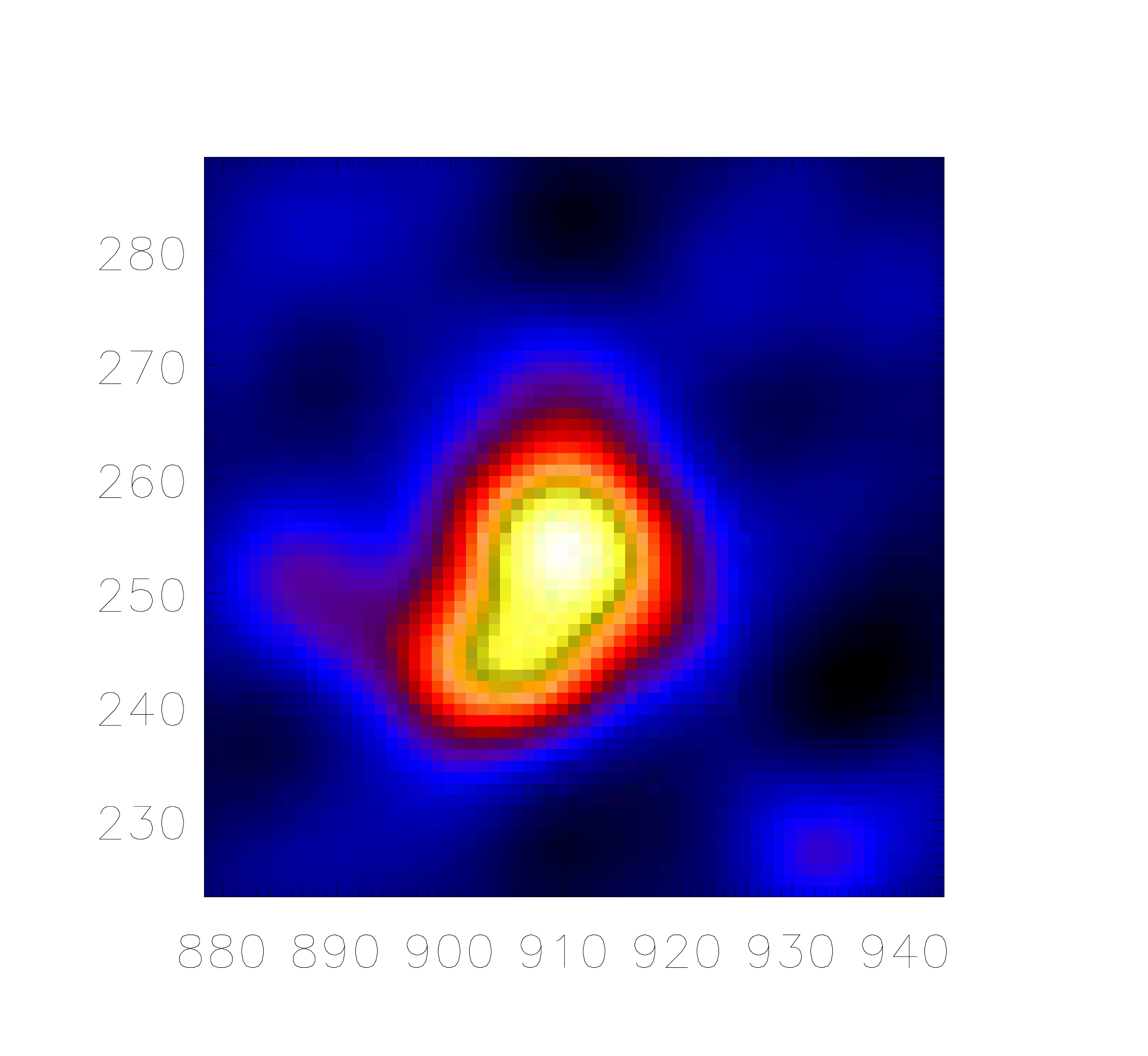}\hspace{-1cm}
    \includegraphics[height=3.7cm]{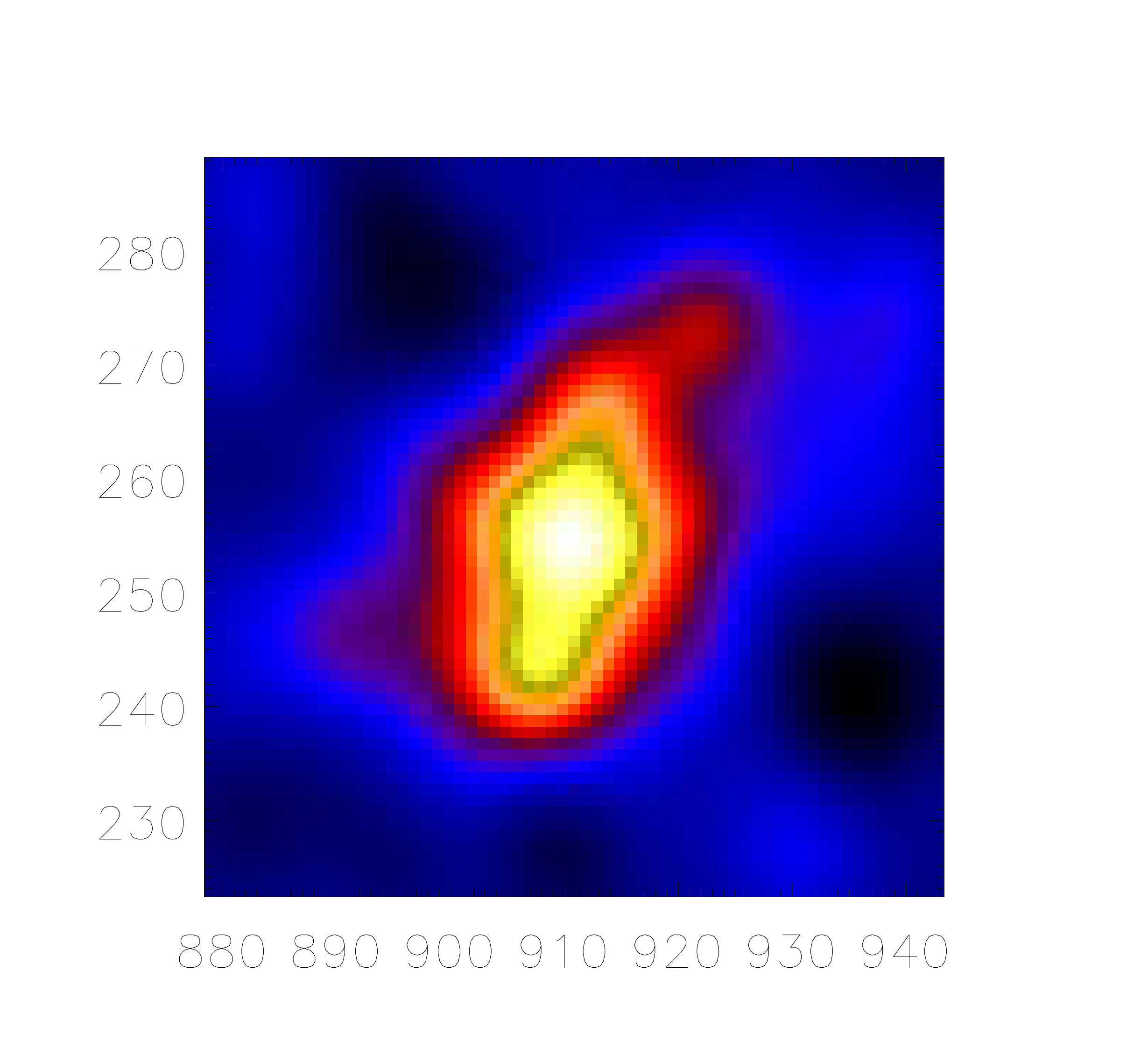}\hspace{-1cm}
    \includegraphics[height=3.7cm]{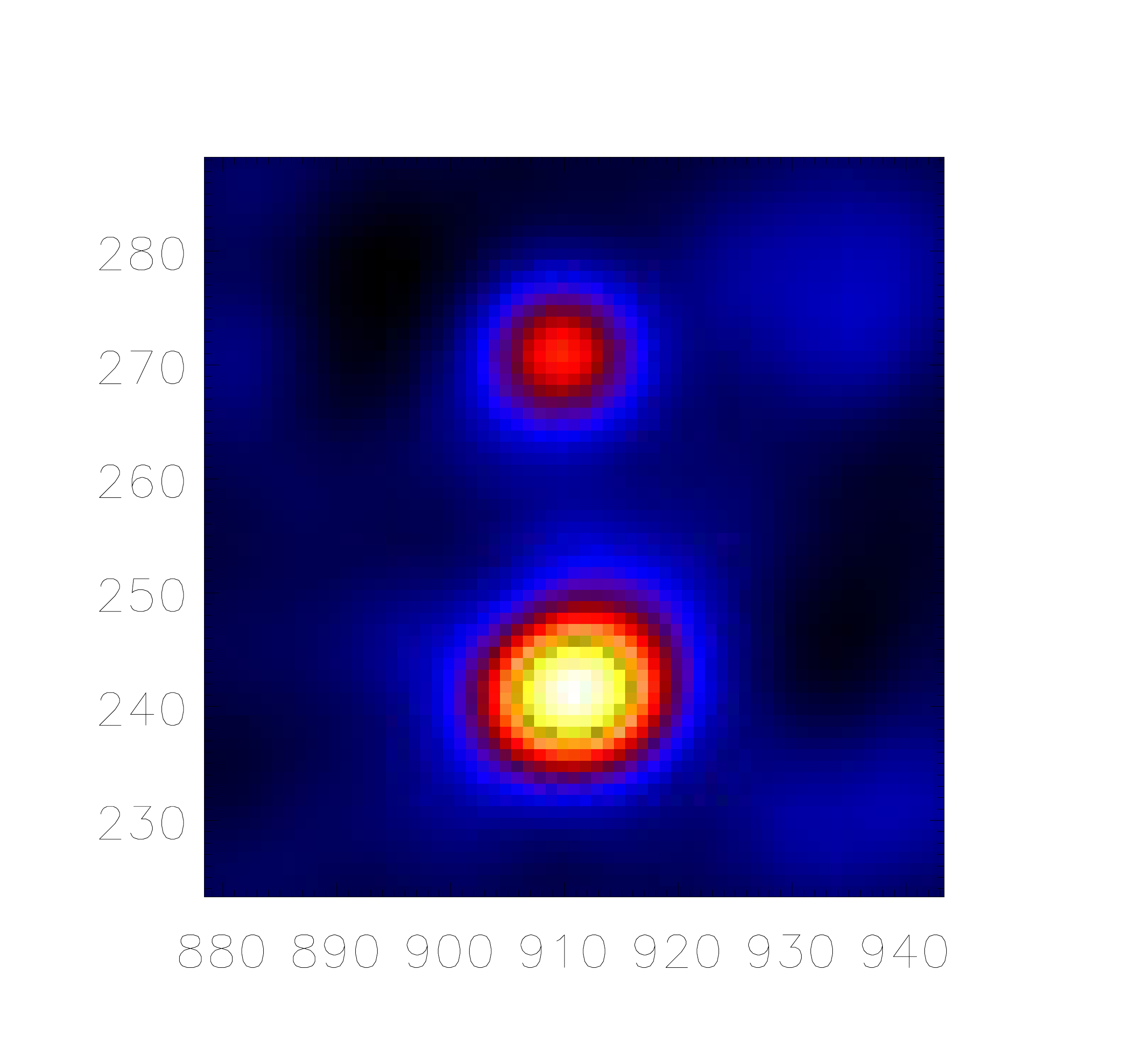}\hspace{-1cm}
    \includegraphics[height=3.7cm]{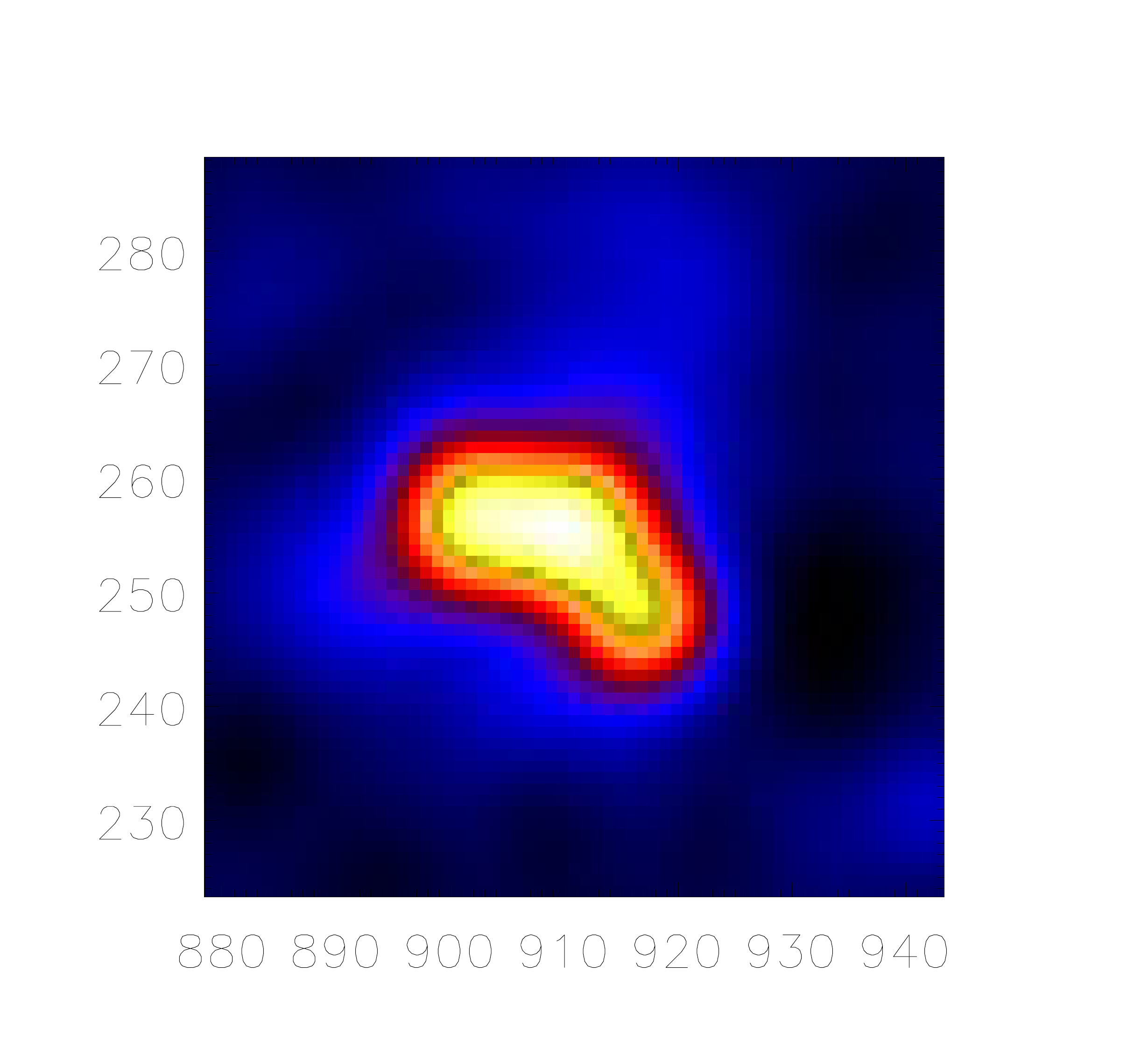}\\ \vspace{-1cm}
    \end{minipage}
        \begin{minipage}{.11\textwidth}
\centering
VIS\texttt{\_} FWDFIT
\end{minipage}
\begin{minipage}{.88\textwidth}
 \includegraphics[height=3.7cm]{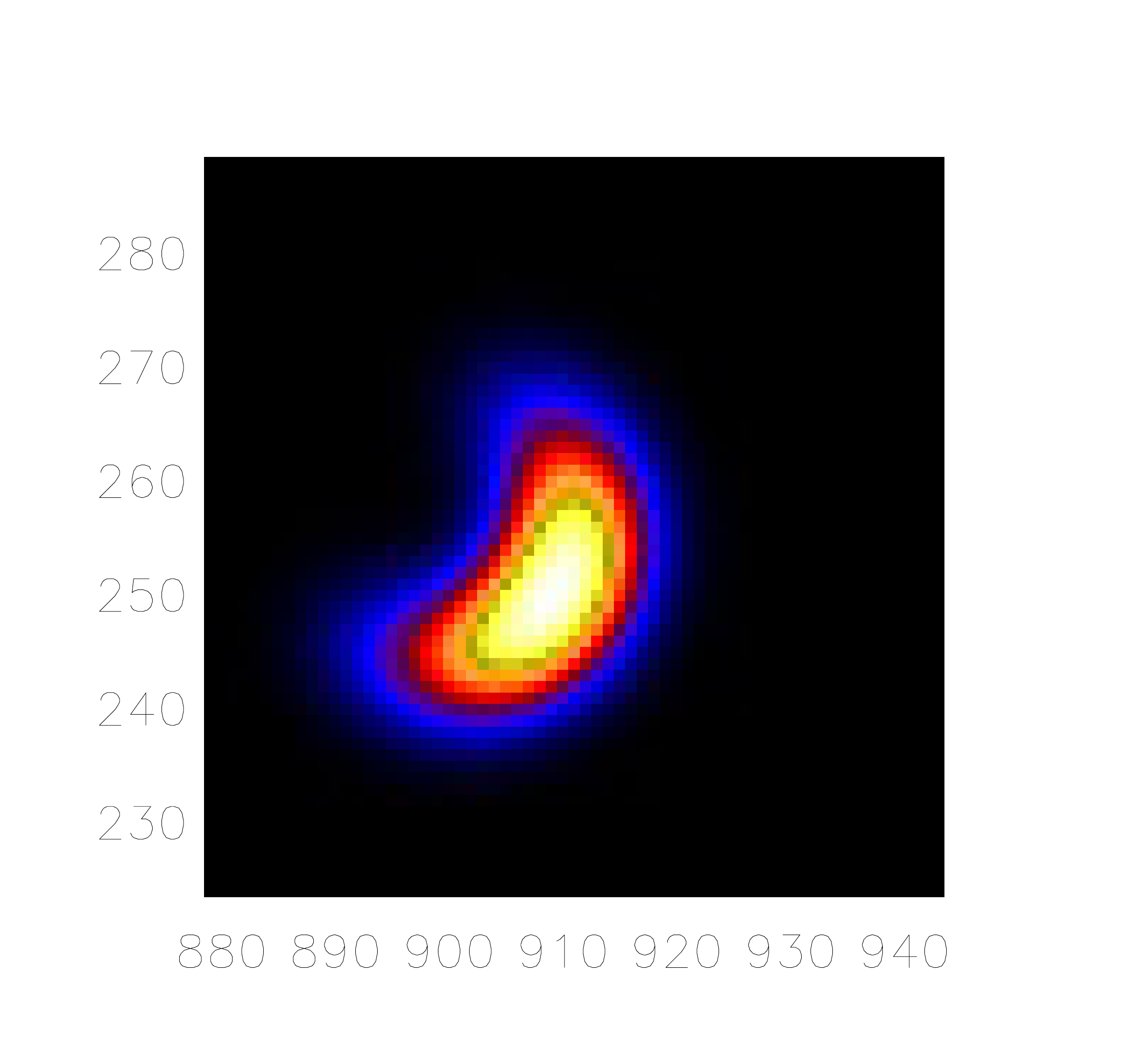}\hspace{-1cm}
    \includegraphics[height=3.7cm]{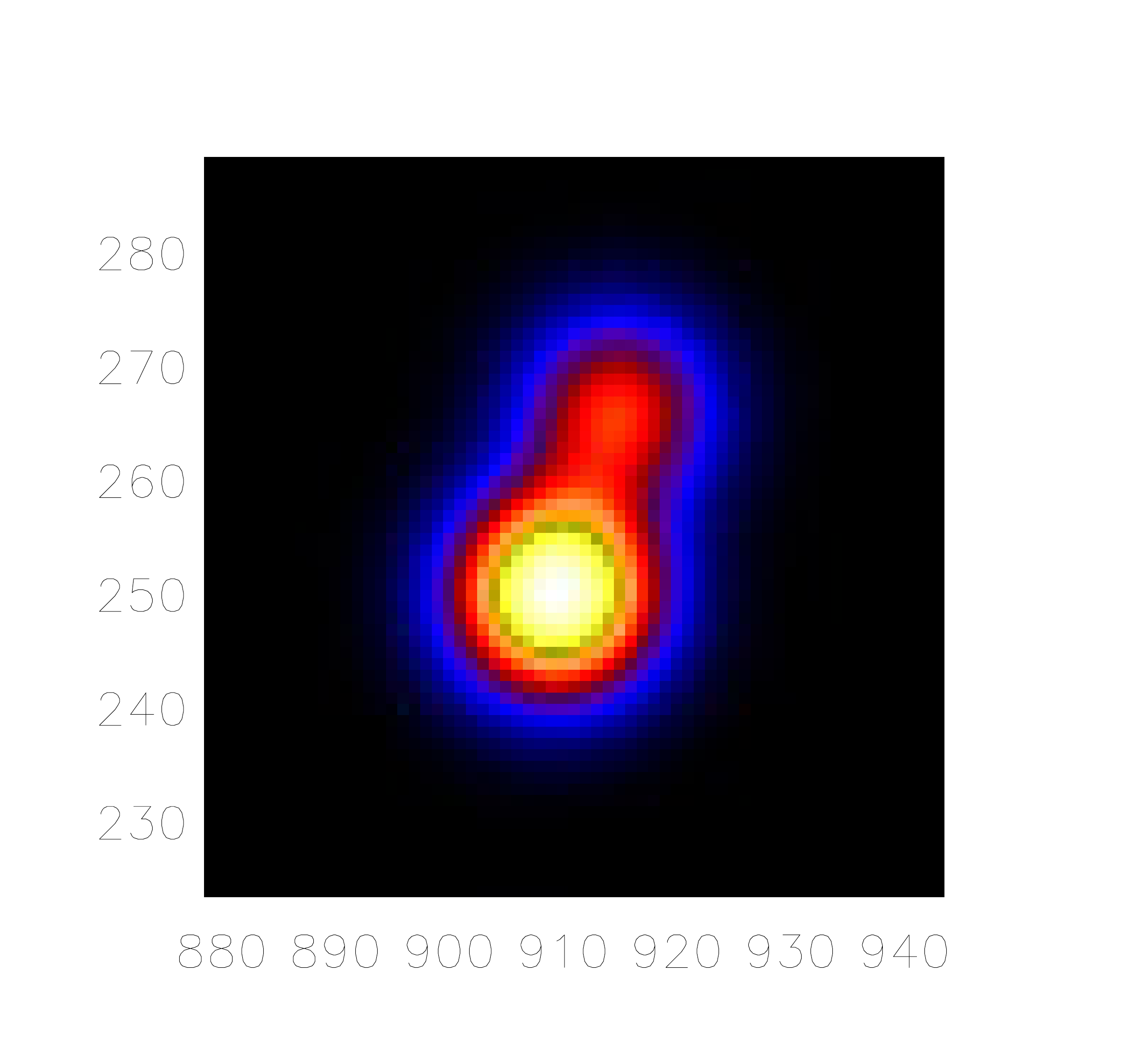}\hspace{-1cm}
    \includegraphics[height=3.7cm]{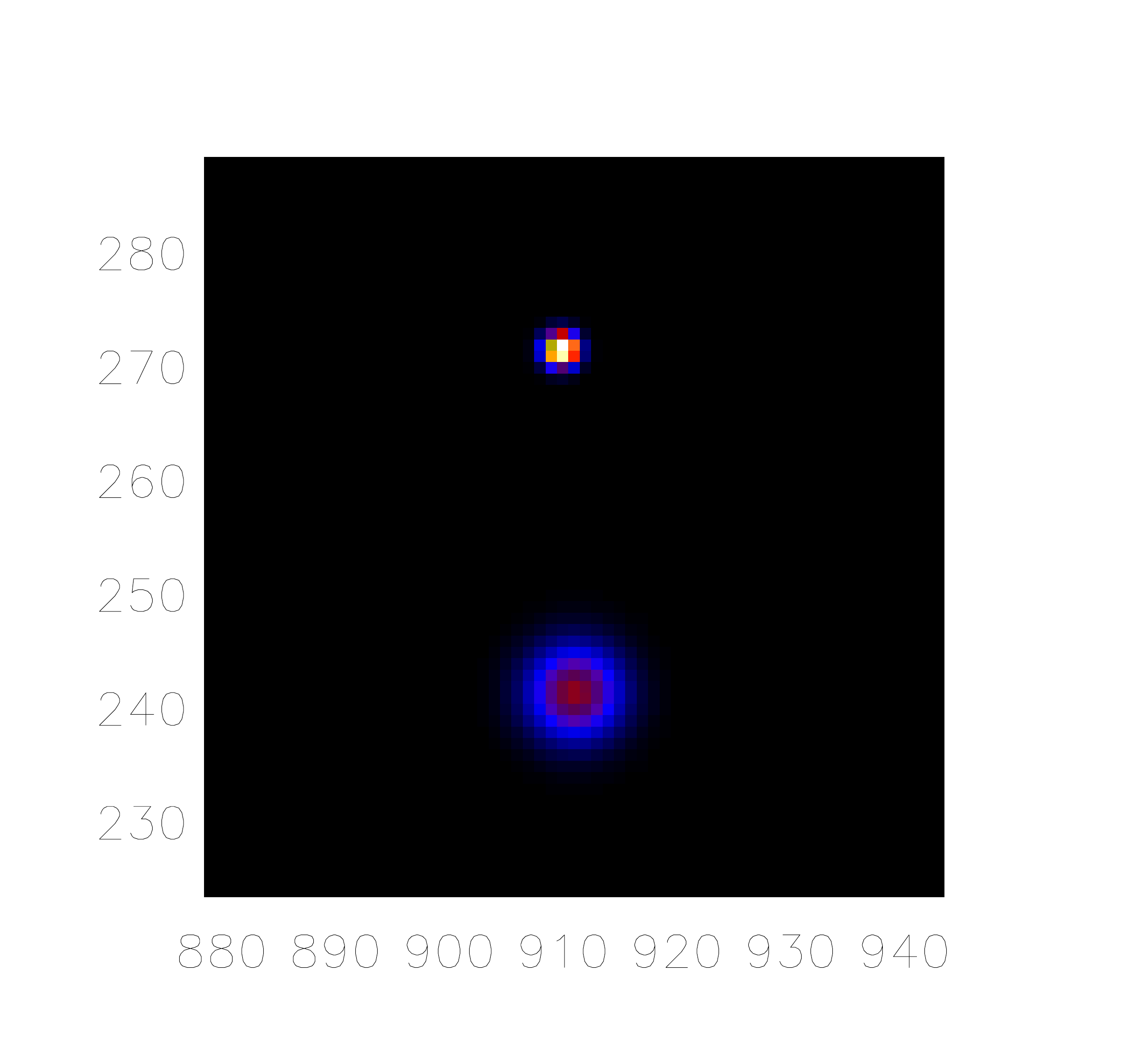}\hspace{-1cm}
    \includegraphics[height=3.7cm]{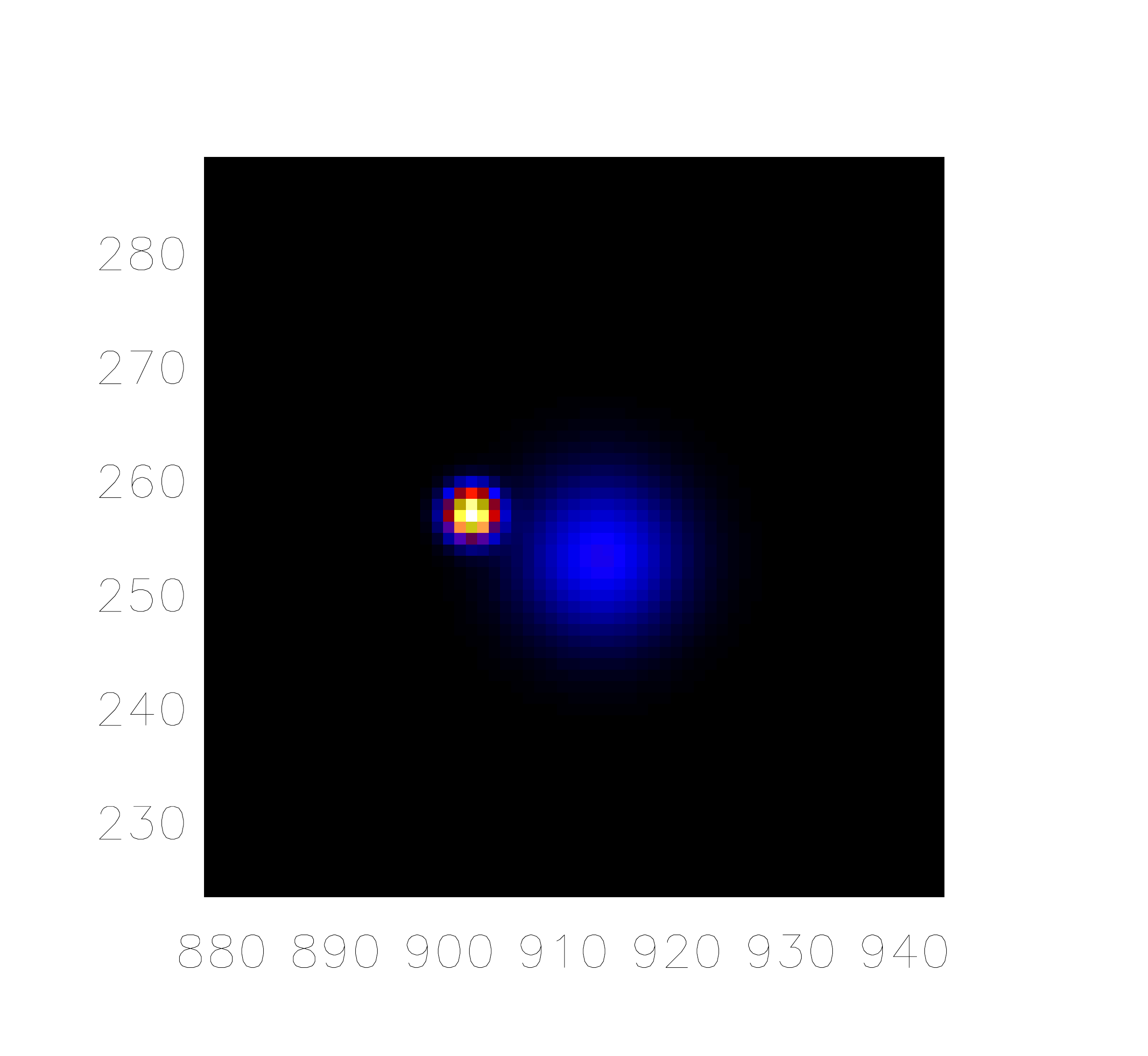}\\ \vspace{-1cm}
    \end{minipage}
        \begin{minipage}{.11\textwidth}
\centering
VIS\texttt{\_}CS
\end{minipage}
\begin{minipage}{.88\textwidth}
 \includegraphics[height=3.7cm]{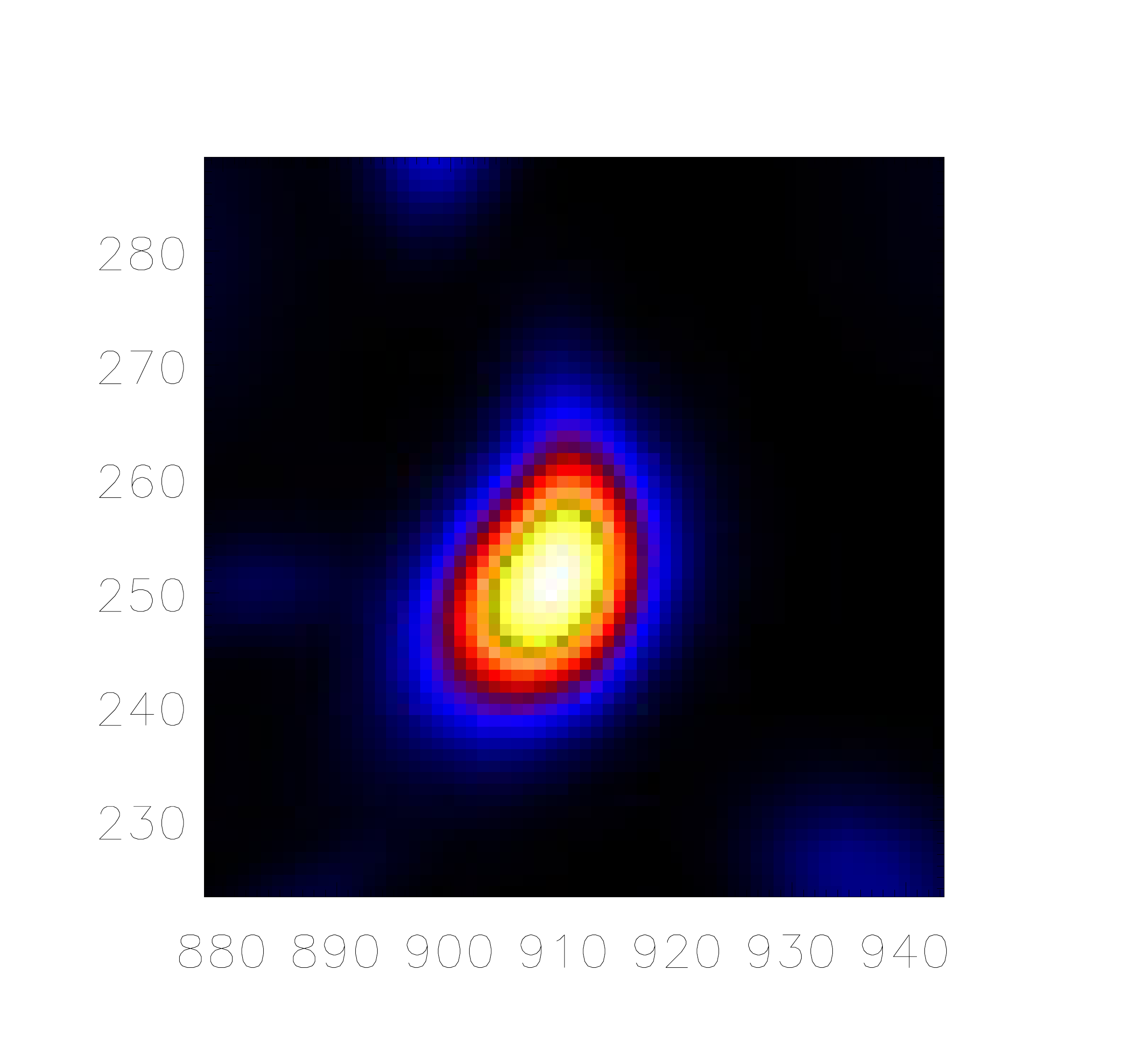}\hspace{-1cm}
    \includegraphics[height=3.7cm]{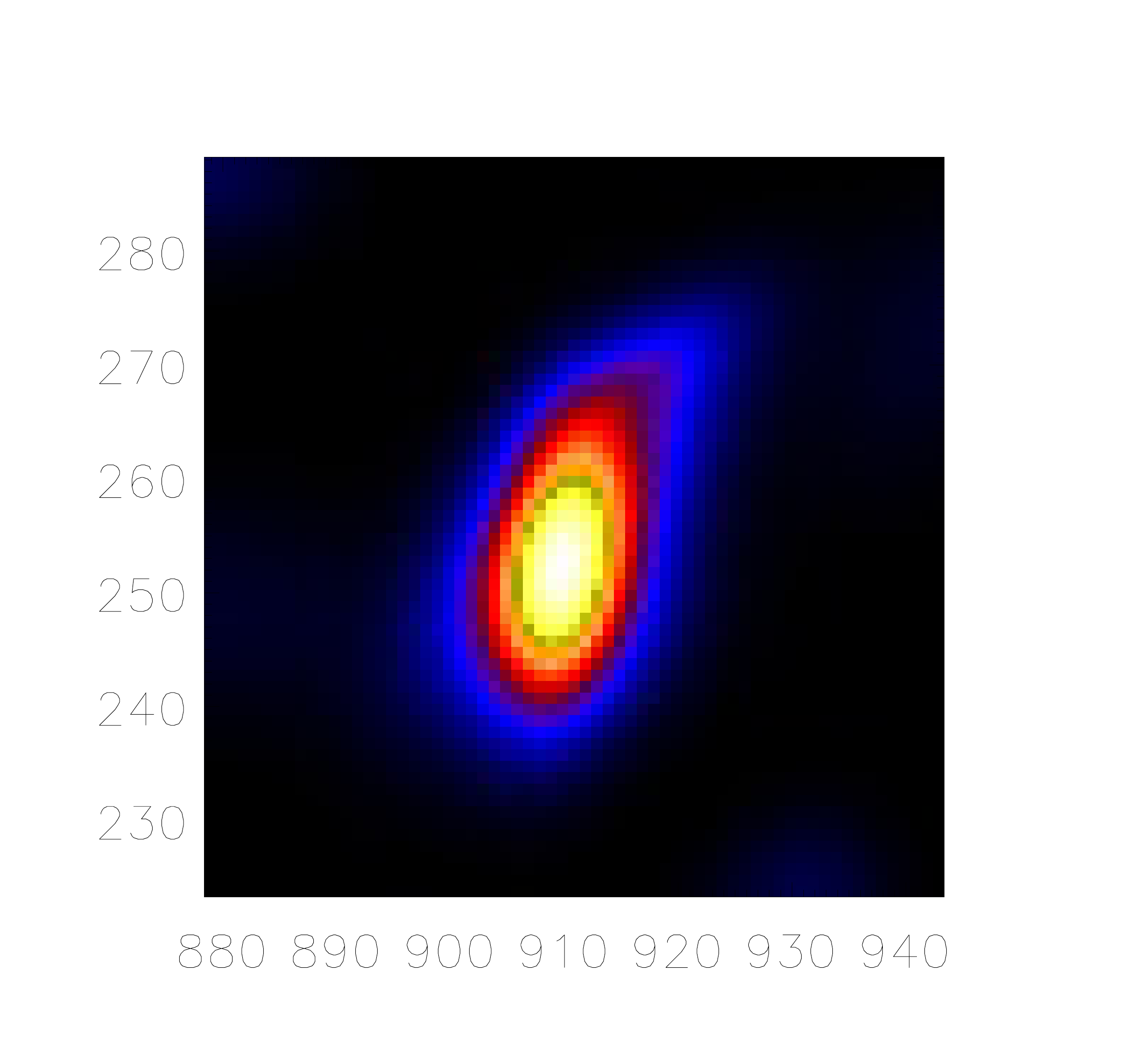}\hspace{-1cm}
    \includegraphics[height=3.7cm]{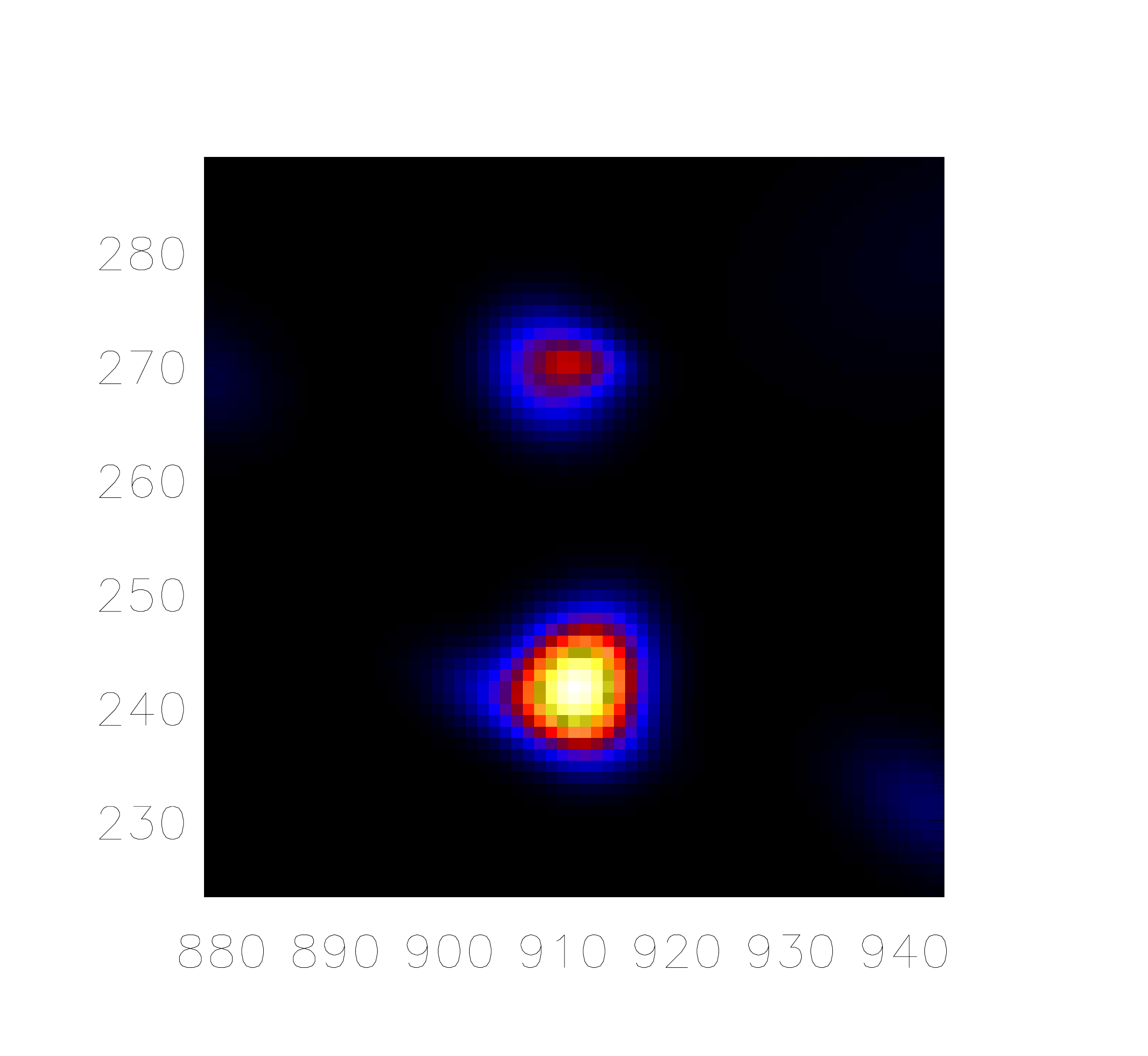}\hspace{-1cm}
    \includegraphics[height=3.7cm]{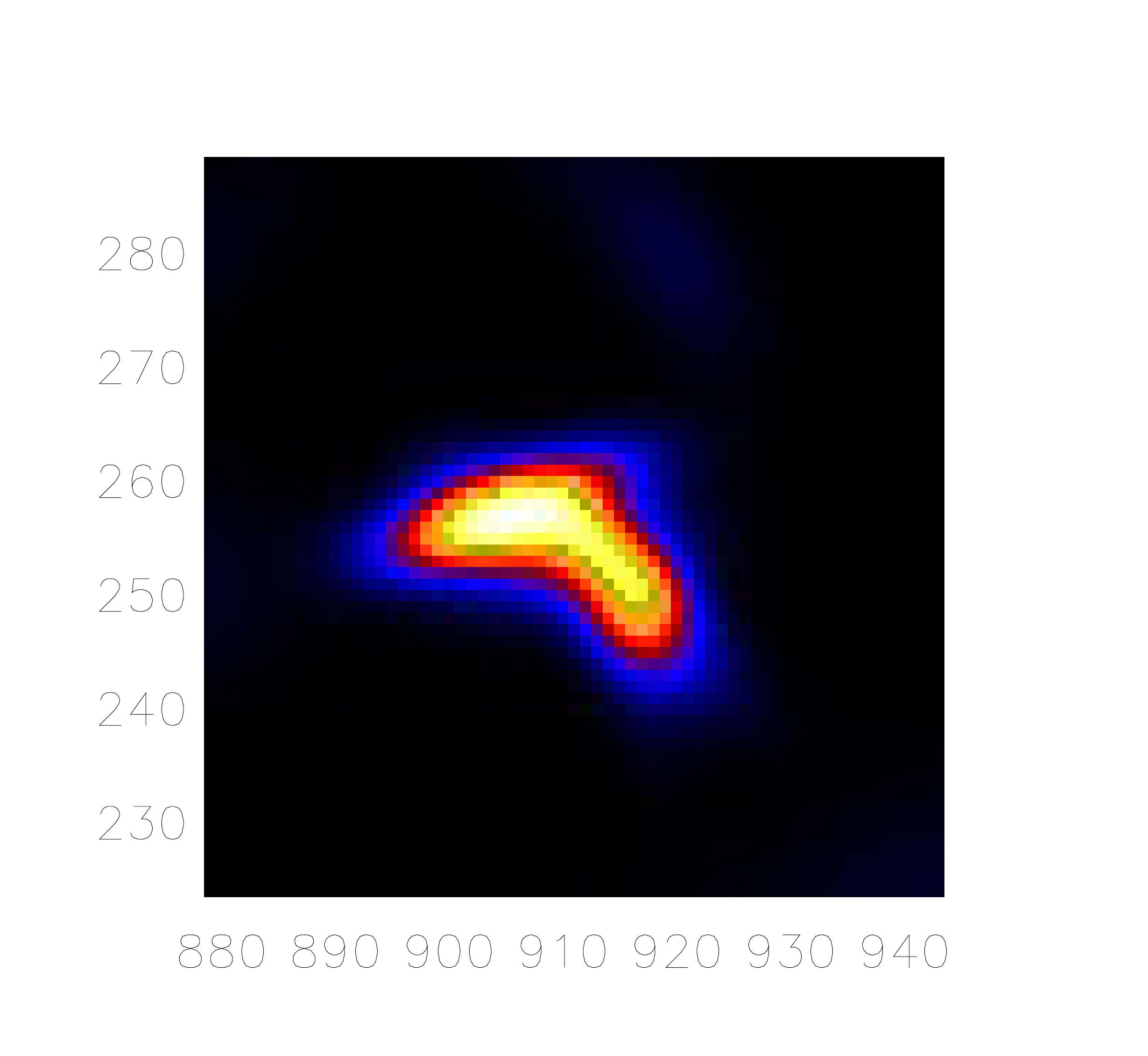}\\ \vspace{-1cm}
    \end{minipage}
    \begin{minipage}{.11\textwidth}
\centering
VIS\texttt{\_}WV 
\end{minipage}
\begin{minipage}{.88\textwidth}
 \includegraphics[height=3.7cm]{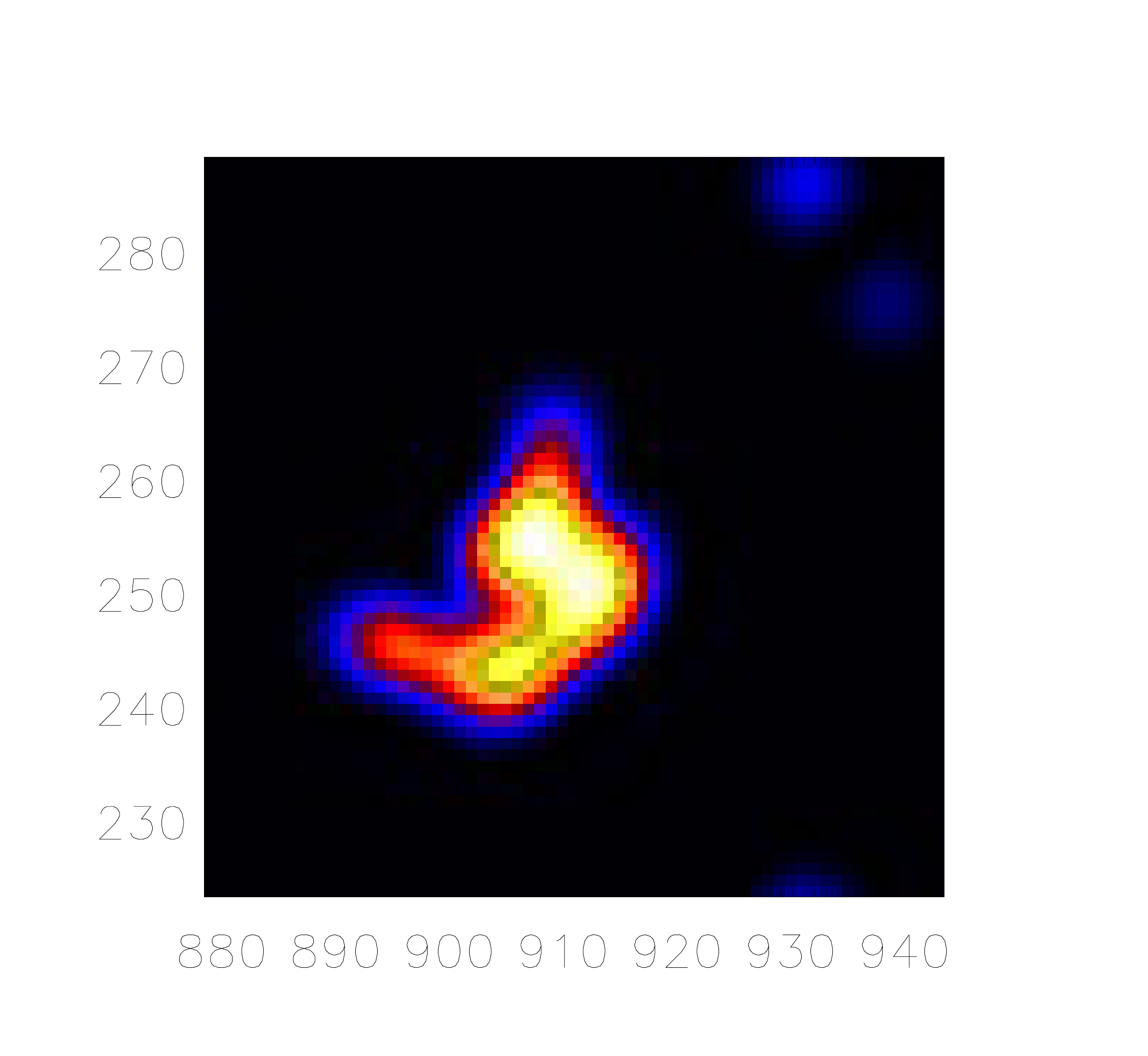}\hspace{-1cm}
    \includegraphics[height=3.7cm]{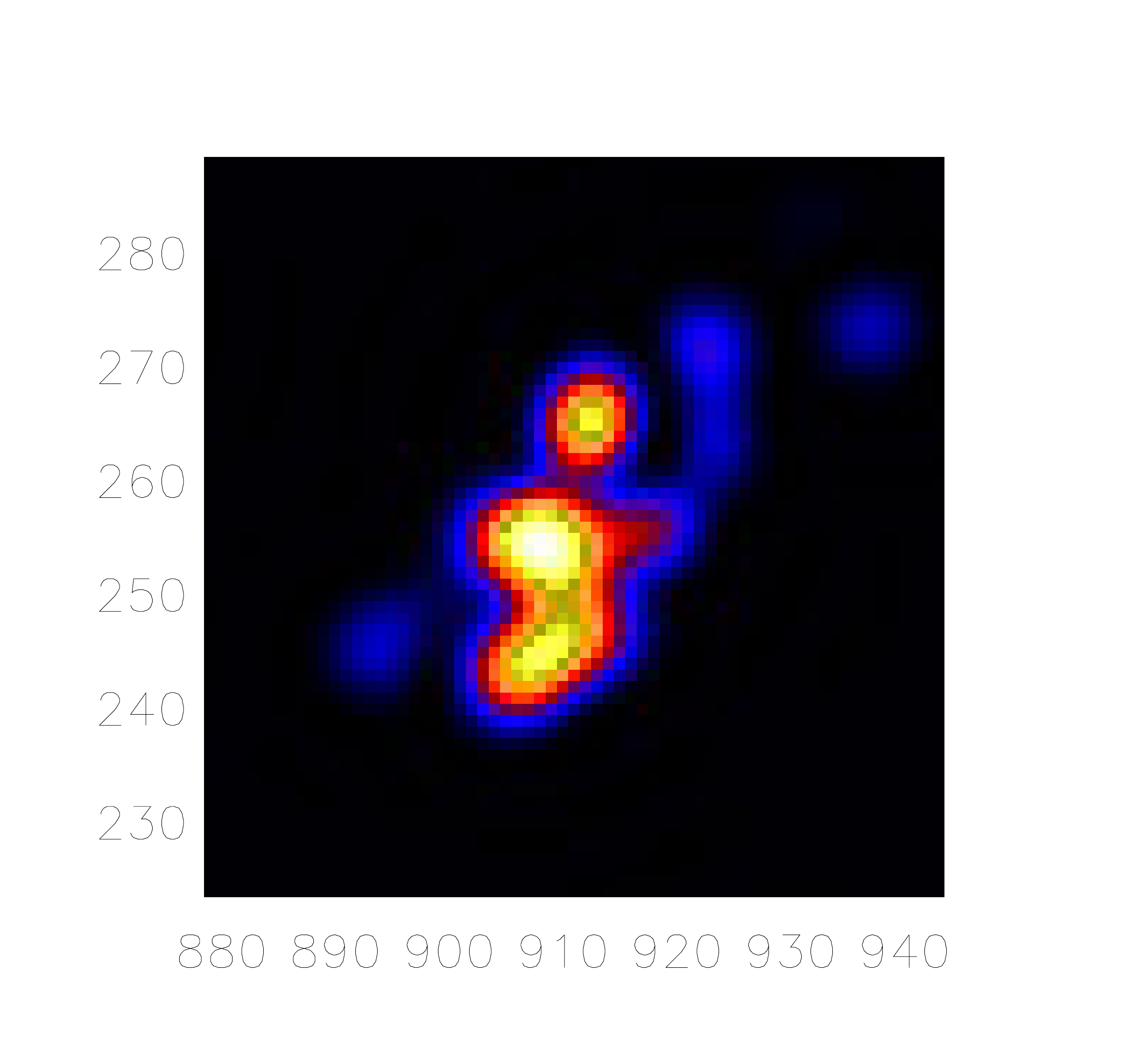}\hspace{-1cm}
    \includegraphics[height=3.7cm]{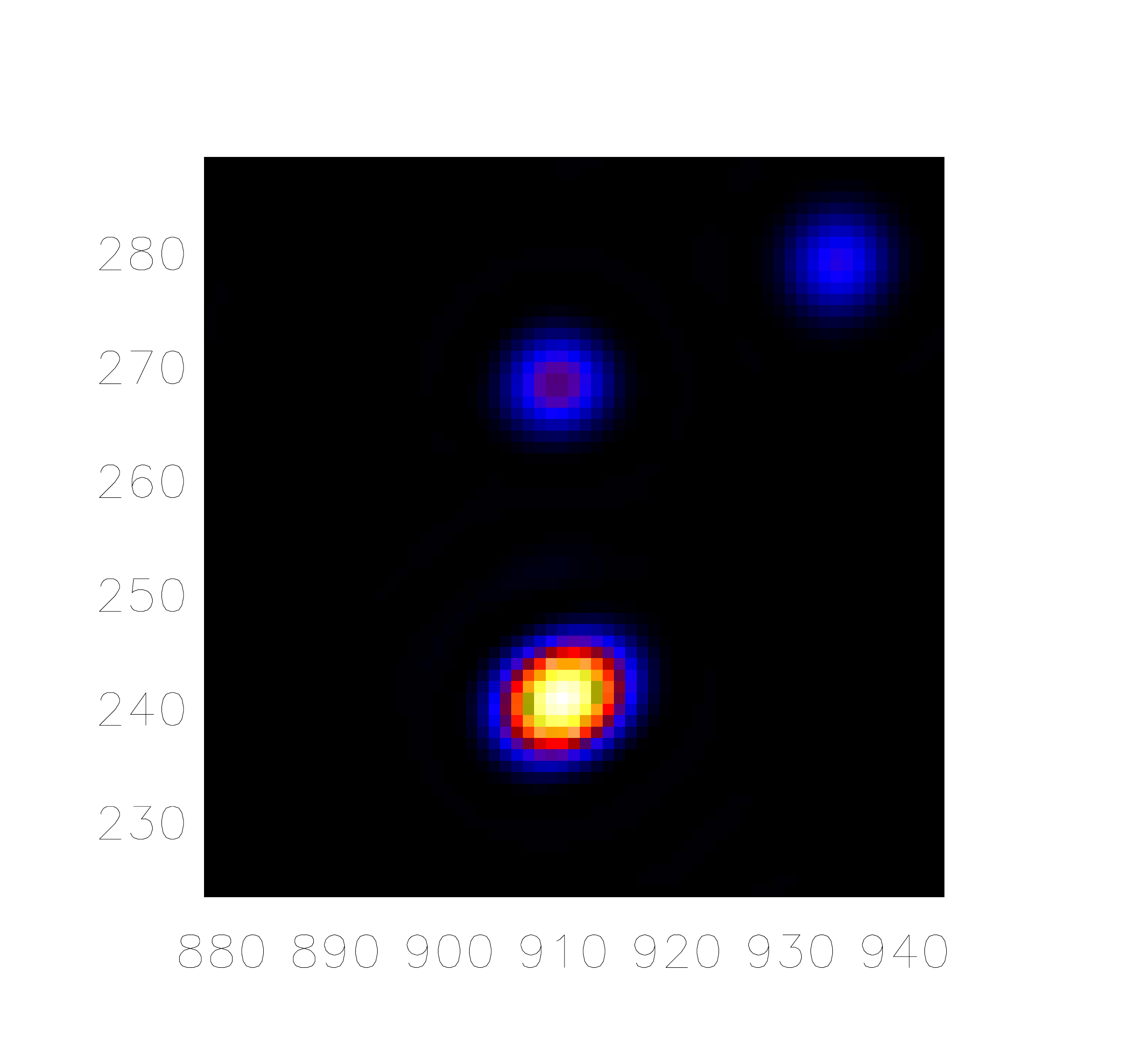}\hspace{-1cm}
    \includegraphics[height=3.7cm]{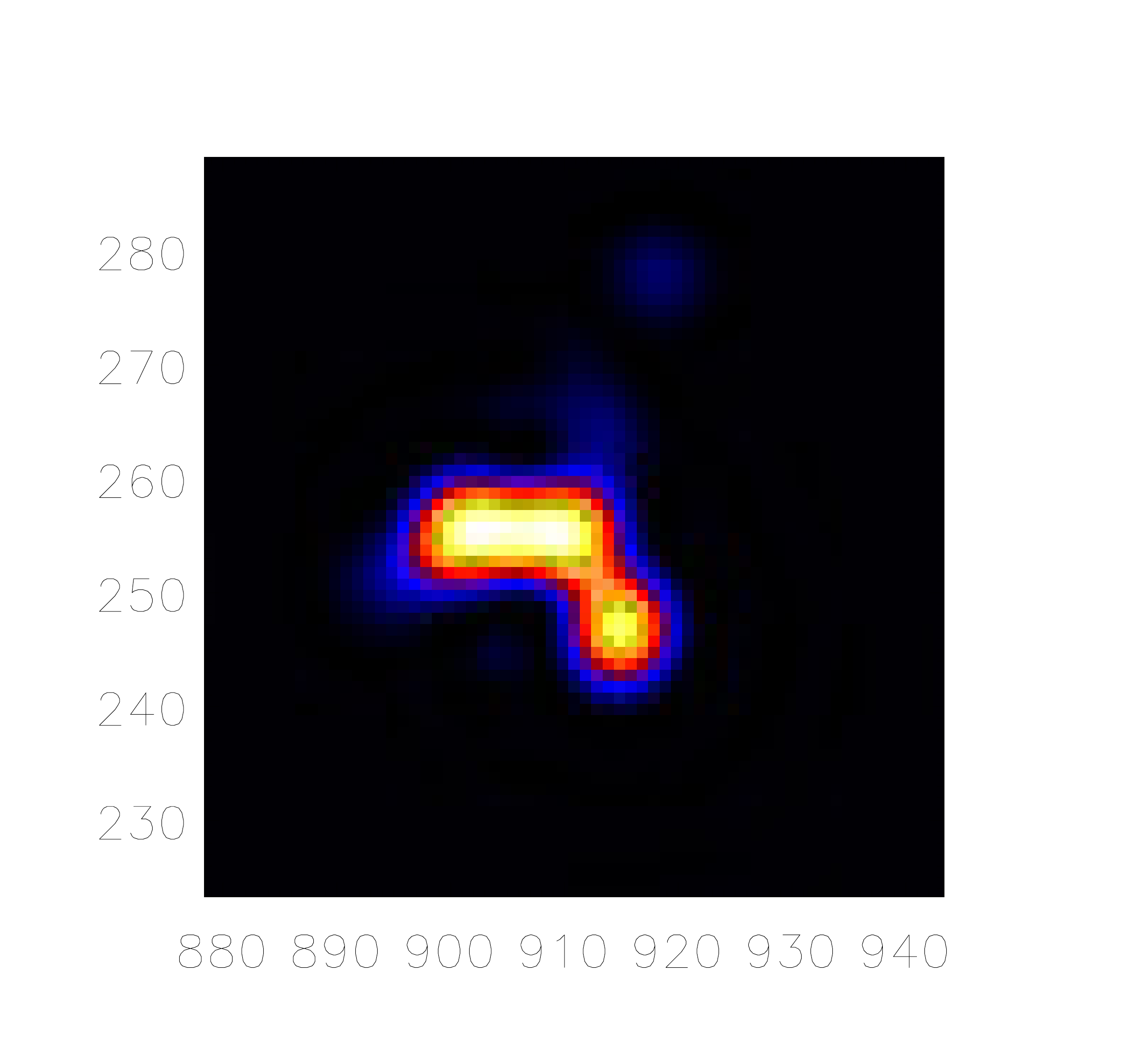}\\ \vspace{-1cm}
    \end{minipage}   
        \begin{minipage}{.11\textwidth}
\centering
ASMC \quad
\end{minipage}
\begin{minipage}{.88\textwidth}
 \includegraphics[height=3.7cm]{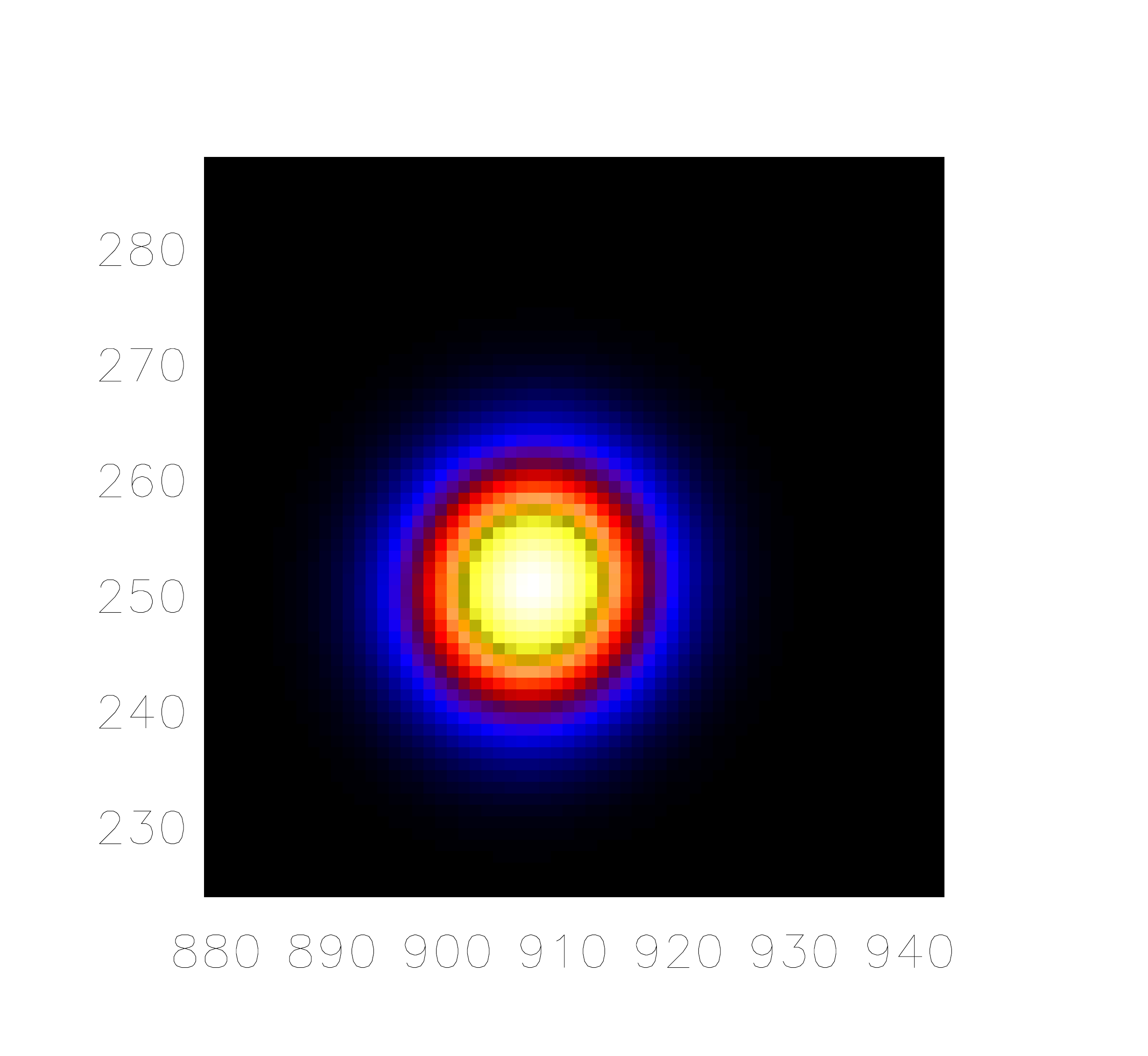}\hspace{-1cm}
    \includegraphics[height=3.7cm]{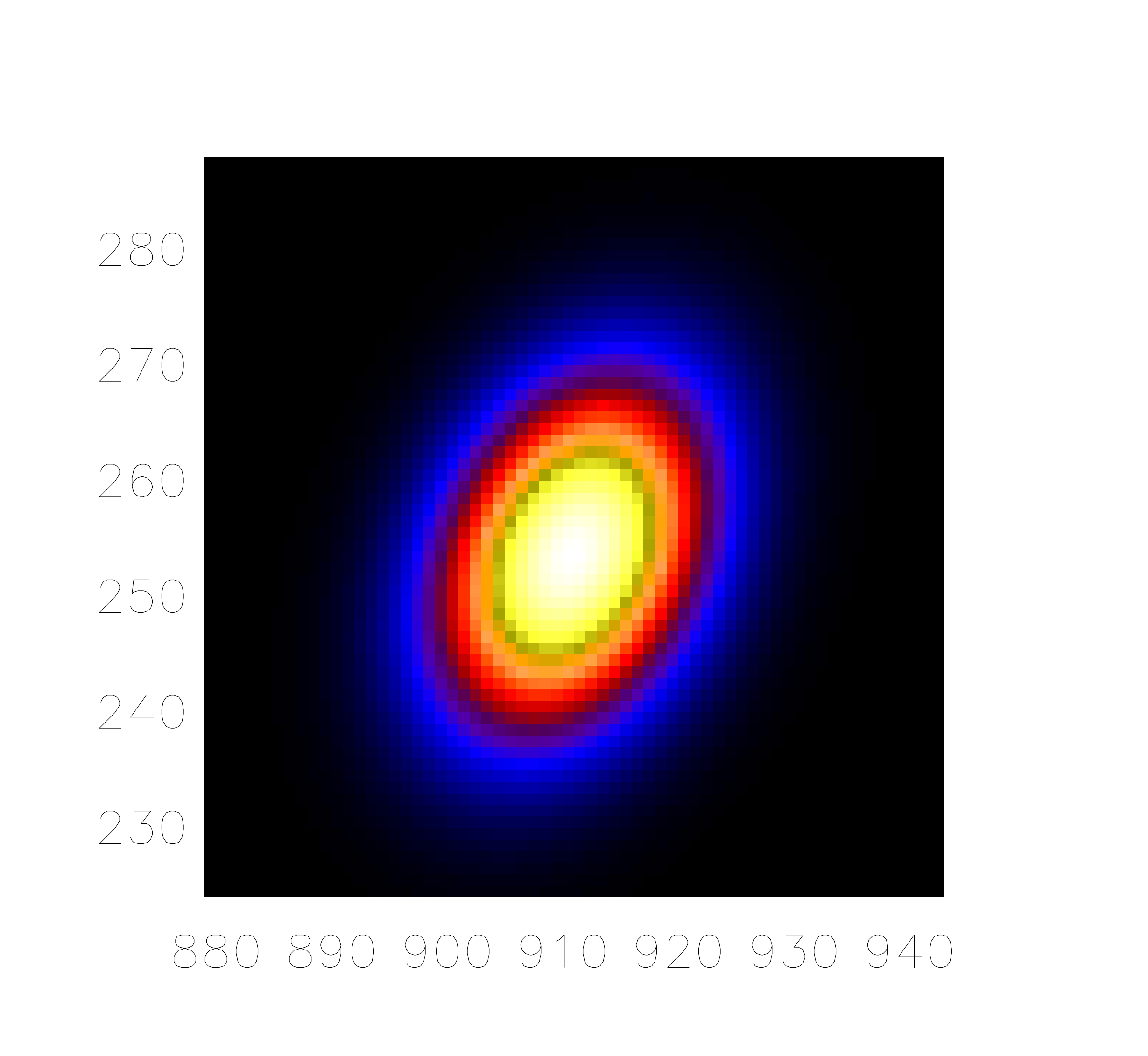}\hspace{-1cm}
    \includegraphics[height=3.7cm]{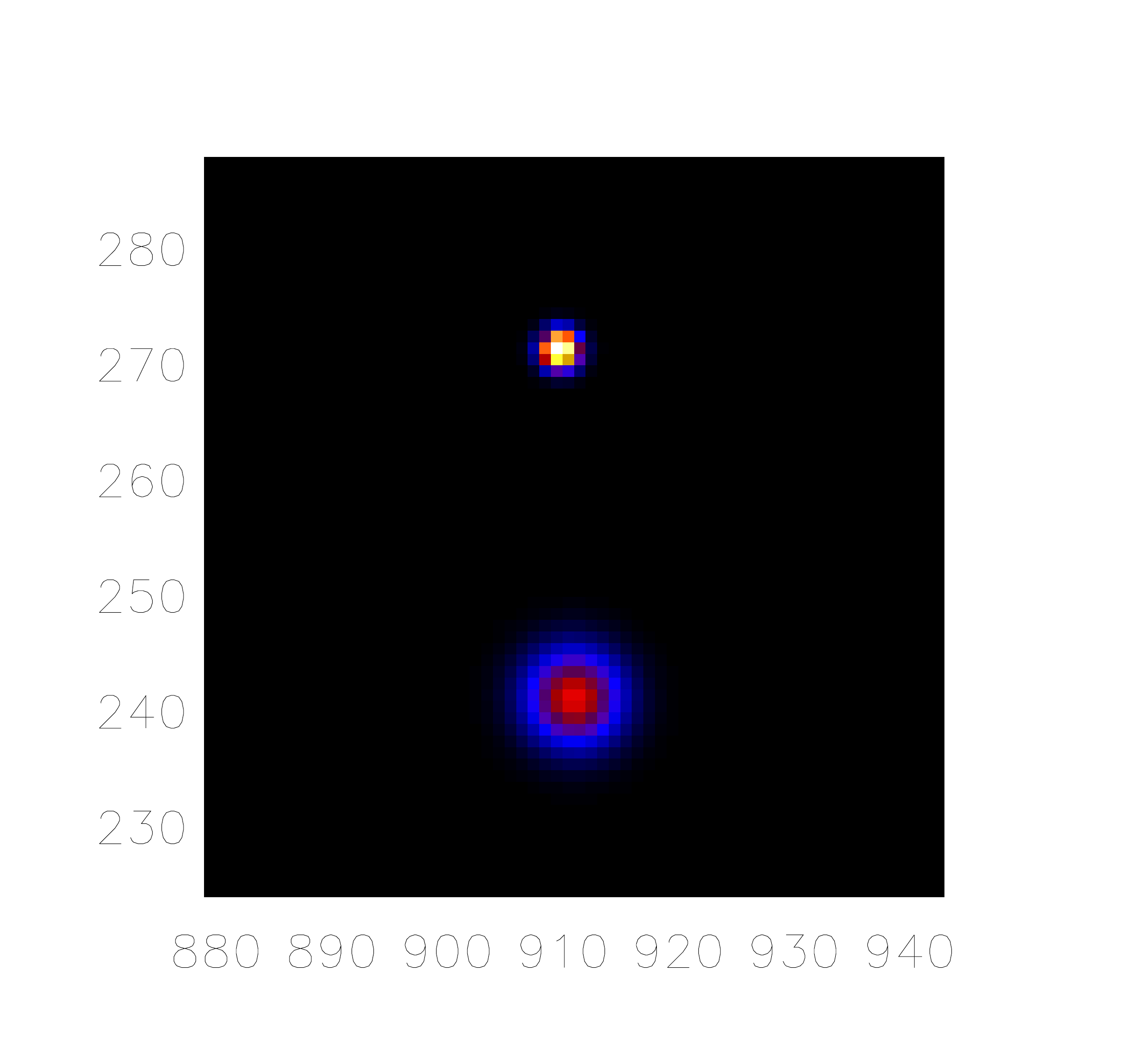}\hspace{-1cm}
    \includegraphics[height=3.7cm]{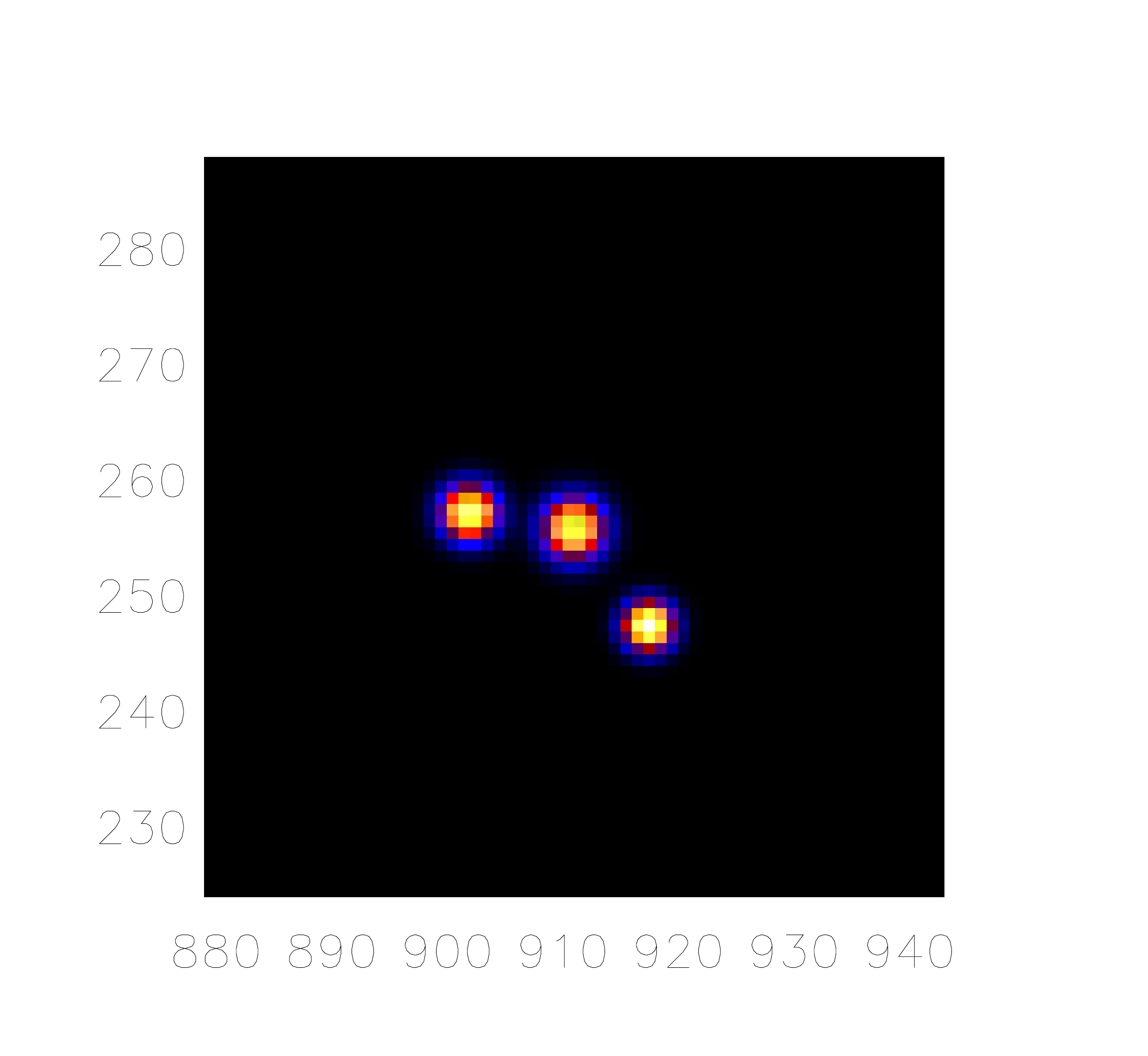}\\ \vspace{-.5cm}
    \end{minipage}
  \caption{High noise case. Reconstructions obtained by VIS{\_}CLEAN (first row), VIS{\_}FWDFIT (second row), VIS\texttt{\_}CS (third row), VIS\texttt{\_}WV (fourth row) and ASMC (fifth row). True sources in \cref{fig:original}.}
  \label{fig:testfig_mm_noise}
\end{figure}

In  \cref{fig:testfig} and \cref{fig:testfig_mm_noise}, the reconstructions given by VIS{\_}CLEAN, VIS{\_}FWDFIT, VIS{\_}CS, VIS\texttt{\_}WV and the ASMC are shown in the case of low and high noise, respectively.  We can clearly see that, even though  VIS{\_}CLEAN has been extensively used, it does not perform very well: the images look too smooth and different sources are recovered as only one (see for instance the S4C). The VIS{\_}FWDFIT has some limitations when there are multiple sources, mainly due to the fact that  in the current version only a single circle, or a single ellipse, or a single loop, or two circles can be recovered, therefore  complex configurations as the ones in  S1E1C and S4C are reconstructed with only two simple circles.   
The compressed sensing and wavelet-based reconstruction methods  work better: most of the reconstructed images are close to the original test images, but some details are lost and in VIS\_CS some spurious sources appear, probably due to the noise in the data (see for instance the top right corner). 

In the low noise case, the reconstructions given by the proposed Bayesian approach (fifth row in \cref{fig:testfig}) are very similar to the original ones (\cref{fig:original}), and the method seems to overcome some of the issues of the VIS\texttt{\_}CS and VIS\texttt{\_}WV. In fact, for the configuration S1E1C (second column in \cref{fig:original}), it is possible to recognize both the ellipse and the circle, while in VIS\texttt{\_}CS and VIS\texttt{\_}WV only one source seems to be recovered. Furthermore, the shape of the loop-source (first column in \cref{fig:original}) is more accurate than with VIS\texttt{\_}CLEAN, VIS\texttt{\_}CS and VIS\texttt{\_}WV. 

When the noise level is high, all the recontructions get worse; the Bayesian reconstructions, in particular, tend to lose some degree of complexity, either missing entire sources or estimating them as simpler geometrical objects (e.g. a circle rather than a loop in S1L). Some of these drawbacks can be overcome by tuning the parameters, as discussed below; we notice however that this behaviour can be simply interpreted as a consequence of the lower information content of the data, due to the presence of noise. 

\begin{figure}[h!]
  \centering
  \includegraphics[height=3.1cm]{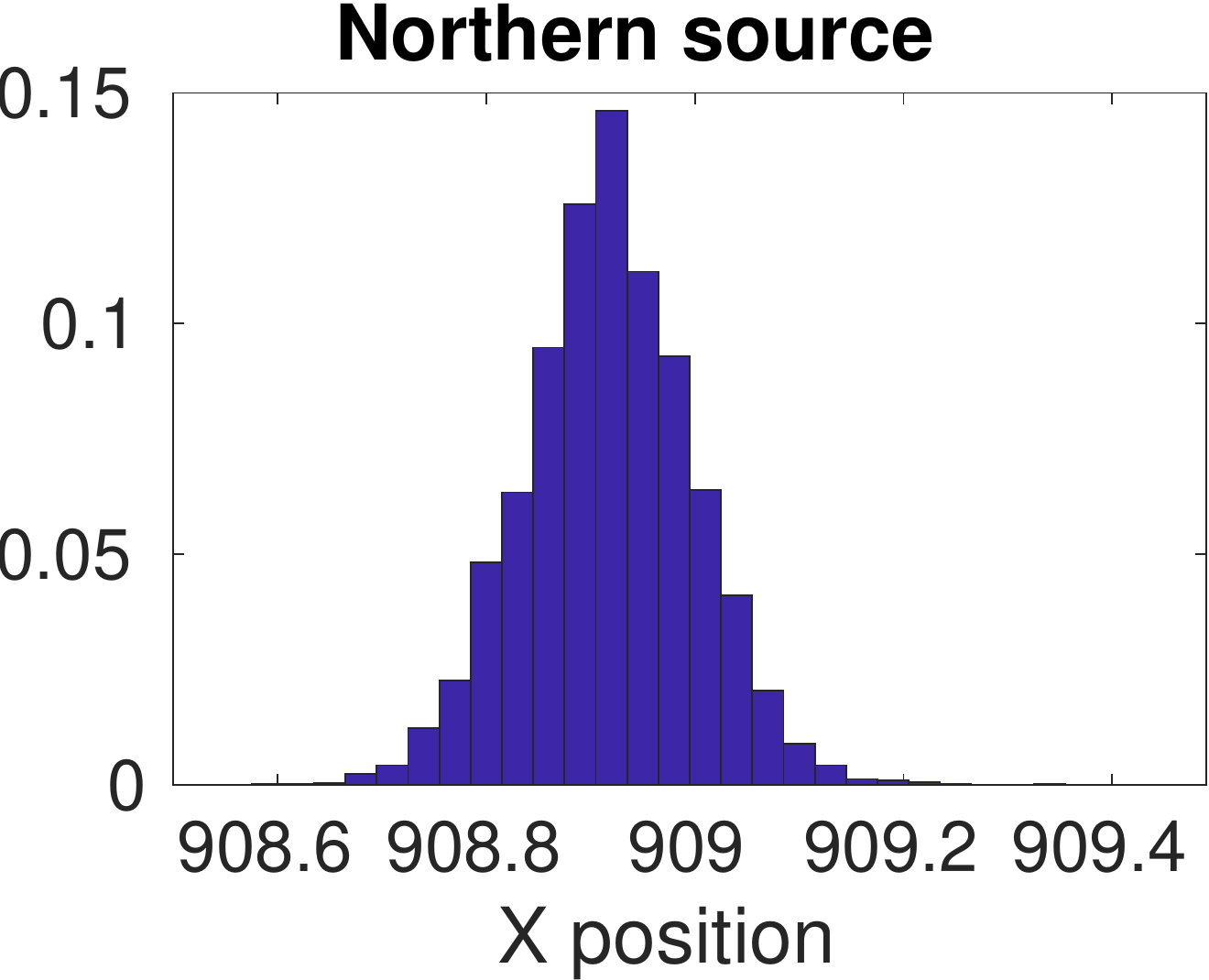}
    \includegraphics[height=3.1cm]{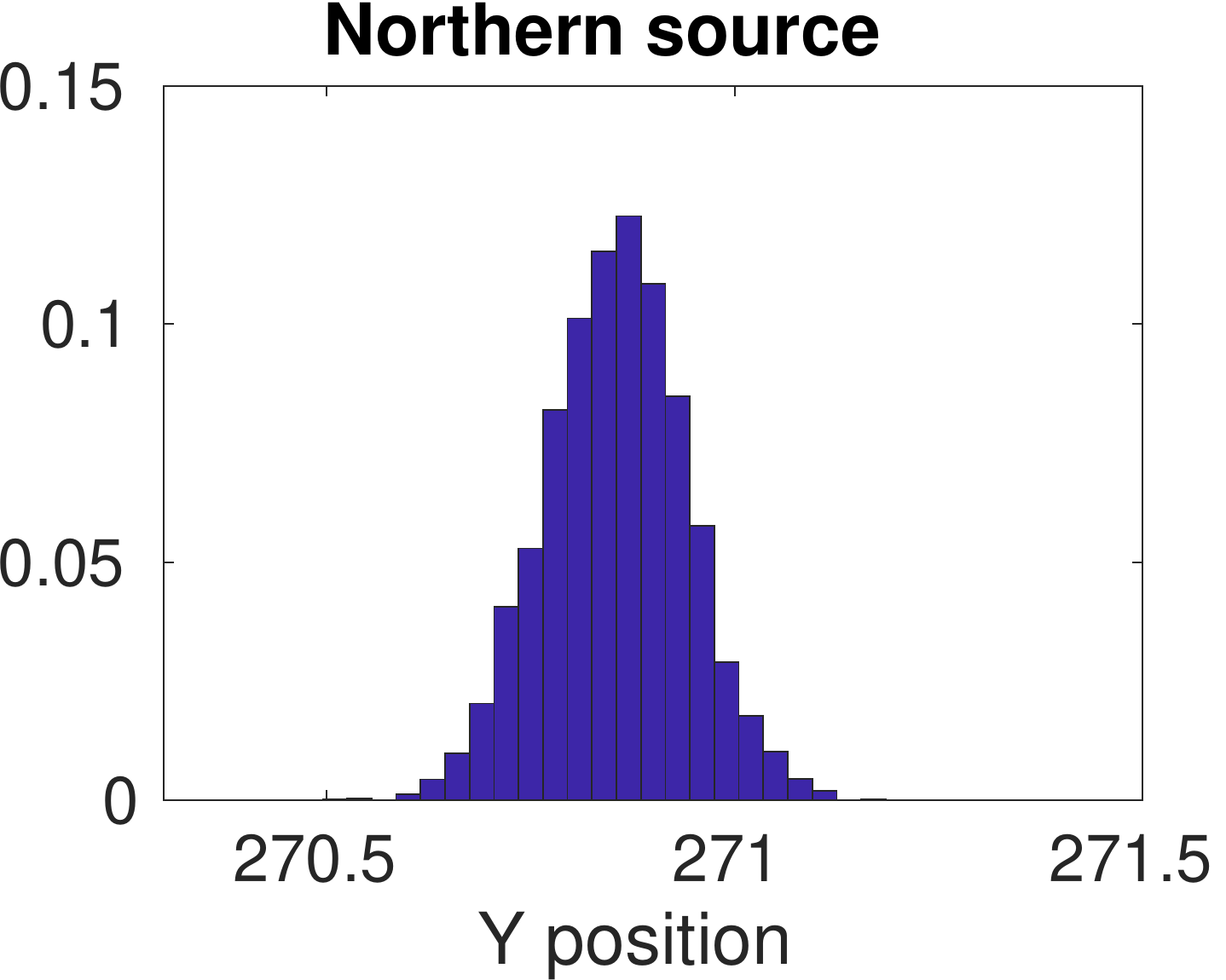}
    \includegraphics[height=3.1cm]{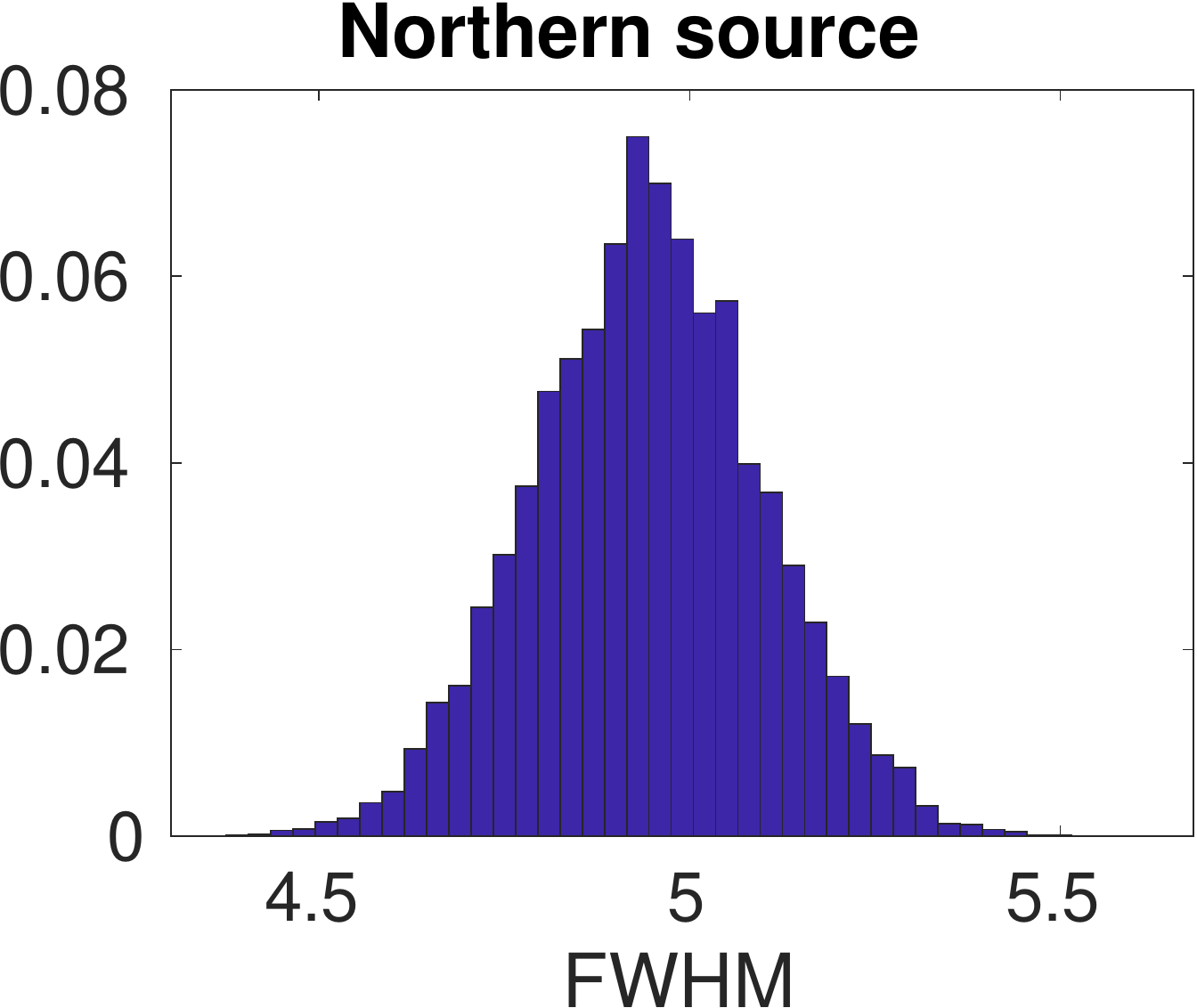}
    \includegraphics[height=3.1cm]{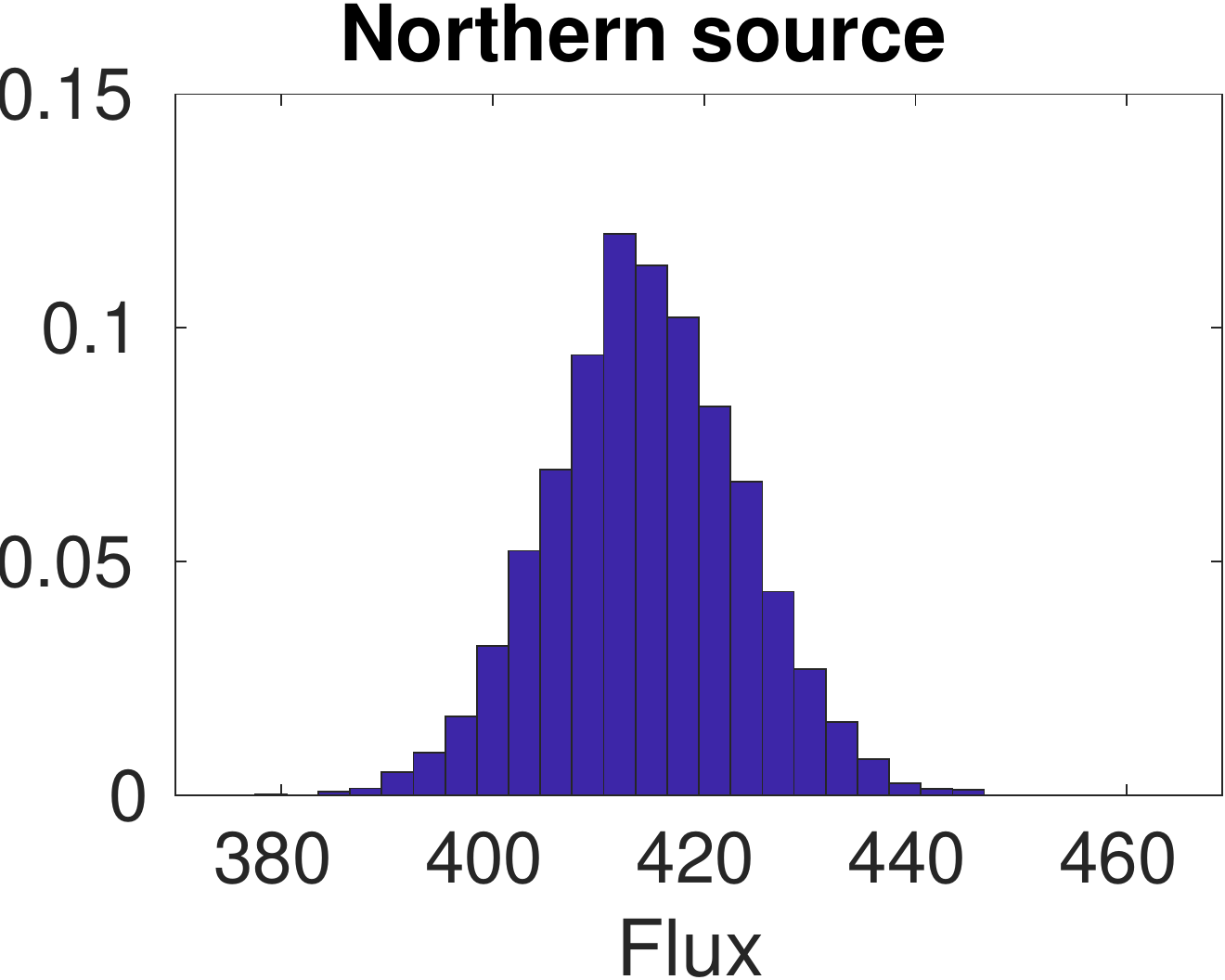}\\ 
 \includegraphics[height=3.1cm]{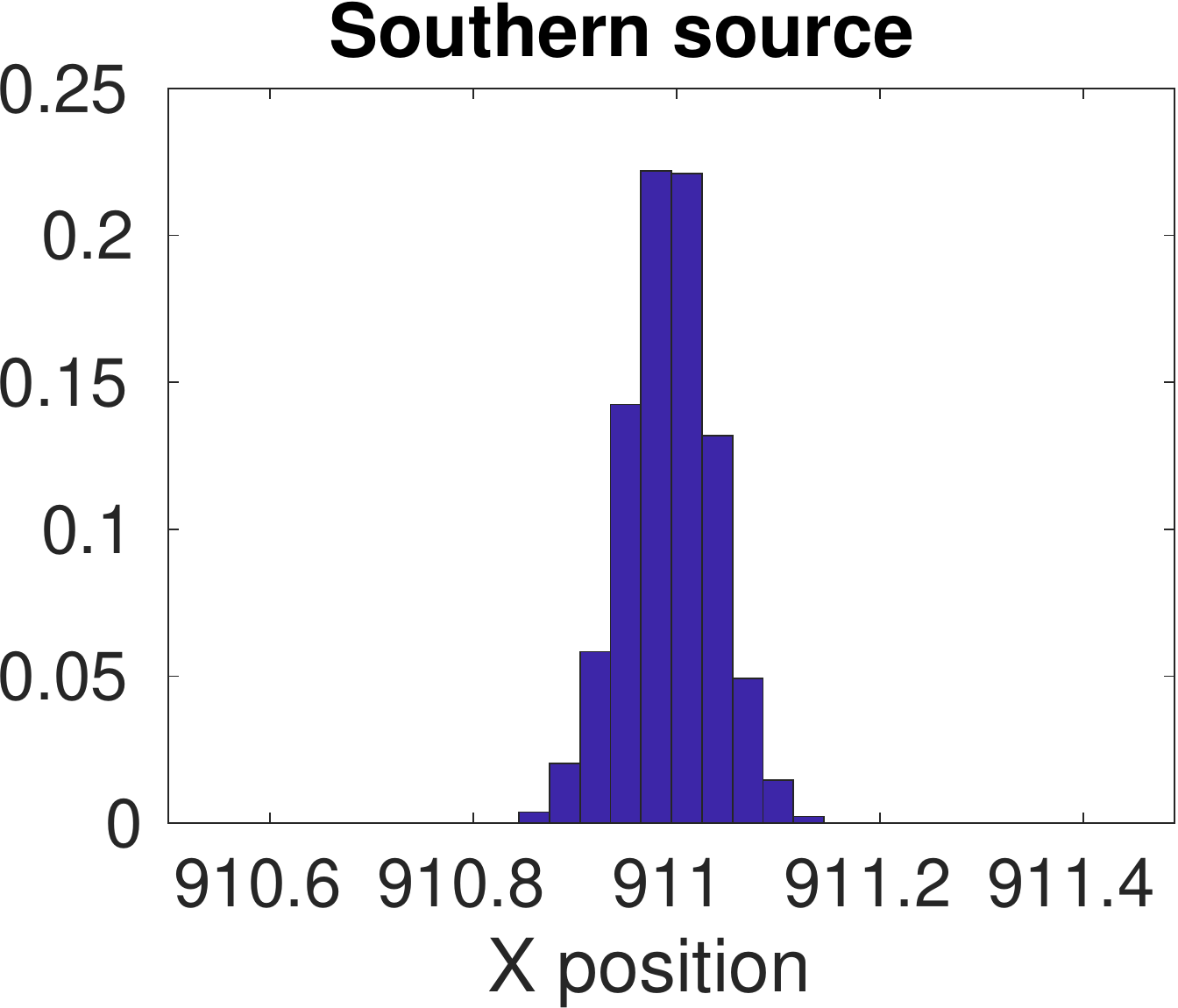}
    \includegraphics[height=3.1cm]{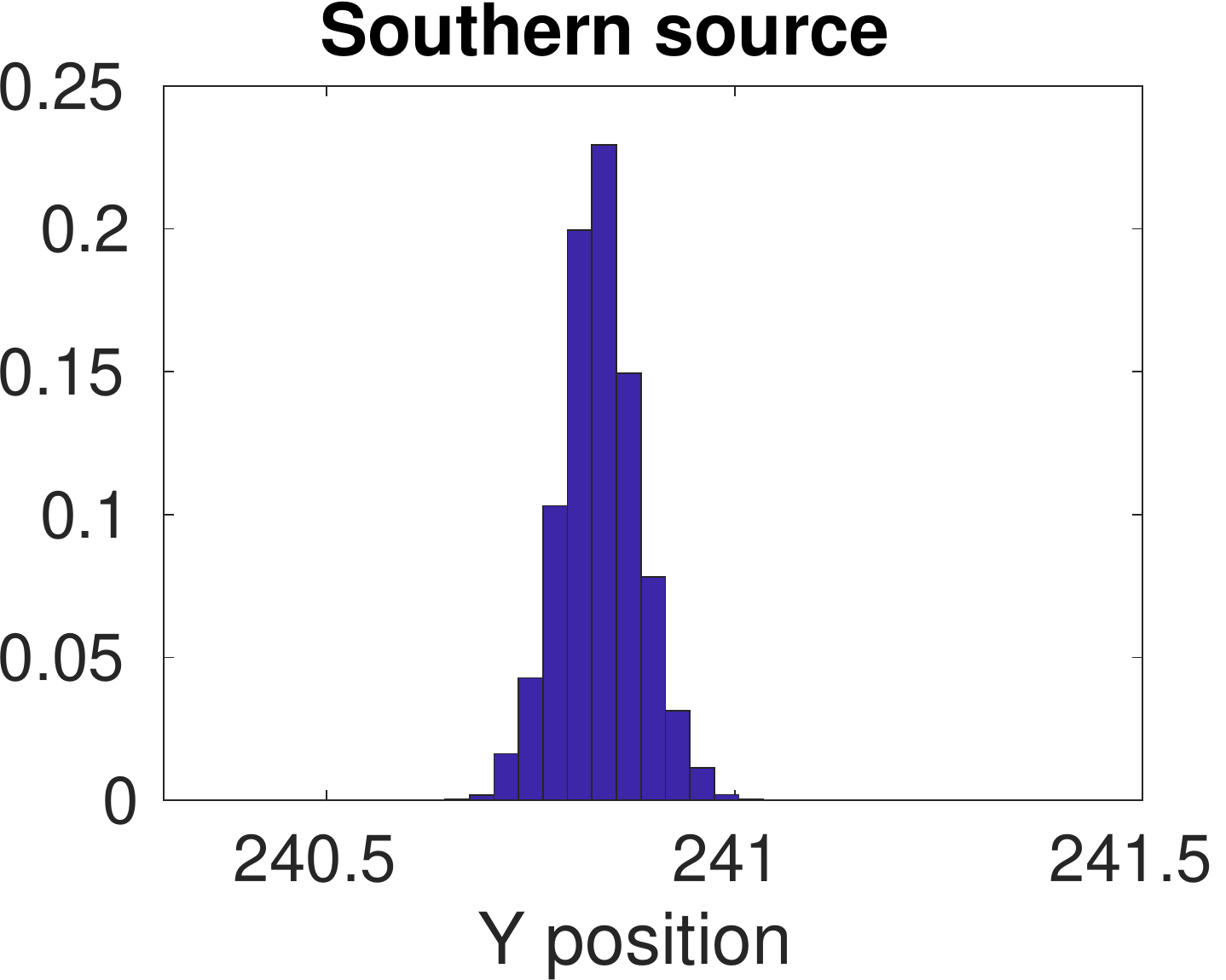}
    \includegraphics[height=3.1cm]{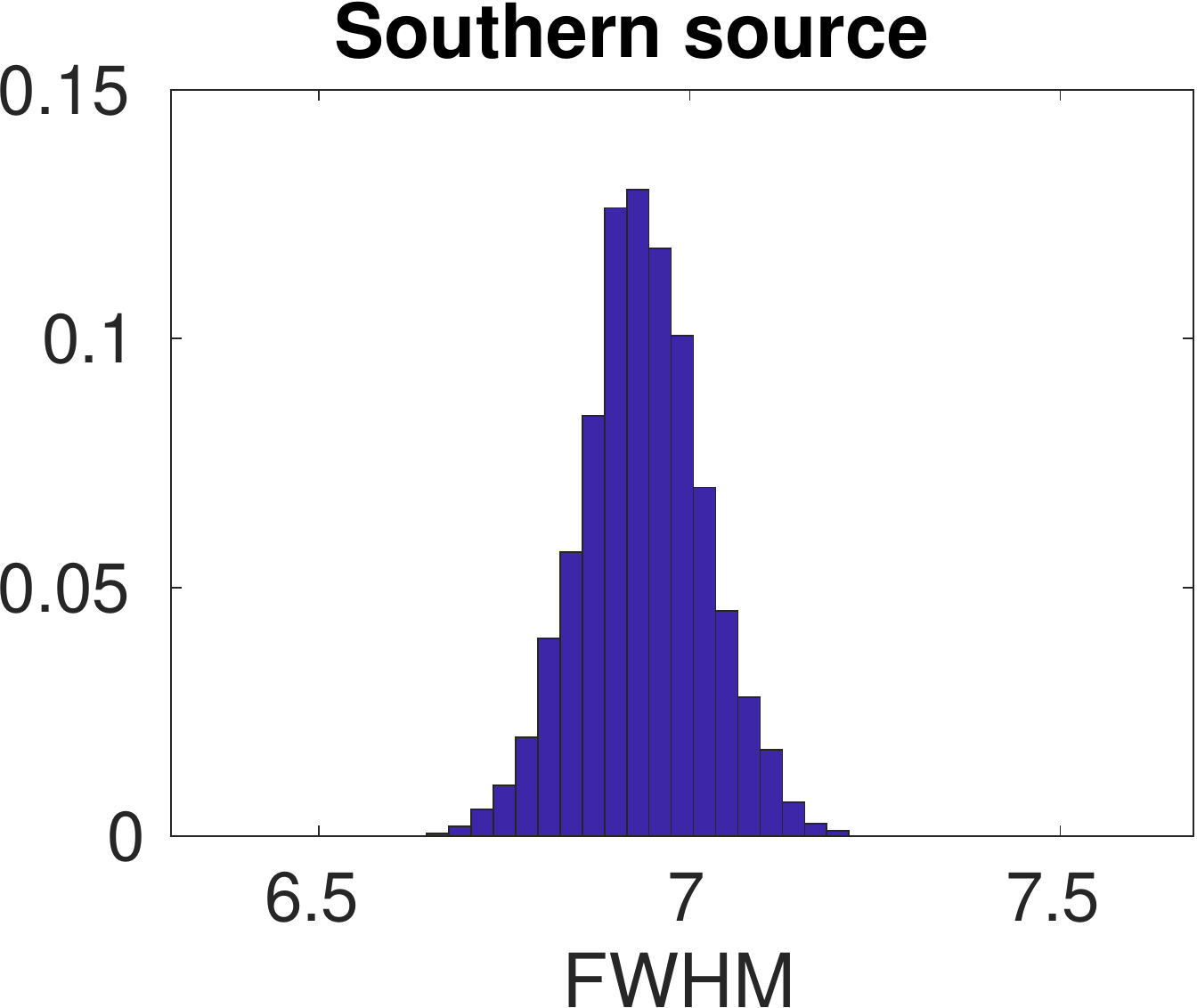}
    \includegraphics[height=3.1cm]{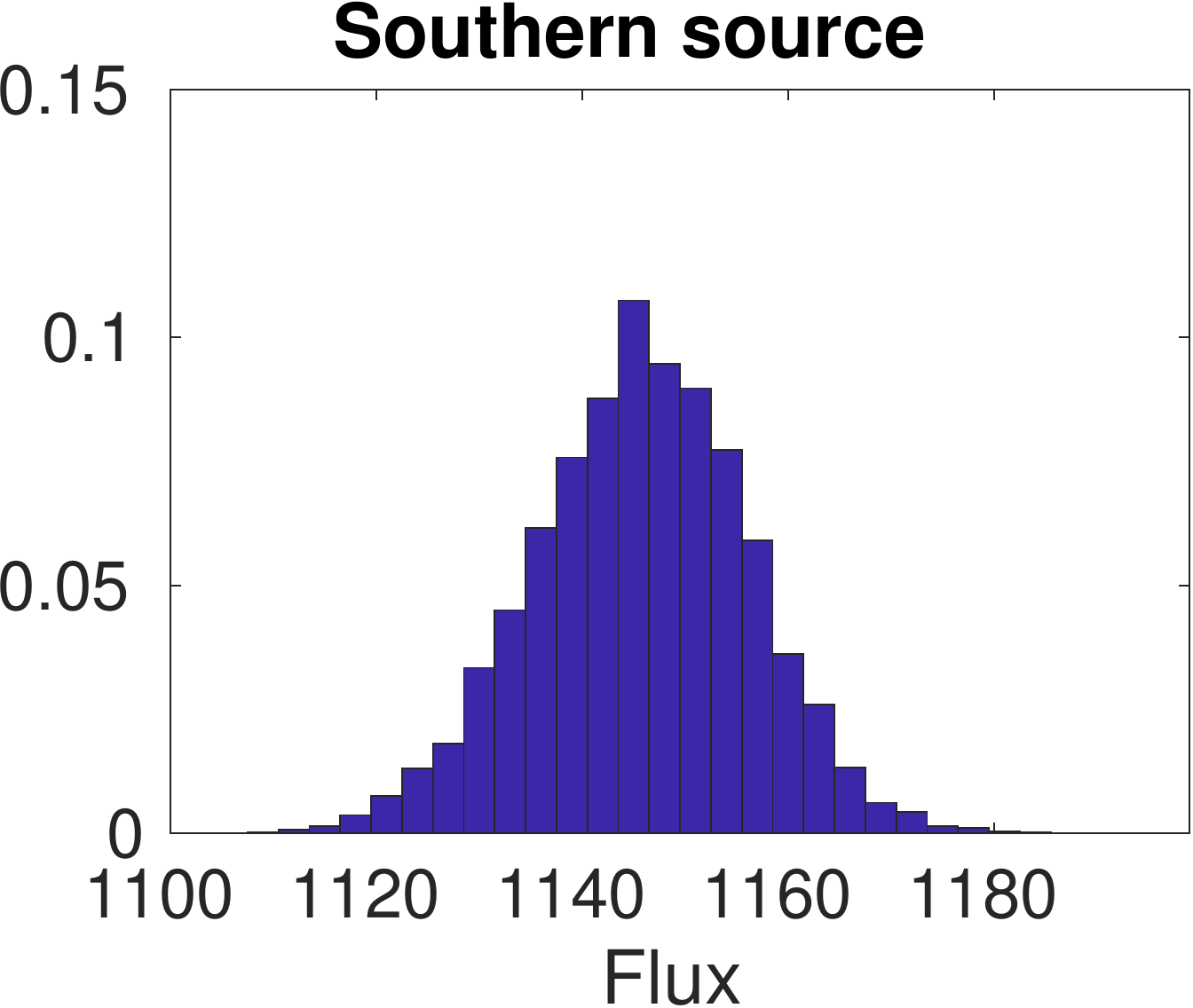}
  \caption{Approximated posterior distributions, obtained as the weighted histograms of the Monte Carlo samples, of the parameters for the S2C, low noise case.}
  \label{fig:hist_22}
\end{figure}

We finally notice that, even though in \cref{fig:testfig} and \cref{fig:testfig_mm_noise} we only compare the estimated images, the posterior distribution approximated by the ASMC contains much more information. First, it contains the estimated number of sources, which turned out to be: 1 for S1L, for both noise levels; 2 for S1E1C in the low noise case, 1 in the high noise case; 2 for S2C, for both noise levels; 4 for S4C in the low noise case, 3 in the high noise case. Furthermore, the posterior distribution also contains uncertainty on the individual parameters, as the empirical variance of the parameter values in the Monte Carlo samples. In order to visualize such variance, some form of re--ordering (or partitioning) of the sources is necessary, because the order of the sources is not necessarily consistent across Monte Carlo samples: here we used a $k$--means \cite{hartigan1979algorithm} algorithm with $\hat{N}_S$ clusters, applied to the set of all individual sources extracted from the Monte Carlo samples. Of course, when applying such post--processing technique care must be taken that sources belonging to the same particle do not fall in the same cluster; this never happened in the simulations below. In \cref{fig:hist_22} we show the posterior distributions for the estimated parameters for the S2C in the low noise case. We observe that both the location and the size of the southern source have smaller standard deviation than those of the northern source; this is expected because the southern source is stronger.

\begin{figure}[h!]
\centering
 \includegraphics[height=5.5cm]{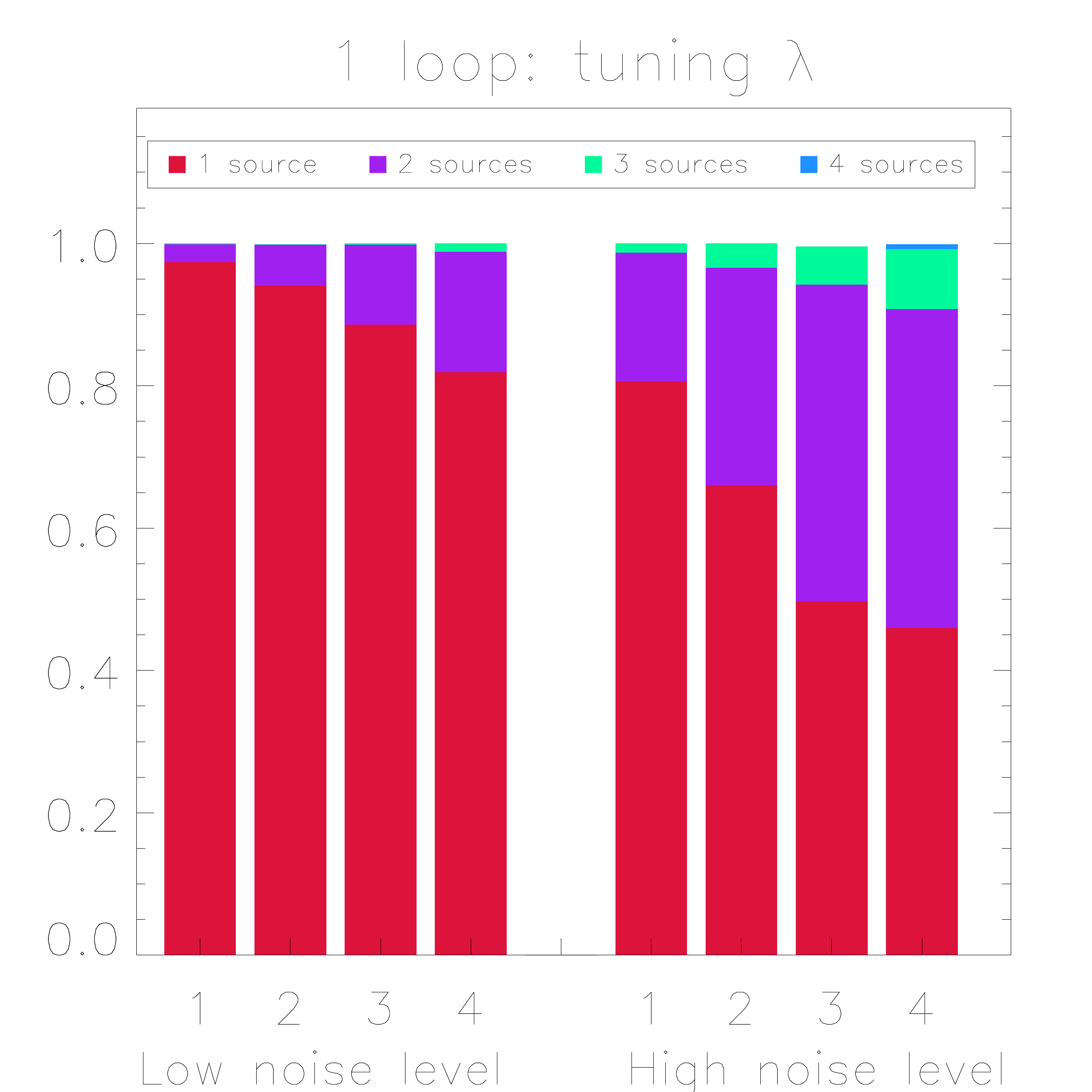}
 \includegraphics[height=5.5cm]{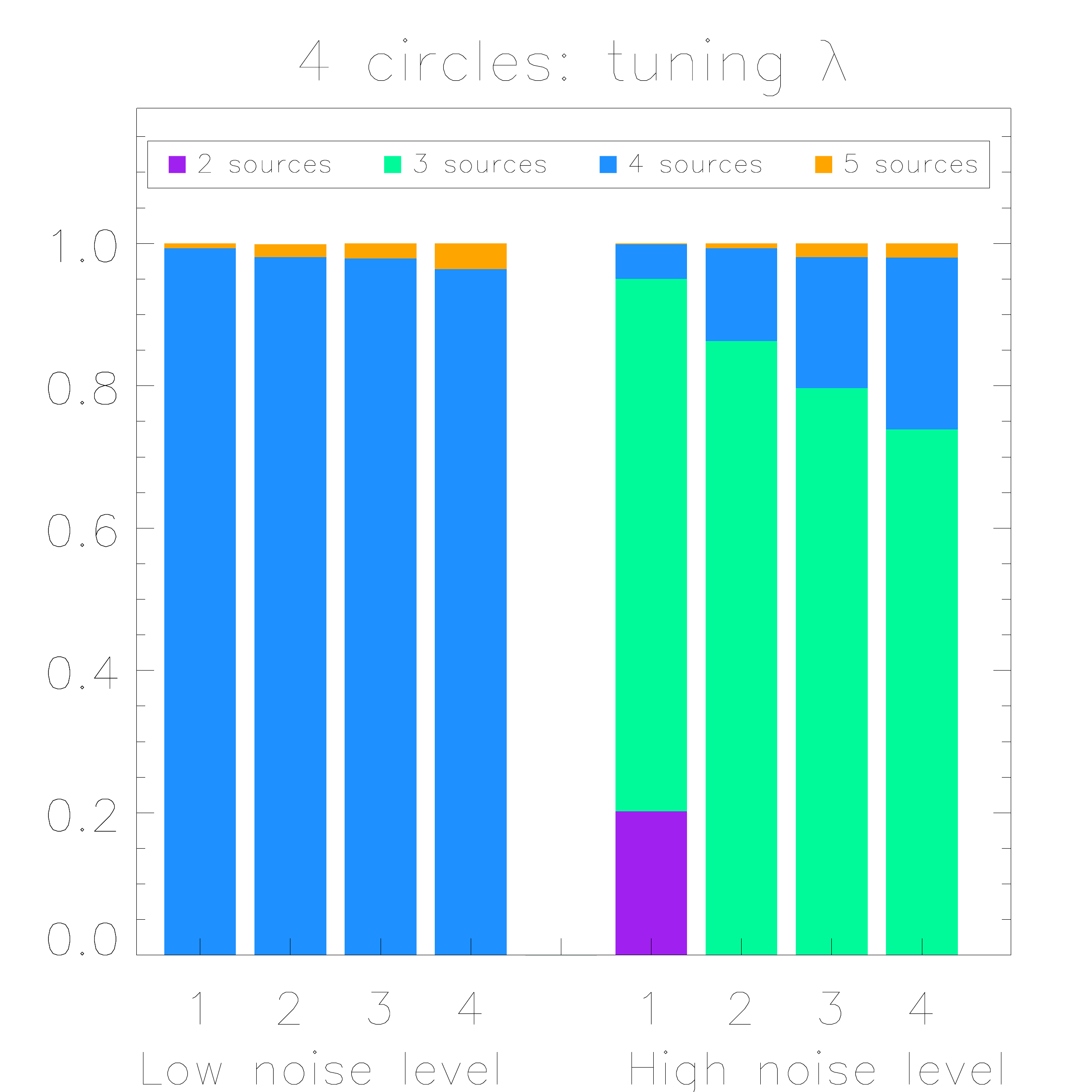} 
 \caption{Posterior probabilities for the number of sources for the single loop (left column) and the four circles (right column), with different values of the prior parameter $\lambda \in [1,4]$. } 
 \label{fig:lambda}
\end{figure}

\begin{figure}[h!]
  \centering
\includegraphics[height=3.2cm]{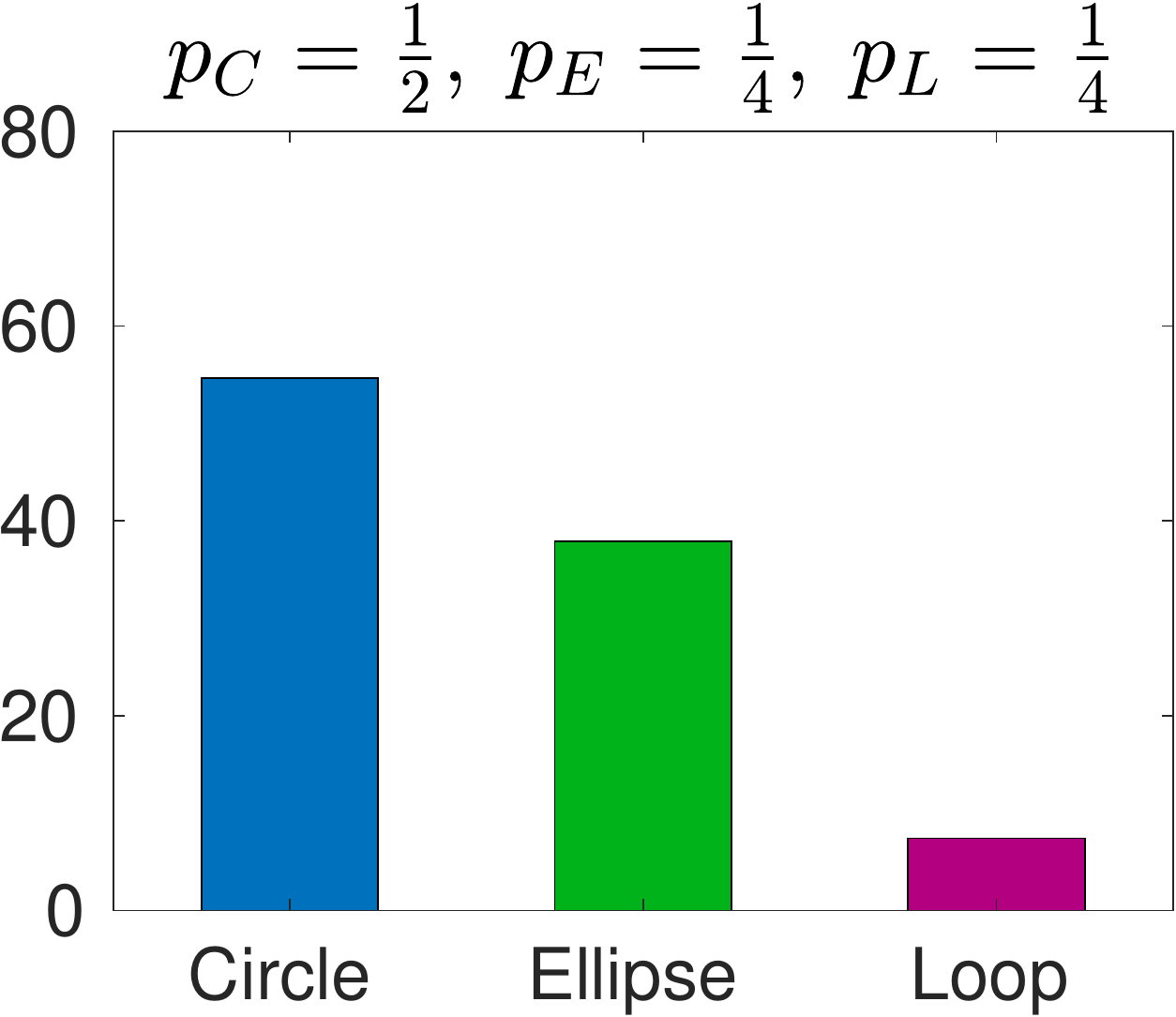}
\includegraphics[height=3.2cm]{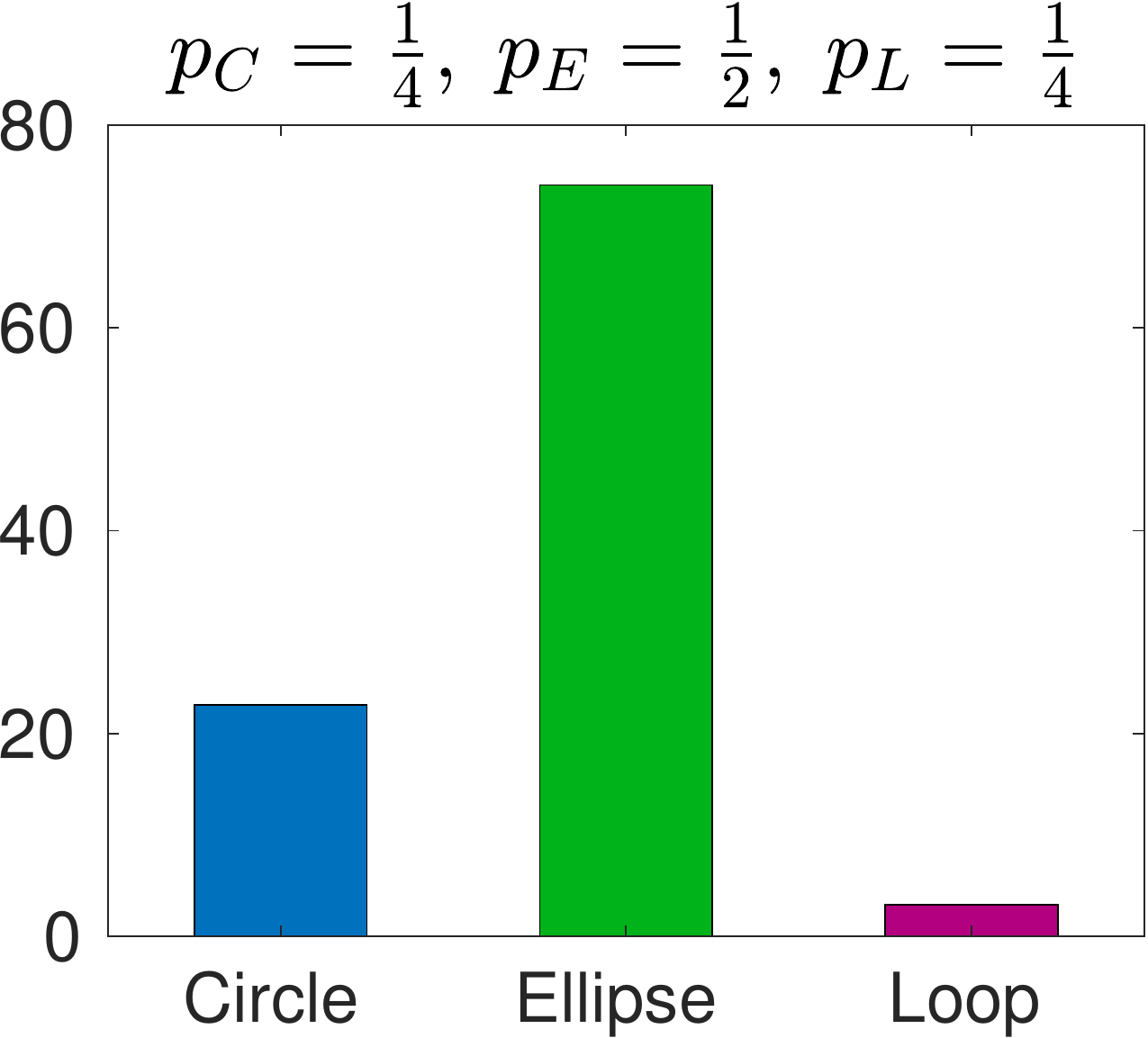}
\includegraphics[height=3.2cm]{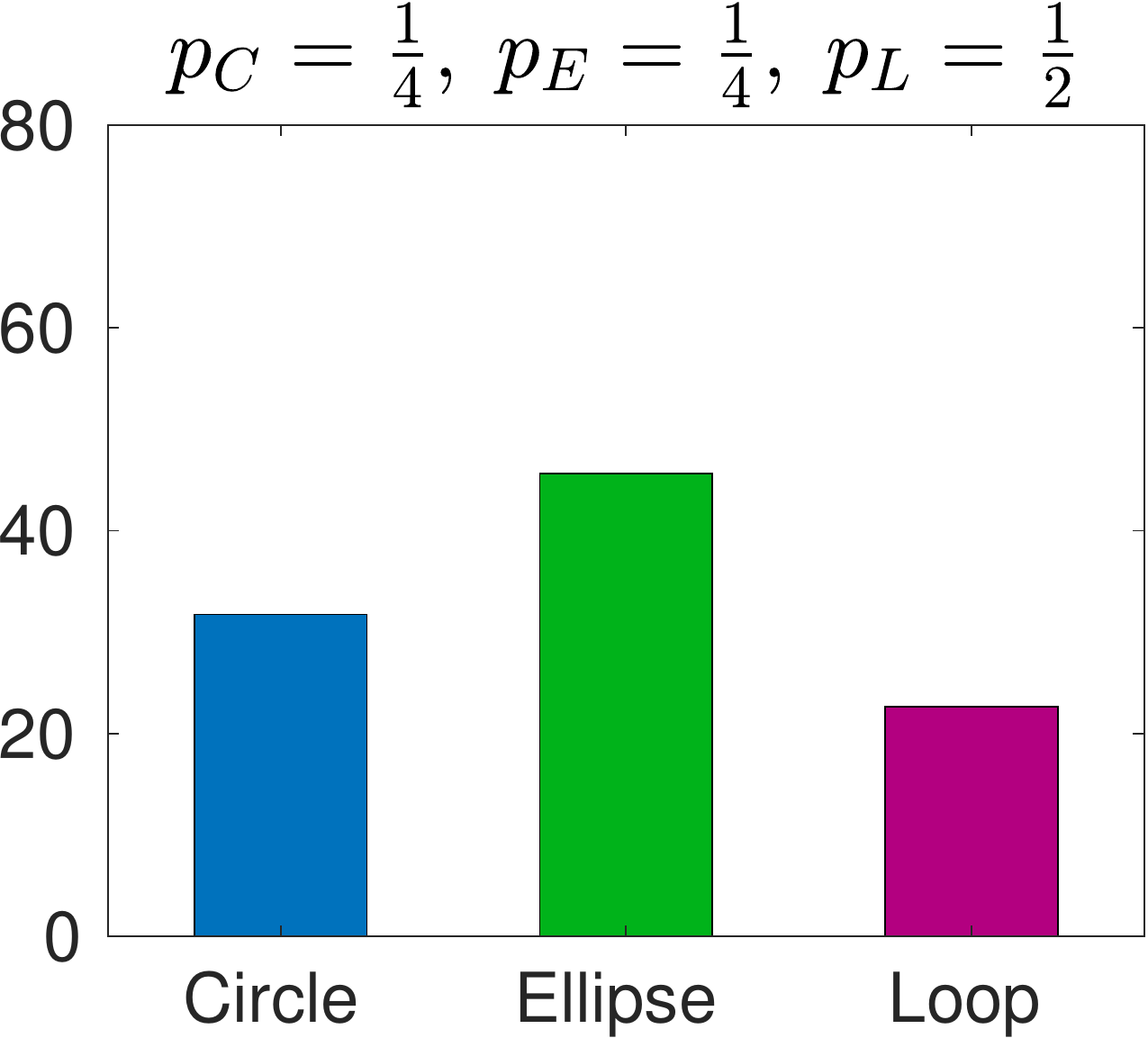}
  \caption{Posterior probabilities for the source types for the single loop configuration with high noise level, using different a priori probability sets: $(p_C,p_E, p_L)=\Big(\frac{1}{2}, \frac{1}{4}, \frac{1}{4}\Big)$ (left);  $(p_C,p_E, p_L)=\Big(\frac{1}{4}, \frac{1}{2}, \frac{1}{4}\Big)$ (middle); $(p_C,p_E, p_L)=\Big(\frac{1}{4}, \frac{1}{4}, \frac{1}{2}\Big)$ (right).}
  \label{fig:pie_loop}
\end{figure}

\subsection{Influence of the parameters}

The results of the Bayesian method presented here can be tuned by changing the parameters in the prior distribution and in the likelihood function, in the same way as the results of a regularization algorithm can be tuned by changing the regularization parameter(s). 
In this Section, we study the impact of (i) the mean number of sources $\lambda$ in the Poisson prior distribution, (ii) the prior probabilities for the source types and (iii) the noise variance. For (i) and (ii) we run the ASMC method with different values of the parameters; for (iii), on the other hand, we exploit the fact that at each iteration the ASMC algorithm approximates a different distribution, that can be interpreted as the posterior distribution under a different noise standard deviation (see the paragraph at the end of  \cref{subsec:SMCsamplers}).

To study the influence of the parameter $\lambda$, in  \cref{fig:lambda} we show the posterior probability distribution of the number of sources for the simulated loop S1L and the simulated four sources S4C with different noise levels, and four different values of $\lambda$. The first  result is that, expectedly, the posterior distribution depends weakly on the value of $\lambda$ for low noise levels, i.e. when the data are highly informative; on the other hand, the posterior probabilities are more affected by the prior when the noise level is high. For example, for the S4C the posterior mode indicates three sources for the high noise case; for increasing values of $\lambda$, however, the posterior probability for the two--source model goes from 20$\%$ to zero, while the posterior probability for the four--source model increases from zero to about 20$\%$. 

To study the influence of the prior on the source type, we considered the S1L case and run the algorithm three times; the prior values were set to favour, each time, a different source type: the preferred source type was given a prior probability of $\frac{1}{2}$, while the other two were given a prior probability of $\frac{1}{4}$. In the low noise case, the results do not depend on the choice of the prior, as the posterior probability assignes more than $99\%$ to the loop type.
The high noise case is more delicate: in \cref{fig:pie_loop} we show the influence of the prior on the source types for the S1L with high noise level. From the histograms, we see that the posterior probabilities are now affected by the priors. This suggests that noise perturbs the data enough to reduce the information on the source shape to such an extent that the prior has a non--negligible impact. This is somewhat confirmed also by visual inspection of the reconstructions obtained by the other methods (see VIS\_CS and VIS\_WV in \cref{fig:testfig_mm_noise}). We notice that even in the case $p_L = \frac{1}{2}$, the posterior indicates the ellipse as the most probable source type; this is coherent with the very little asymmetry of the reconstructions obtained by the other methods in this high--noise scenario.


%
\begin{figure}[h!]
  \centering
\hspace{-1cm}\includegraphics[height=2.8cm]{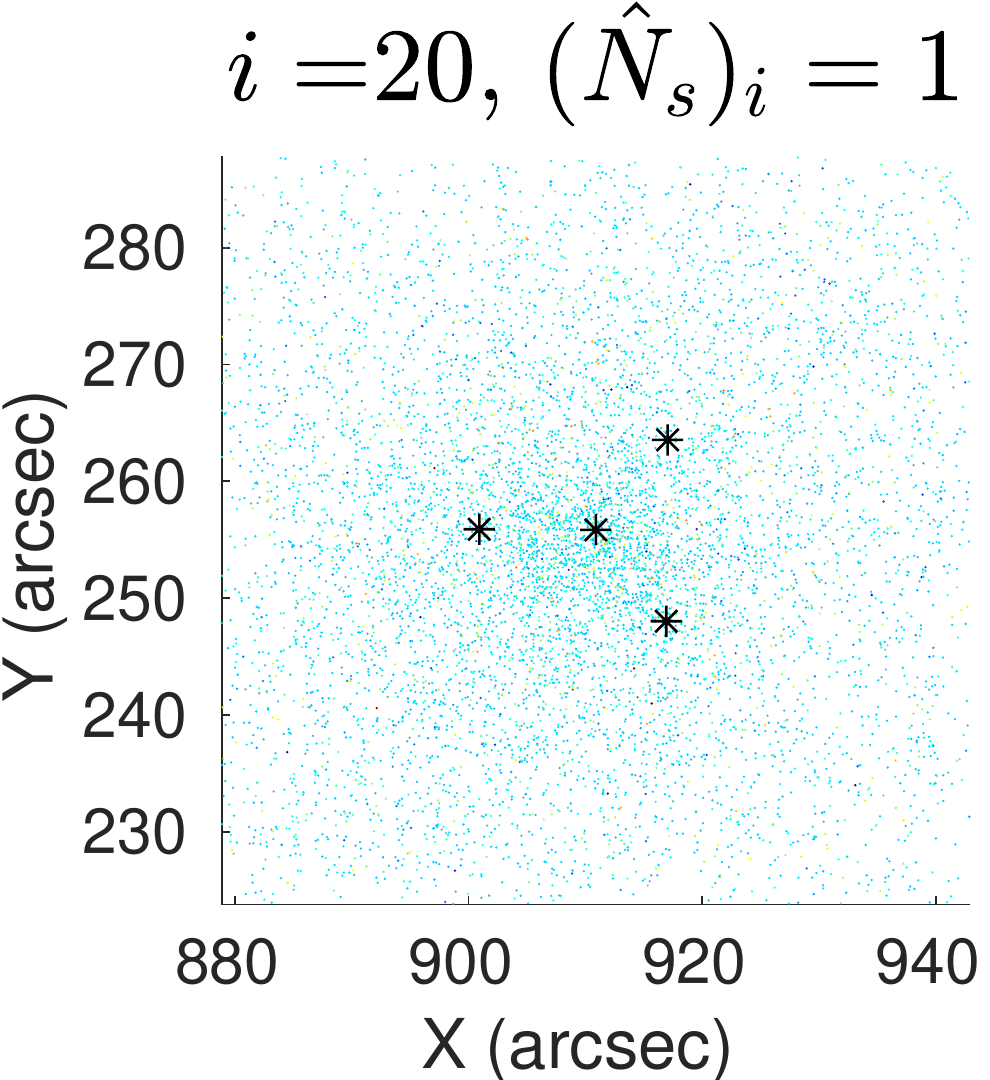}\hspace{0.45cm}
 \includegraphics[height=2.8cm]{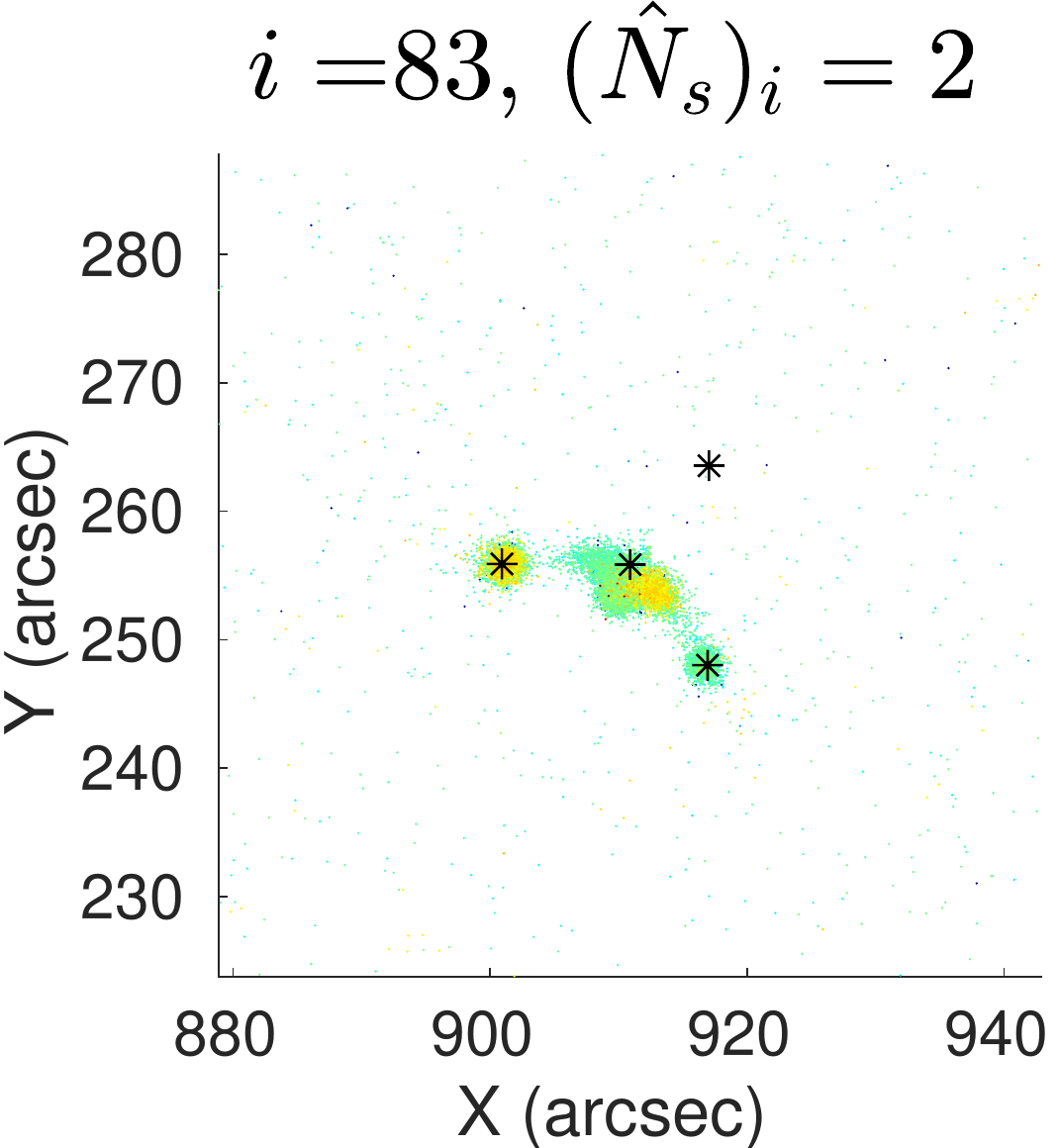}
 \hspace{0.25cm}
 \includegraphics[height=2.8cm]{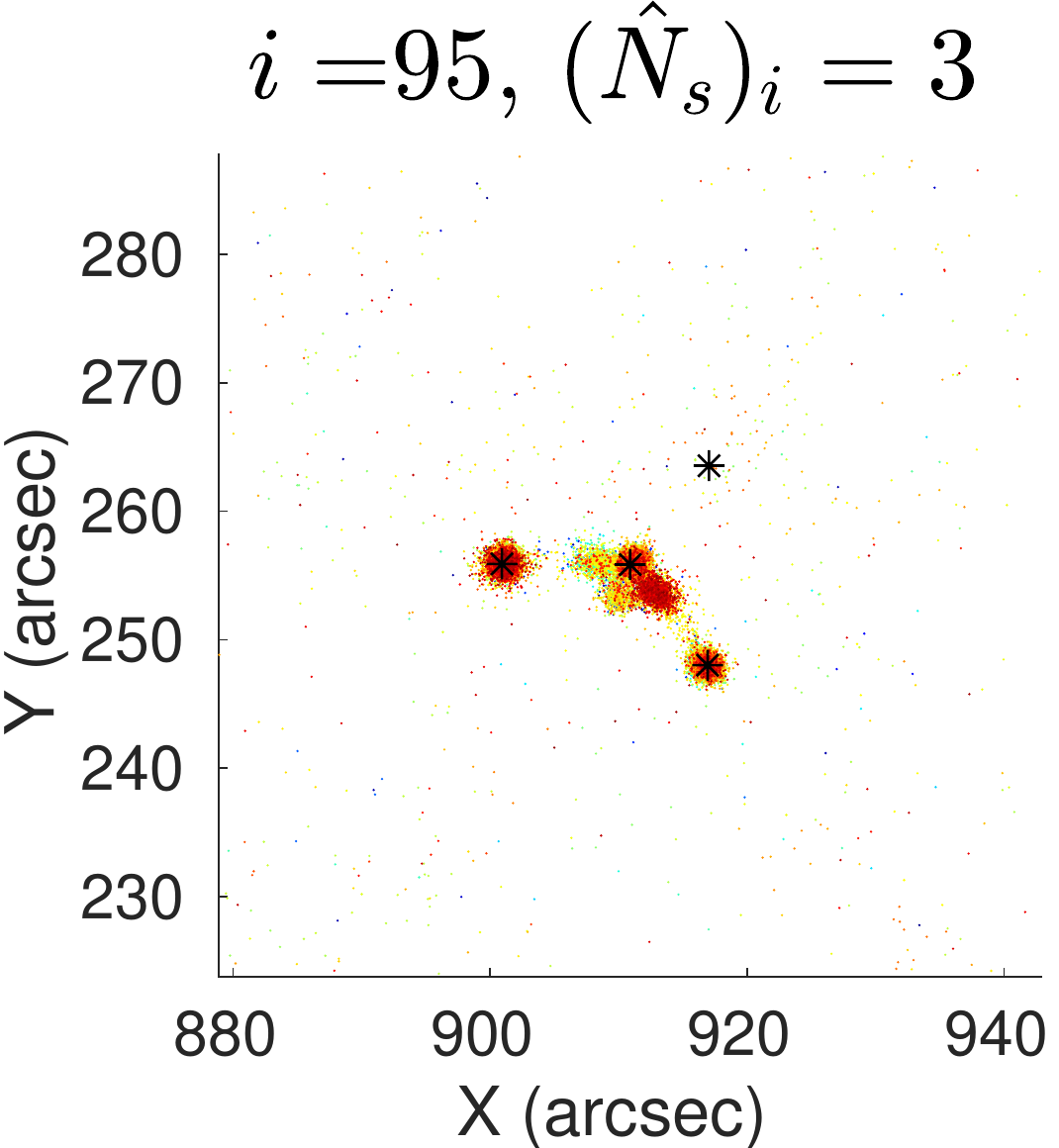}
 \hspace{0.25cm}
  \includegraphics[height=2.8cm]{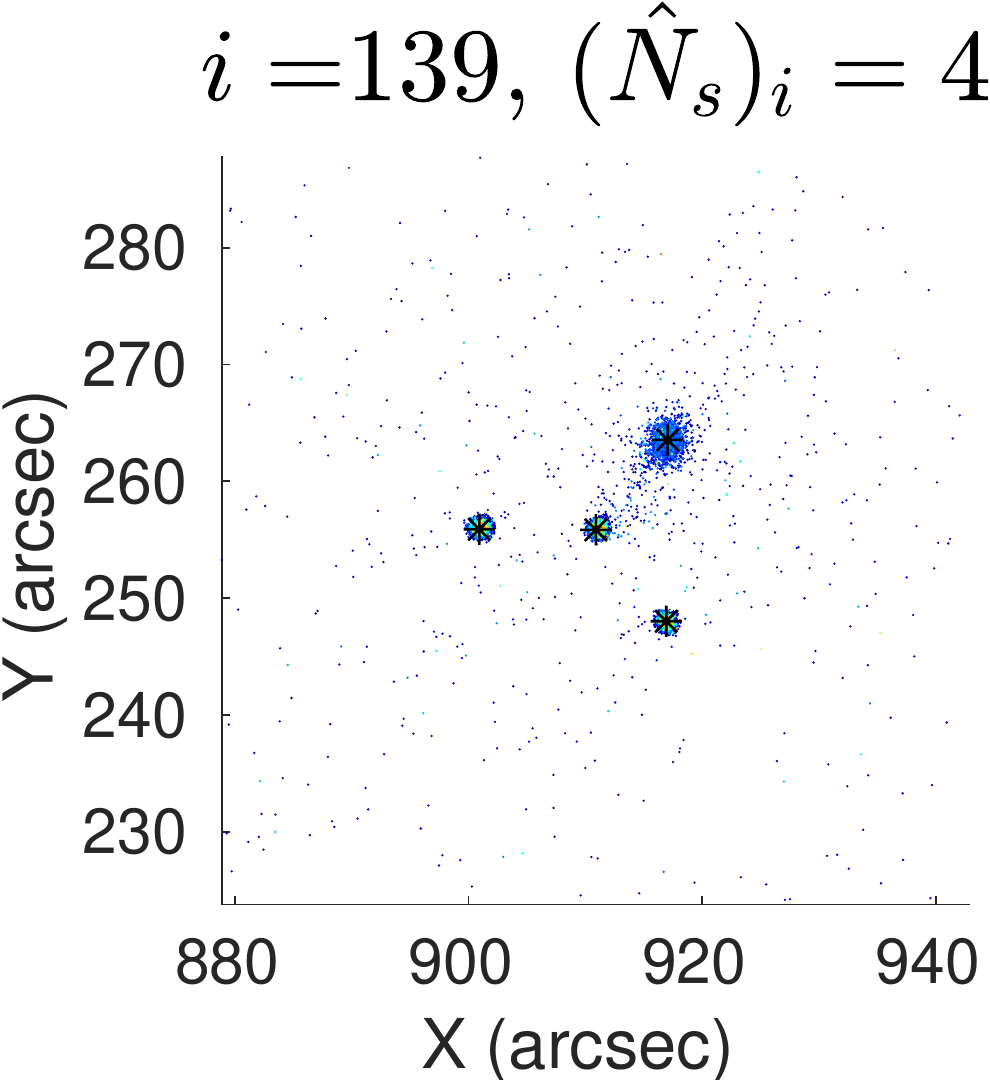}
  \hspace{0.25cm}
    \includegraphics[height=2.8cm]{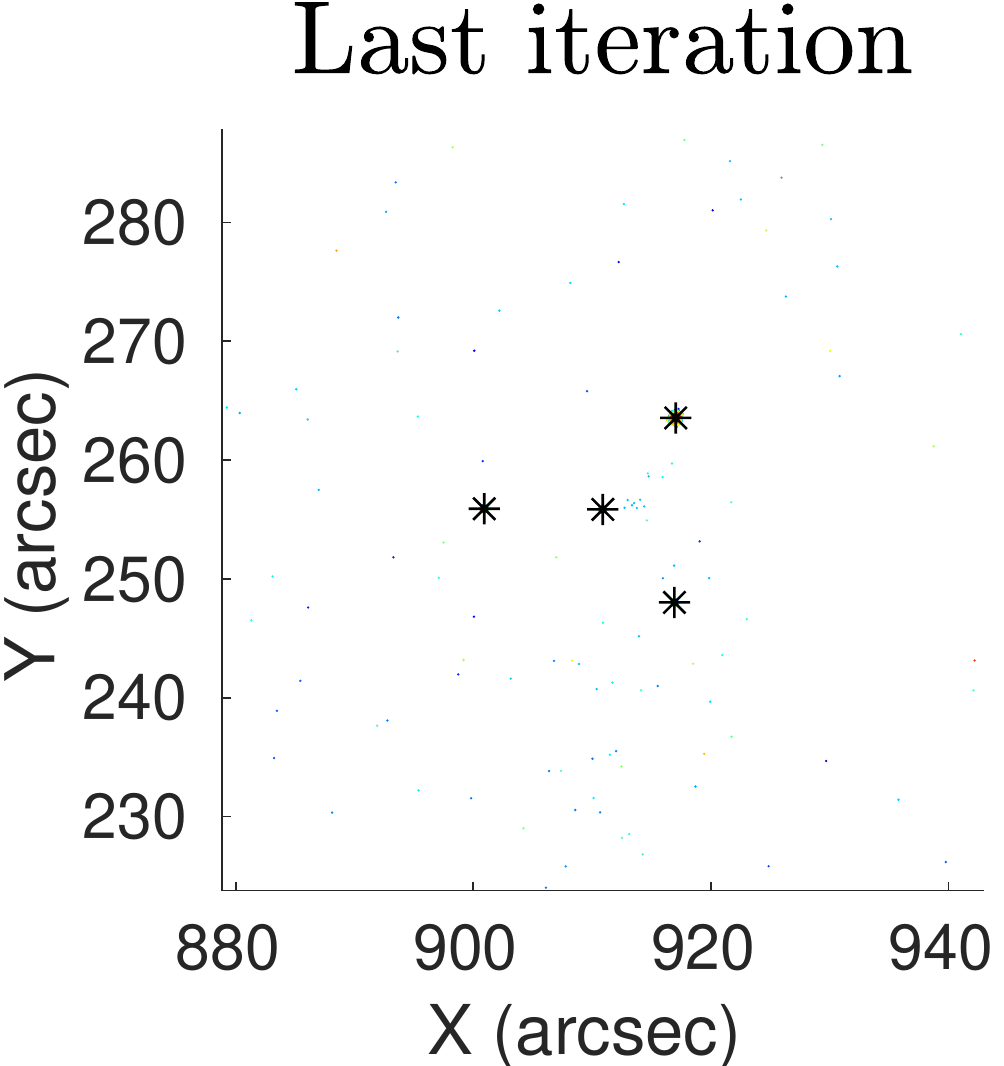}\\
    \hspace{-0.3cm}
  \includegraphics[height=2.9cm]{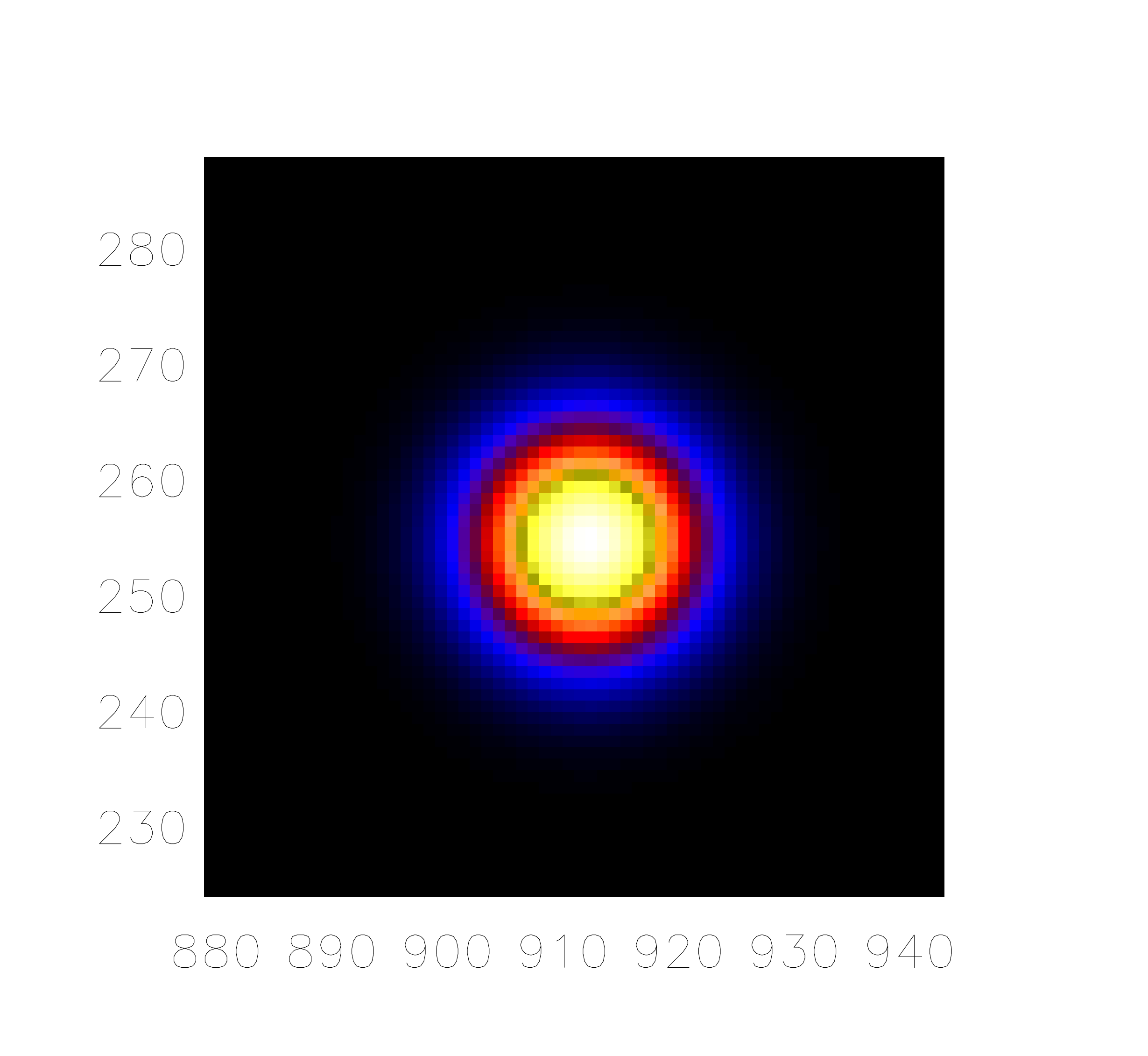} \hspace{-0.3cm}
 \includegraphics[height=2.9cm]{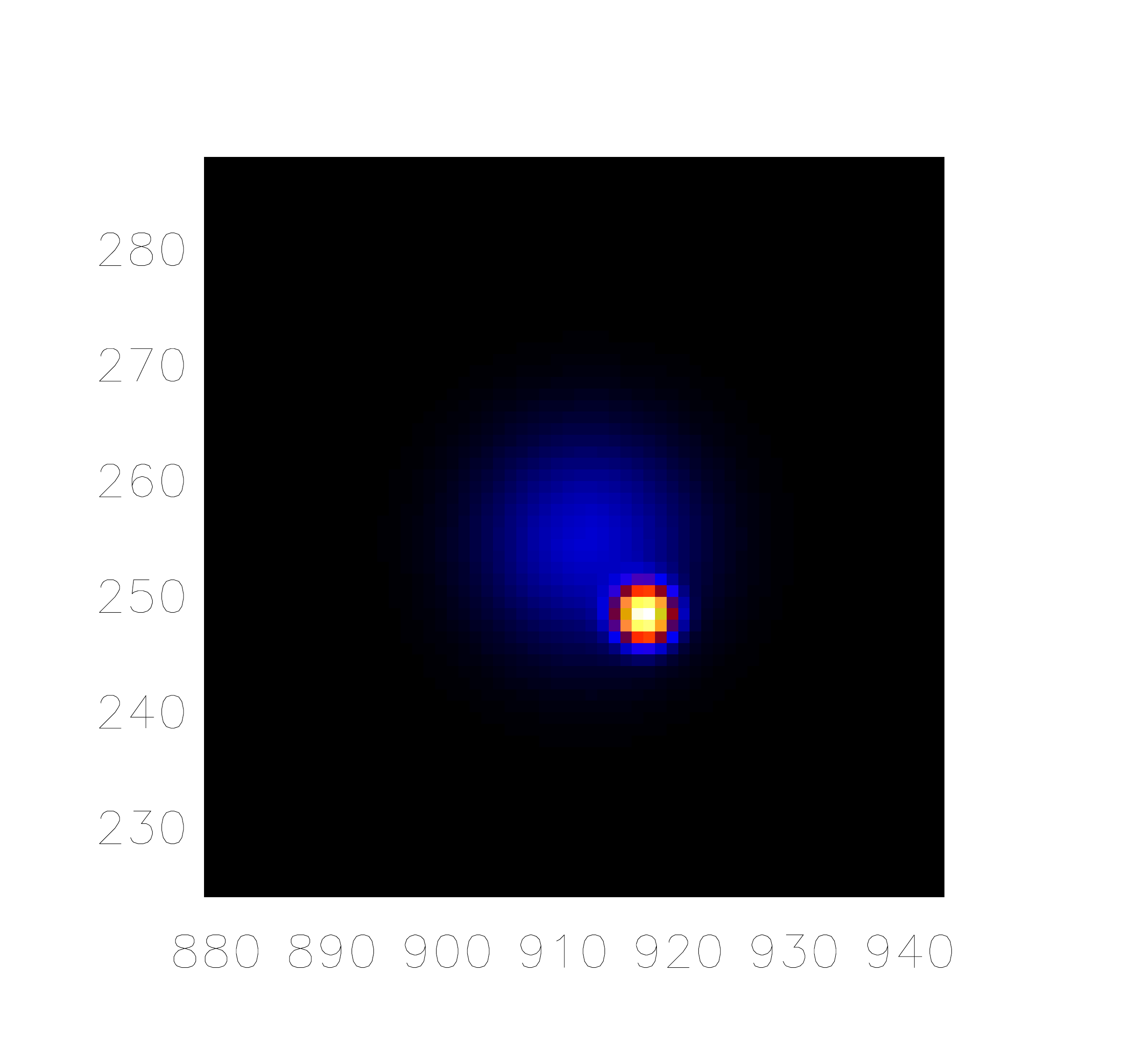}
 \hspace{-0.3cm}
 \includegraphics[height=2.9cm]{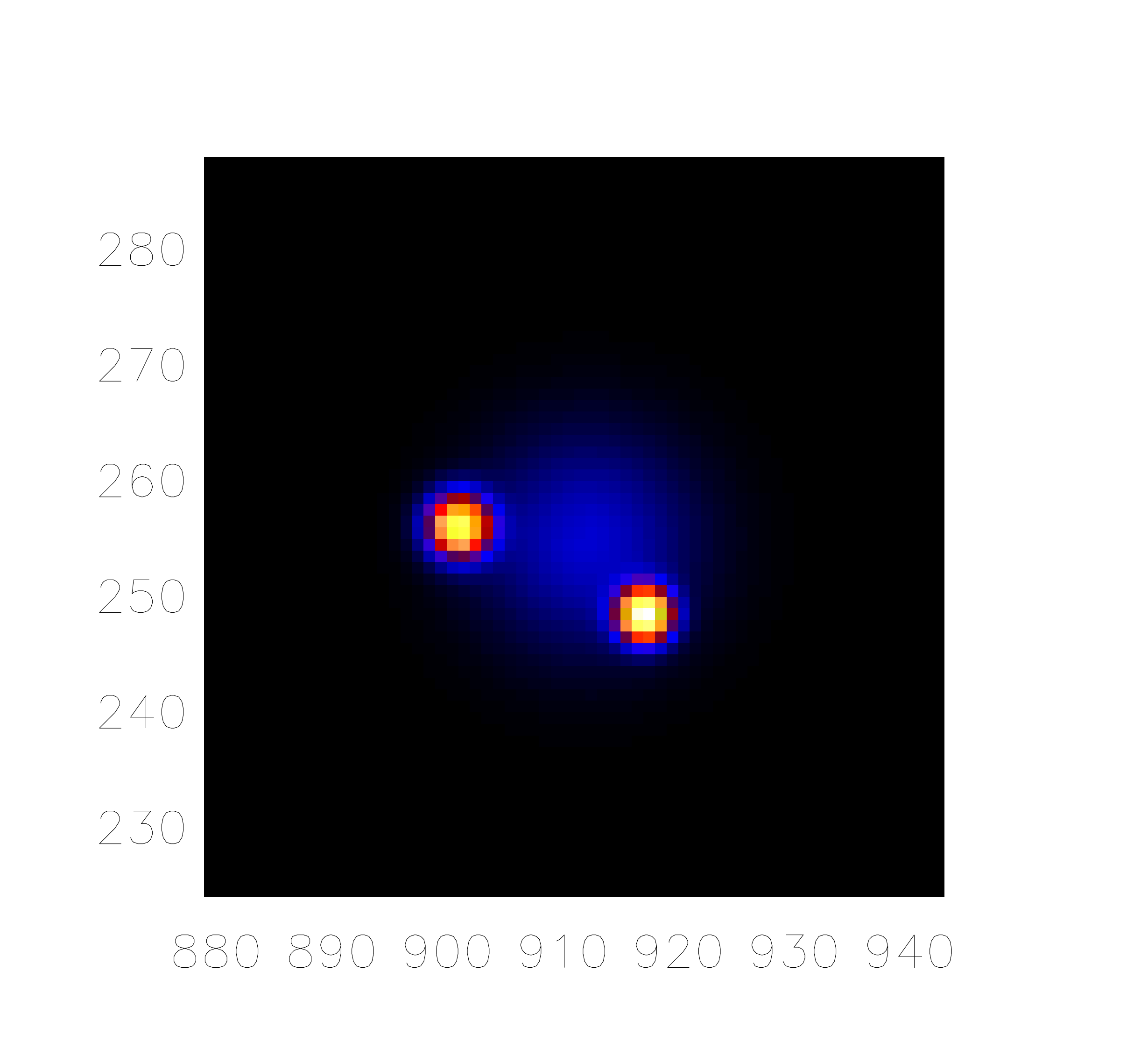}
 \hspace{-0.3cm}
  \includegraphics[height=2.9cm]{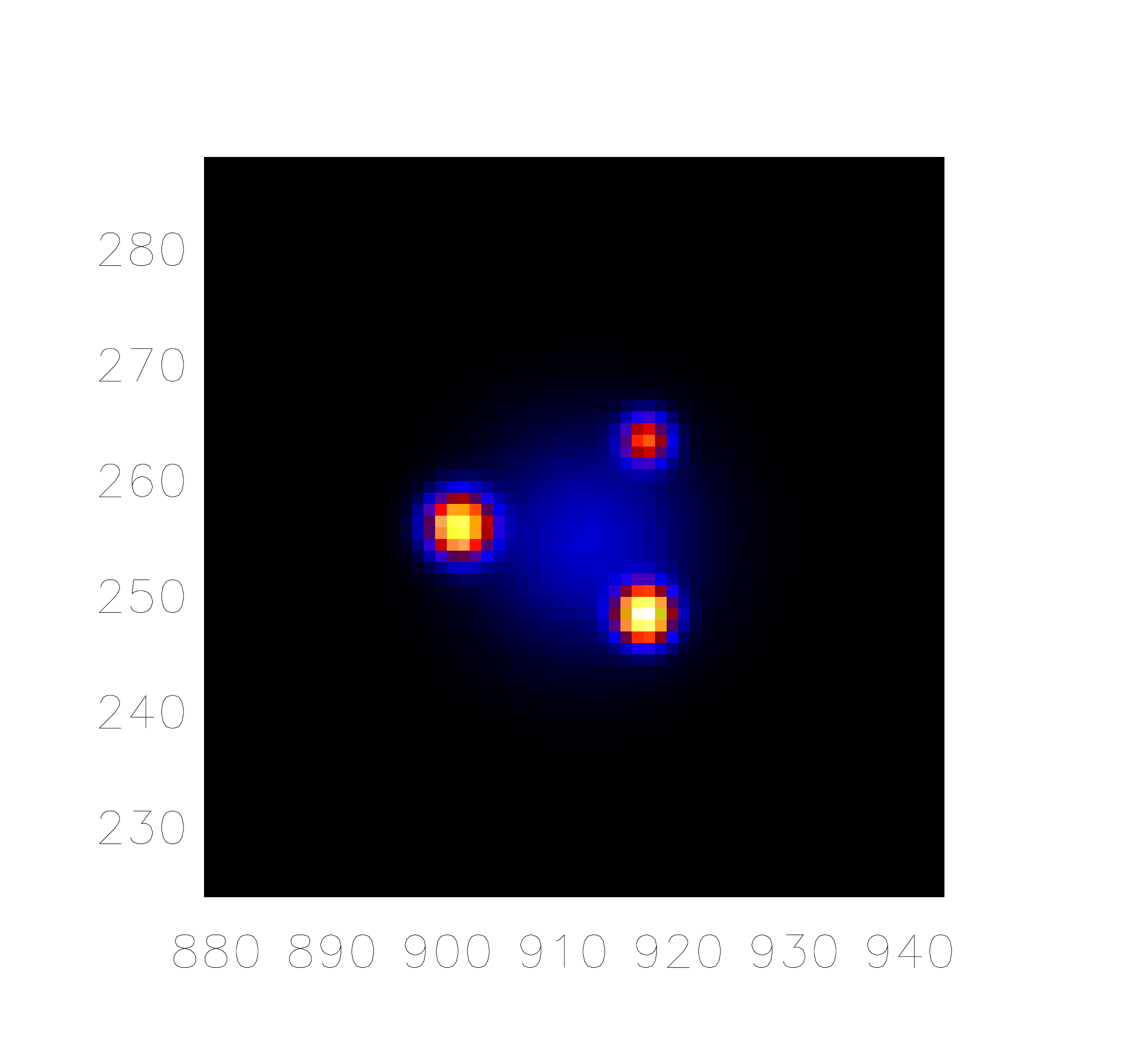}
  \hspace{-0.3cm}
  \includegraphics[height=2.9cm]{fig/ours32.pdf}
  \caption{Distribution of the particles (first row) and their respective X--ray reconstructions (second row) at different iteration. The black stars represent the position of the true sources. }
  \label{fig:iter_config_32}
\end{figure}

We now show the impact of changing the noise variances in \cref{eq:likelihood} by a common factor, which is an other way of tuning regularization by ``trusting'' the data more or less. As previously stated, we do this simply by showing the posterior estimates computed at different, increasing iteration numbers of one single run of the ASMC sampler: this corresponds to decreasing values of the noise standard deviations, i.e. to trusting the data more. As an example, in \cref{fig:iter_config_32} we show the iterations of the ASMC applied to the data generated by the S4C configuration, and using $\lambda=1$ in the prior; we notice that, since the algorithm starts from a sampling of the prior distribution and moves smoothly to reach a sampling of the posterior distribution, the estimated number of sources increases (almost monotonically) with the iterations. In \cref{fig:iter_config_32} we show, in the top row, the spatial distribution of particles at different iterations, more specifically: we show the first iteration $i$ where $(\hat{N}_S)_i=1,2,3,4$, and then the final iteration; in the row below we show the corresponding estimated configurations. At first (left column), a large source is estimated around the true location of the central source, and its location is still uncertain (particles spread relatively high). Then particles begin to concentrate around the two slightly weaker sources (left and bottom right); however, the posterior mode for the number of sources is two, and therefore the final estimate contains only the central and the bottom right source. Subsequently, the third source is reconstructed, and finally also the weaker top right source is recovered, while the uncertainties on the location of the other sources get smaller and smaller: at the last iteration, the variance of the source locations is so small that it is almost invisible in these images, covered by the stars representing the true source locations.

\begin{figure}[h!]
  \centering
    \includegraphics[height=4.8cm]{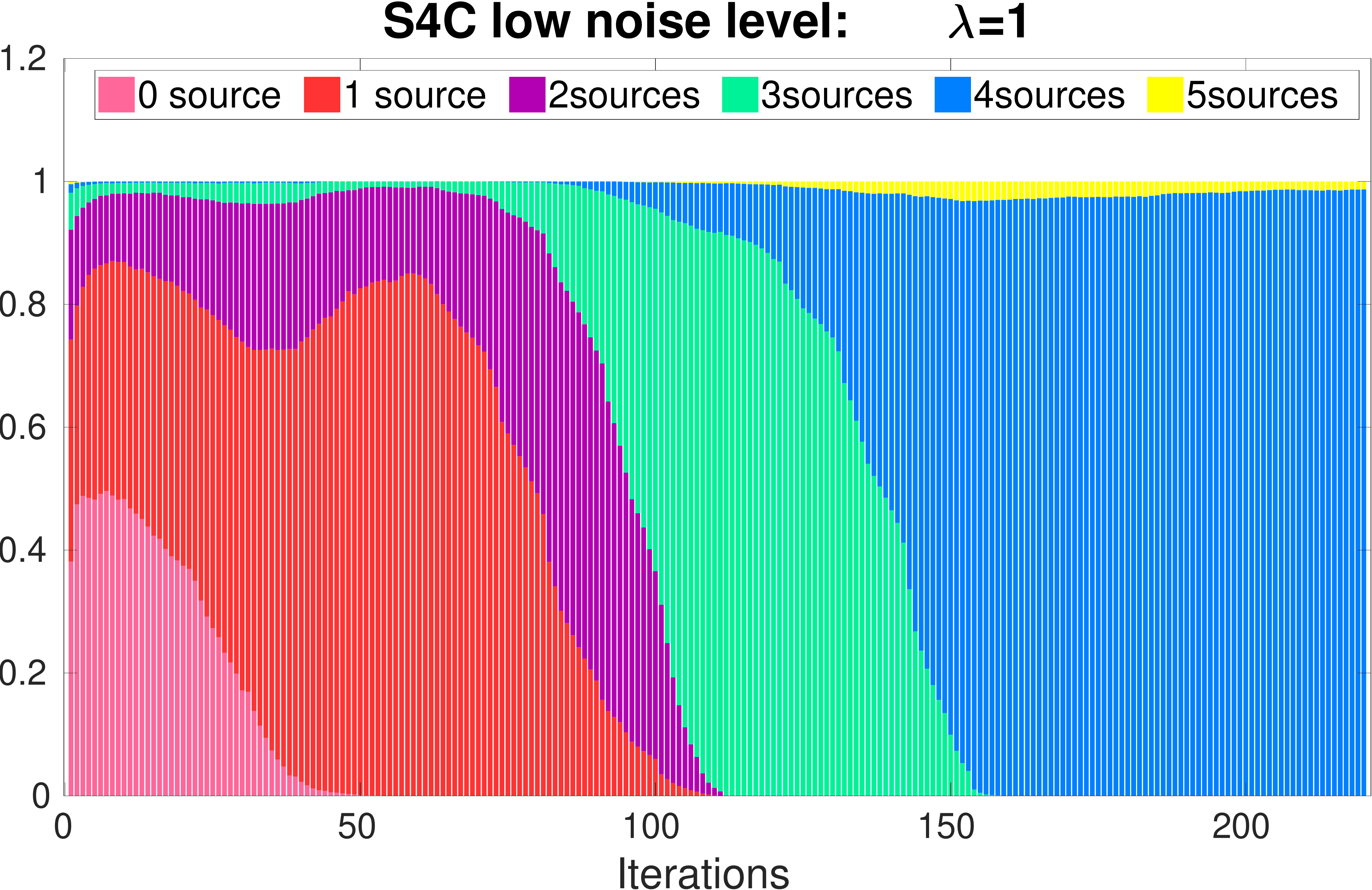} 
    \includegraphics[height=4.8cm]{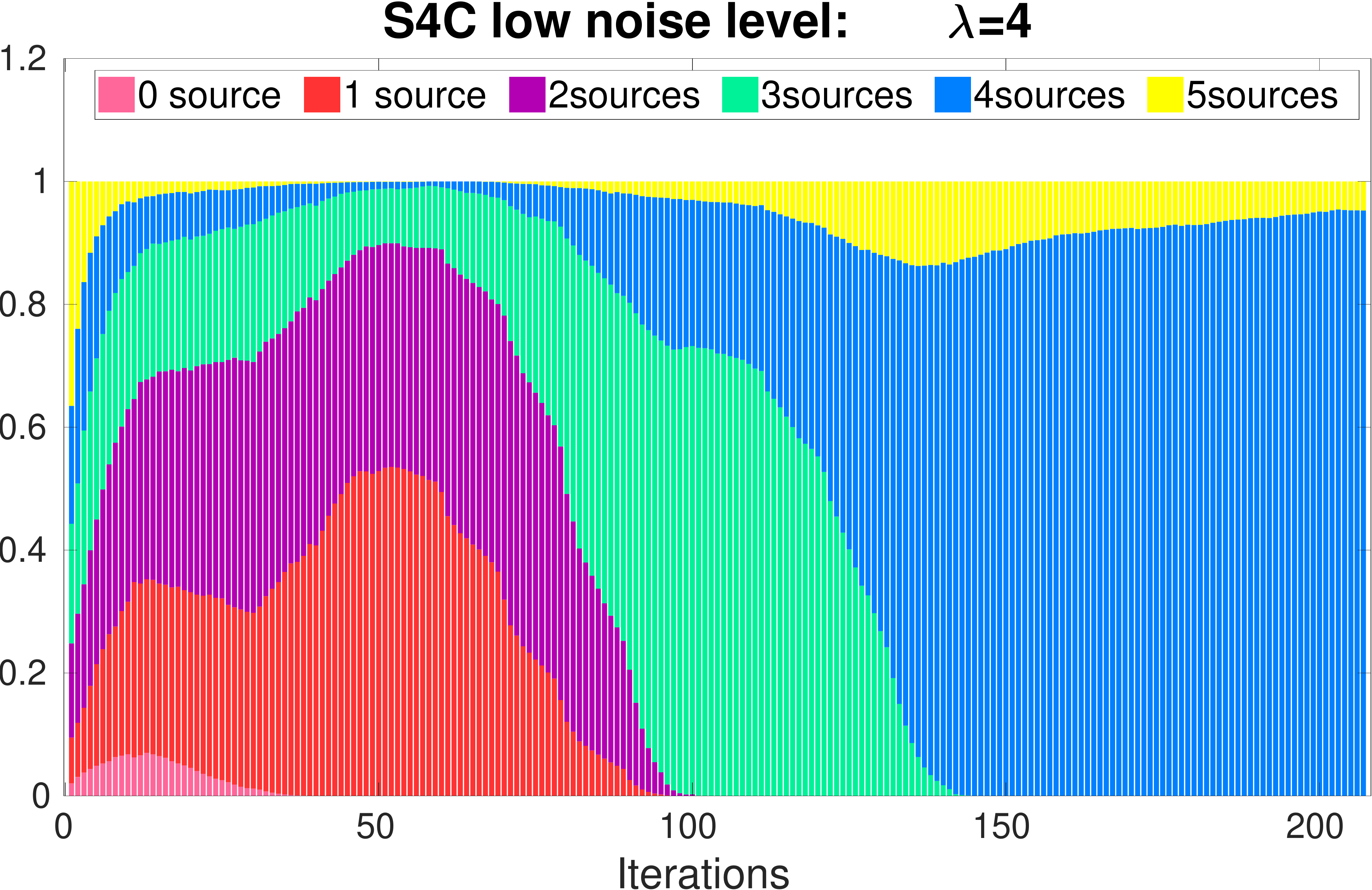}
  \caption{Behaviour of the posterior probability for the number of the sources for the S4C, low noise level, with  $\lambda=1$ (left)  and $\lambda=4$ (right). }
  \label{fig:lambda_hist}
\end{figure}

We finally provide an insight about how different values of $\lambda$ impact the iterative procedure: in \cref{fig:lambda_hist} we show the posterior probability of different number of sources, as a function of the iteration number for $\lambda=1$ (left) and $\lambda=4$ (right).
At the beginning, the number of sources is distributed according to the prior, and then it smoothly shifts towards the final posterior distribution: when $\lambda=1$, this translates in monotonically increasing complexity; when $\lambda=4$, in the first iterations the complexity actually decreases, because most sampled high--dimensional solutions will have a poor fit with the data; around iteration 100, however, the two cases become very similar, as expected because of the low noise level/high information content in the data.

\subsection{Acceptance rates}

\begin{figure}[h!]
  \centering
\includegraphics[height=4.cm]{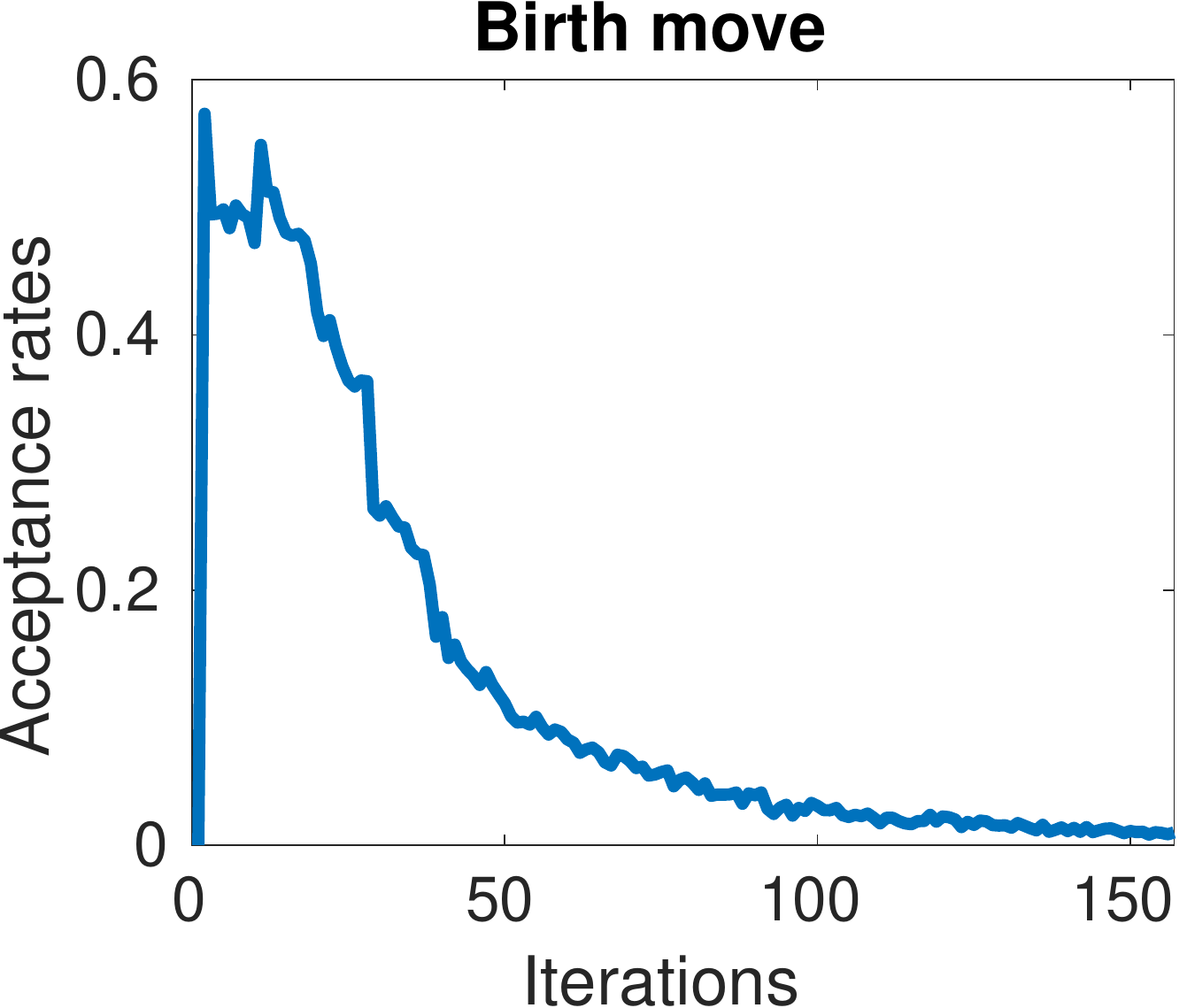}
 \includegraphics[height=4.cm]{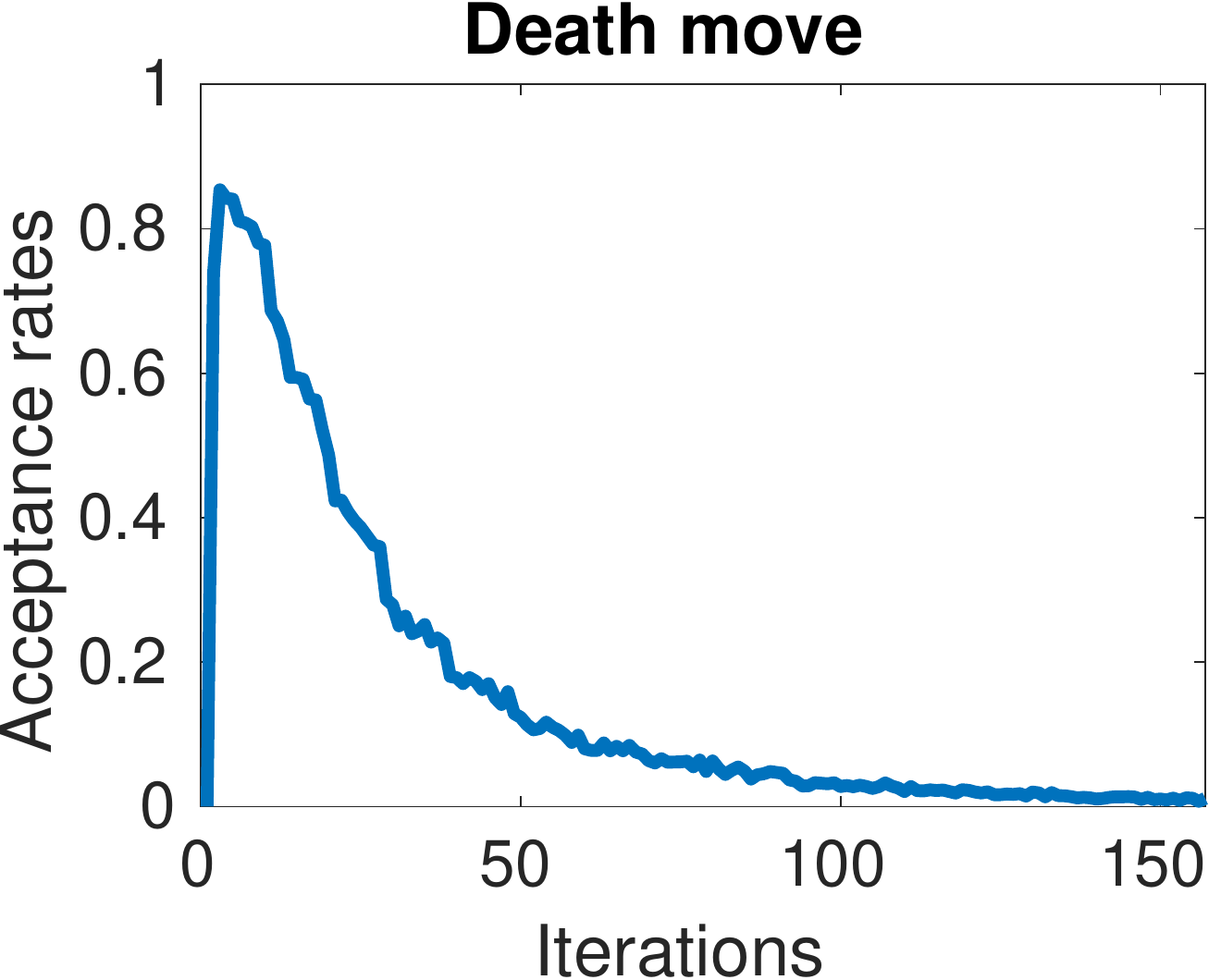}
\includegraphics[height=4.cm]{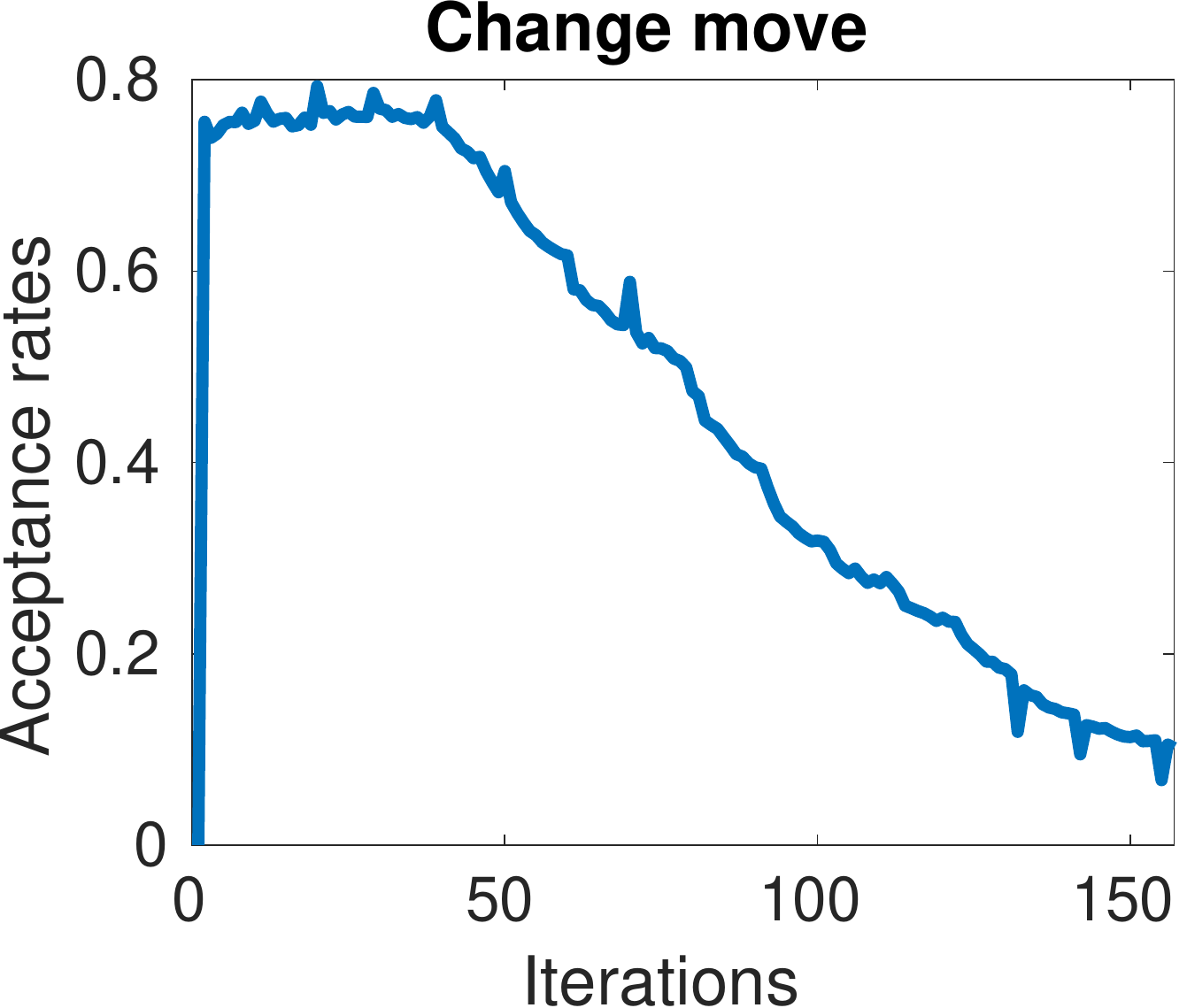}\\
\vspace{0.5cm}
\includegraphics[height=4.cm]{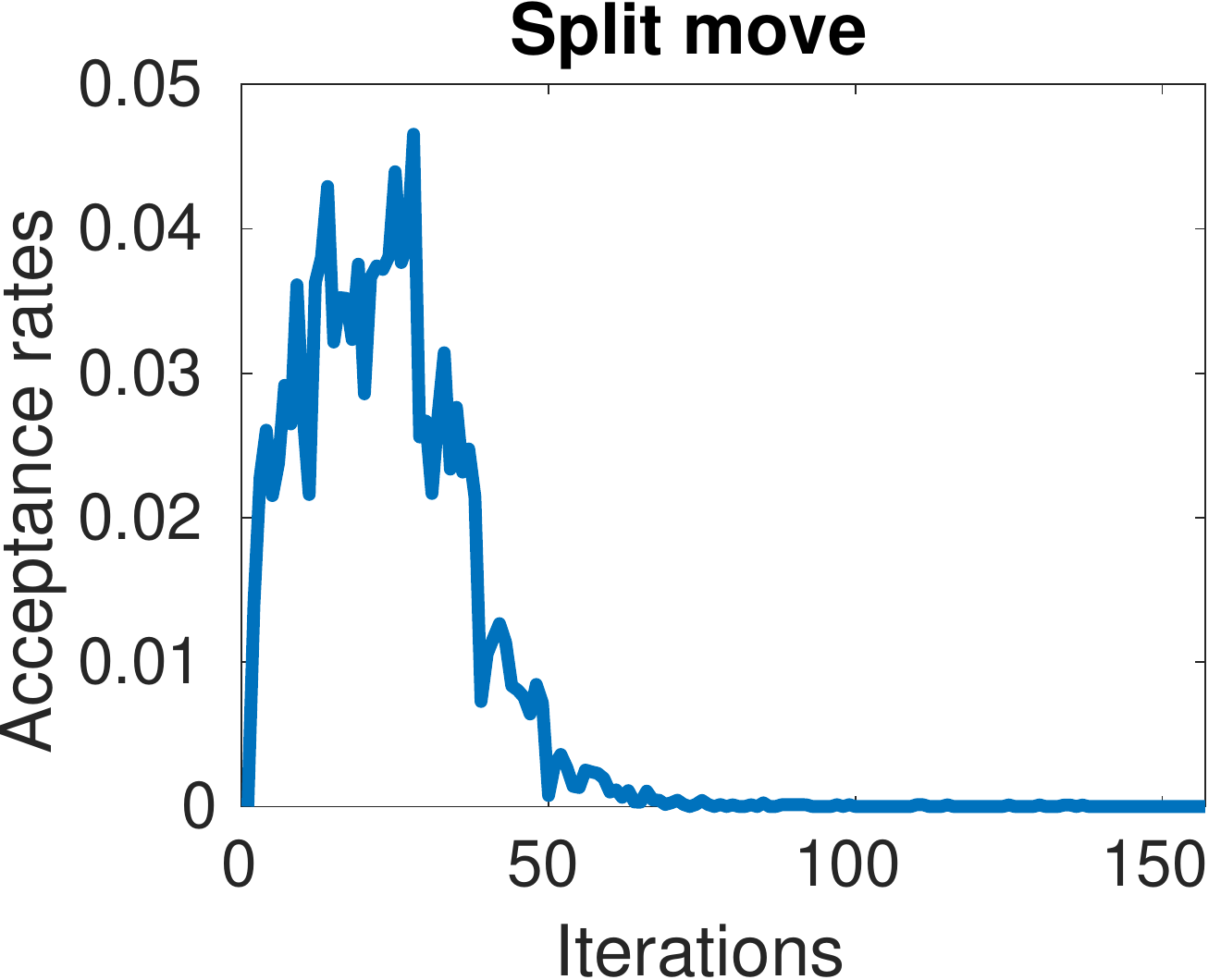}
\includegraphics[height=4cm]{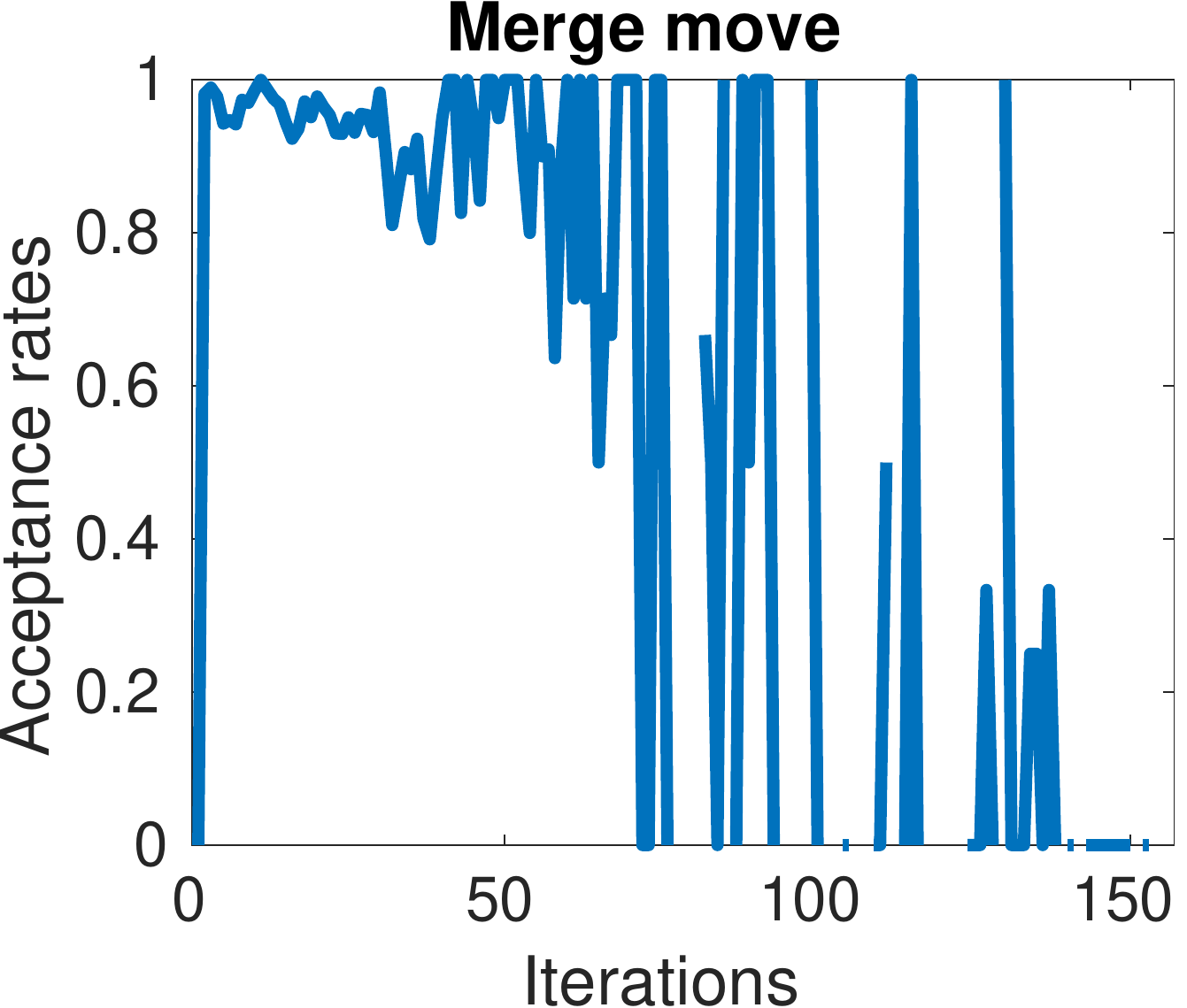}
\includegraphics[height=4.cm]{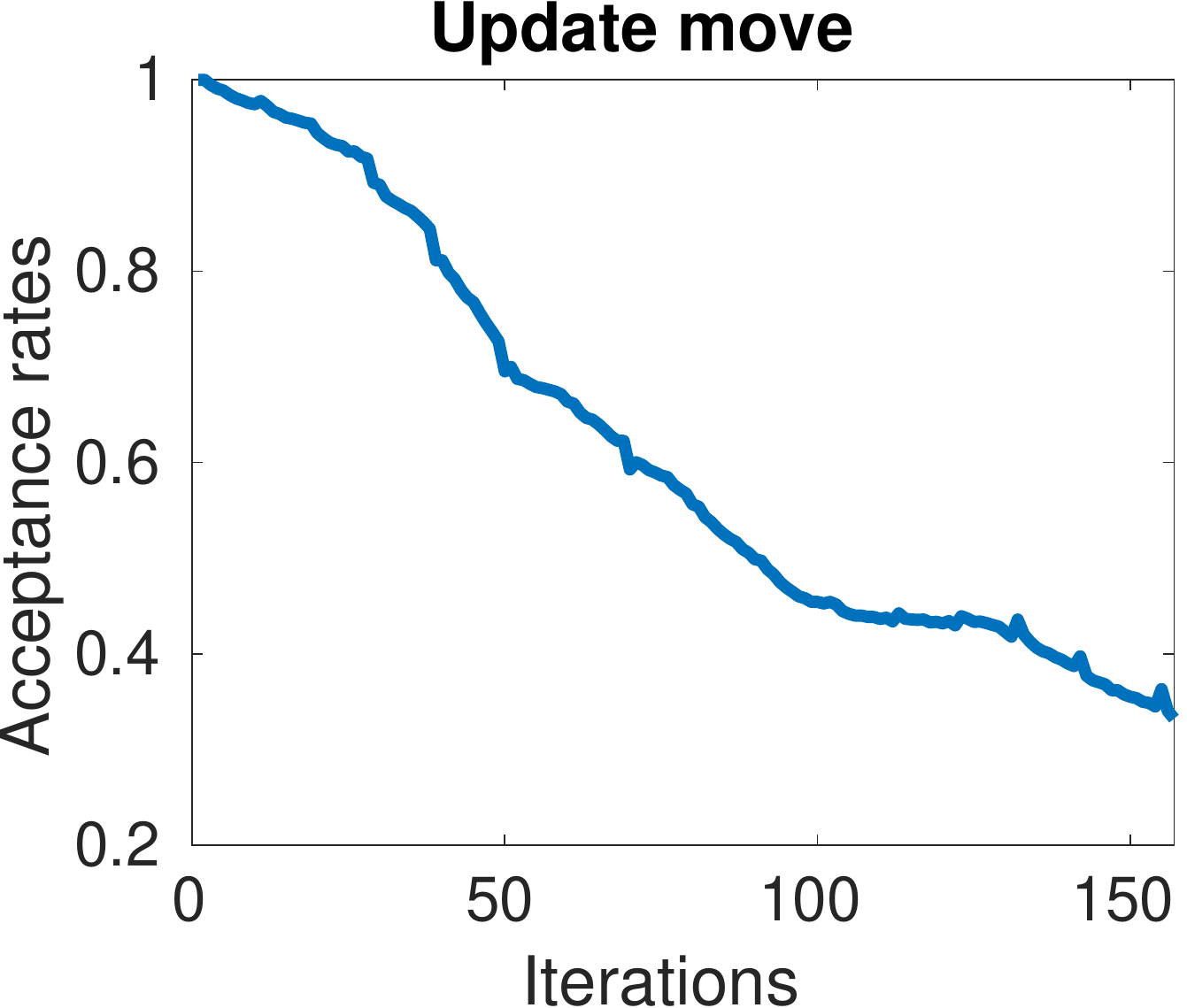}
  \caption{Acceptance rates of the birth, death, change, split, merge and update moves on the data generated by S1E1C.}
  \label{fig:acc_rates}
\end{figure}

Before we proceed with the analysis of experimental data, we provide additional insight into the working of the ASMC algorithm by showing how the acceptance rates behave with the iterations. In \cref{fig:acc_rates} we show the acceptance rates of the birth, death, change, split, merge and update moves obtained when running the ASMC on the data generated by the S1E1C configuration; other configurations may produce slightly different results, but the general behaviour is consistent across all simulations. Specifically: all rates are higher at the beginning and decrease with the iterations, as the samples find (or are moved to, by the resampling) the high--probability region. Birth, death, change and update moves have smoothly decreasing rates, because they are applied to all particles at each iteration; split and merge moves present a less smooth behaviour, due to the fact that they are only attempted on a fraction of samples.

\subsection{Application to real data}
In order to show that the proposed method can work in practice, we finally show an application to experimental data.
We select four real flare events:  the August 23, 2005 (14:27:00 - 14:31:00 UT, 14 - 16 keV), the December 02, 2003 (22:54:00 - 22:58:00 UT, 18-20 KeV), the  February 20, 2002 (11:06:02 - 11:06:34 UT, 22-26 KeV) and the July 23, 2002 events (00:29:10 - 00:30:19 UT, 36-41 KeV). 
Over the years, these events have been widely studied, and used as test bed for image reconstruction methods (see e.g. \cite{duval2018solar, emslie2003rhessi, prato2009regularized, sciacchitano2018identification} and references therein). In addition, these four real events have been used to inspire the generation of the synthetic data analyzed in the previous subsection. Therefore, this analysis represents the perfect complement to the previous subsections, as we study similar source configurations immersed in true solar background noise.

\begin{figure}[h!]
\begin{minipage}{.11\textwidth}
\centering
VIS{\_} CLEAN
\end{minipage}
\begin{minipage}{.88\textwidth}
 \includegraphics[height=3.7cm]{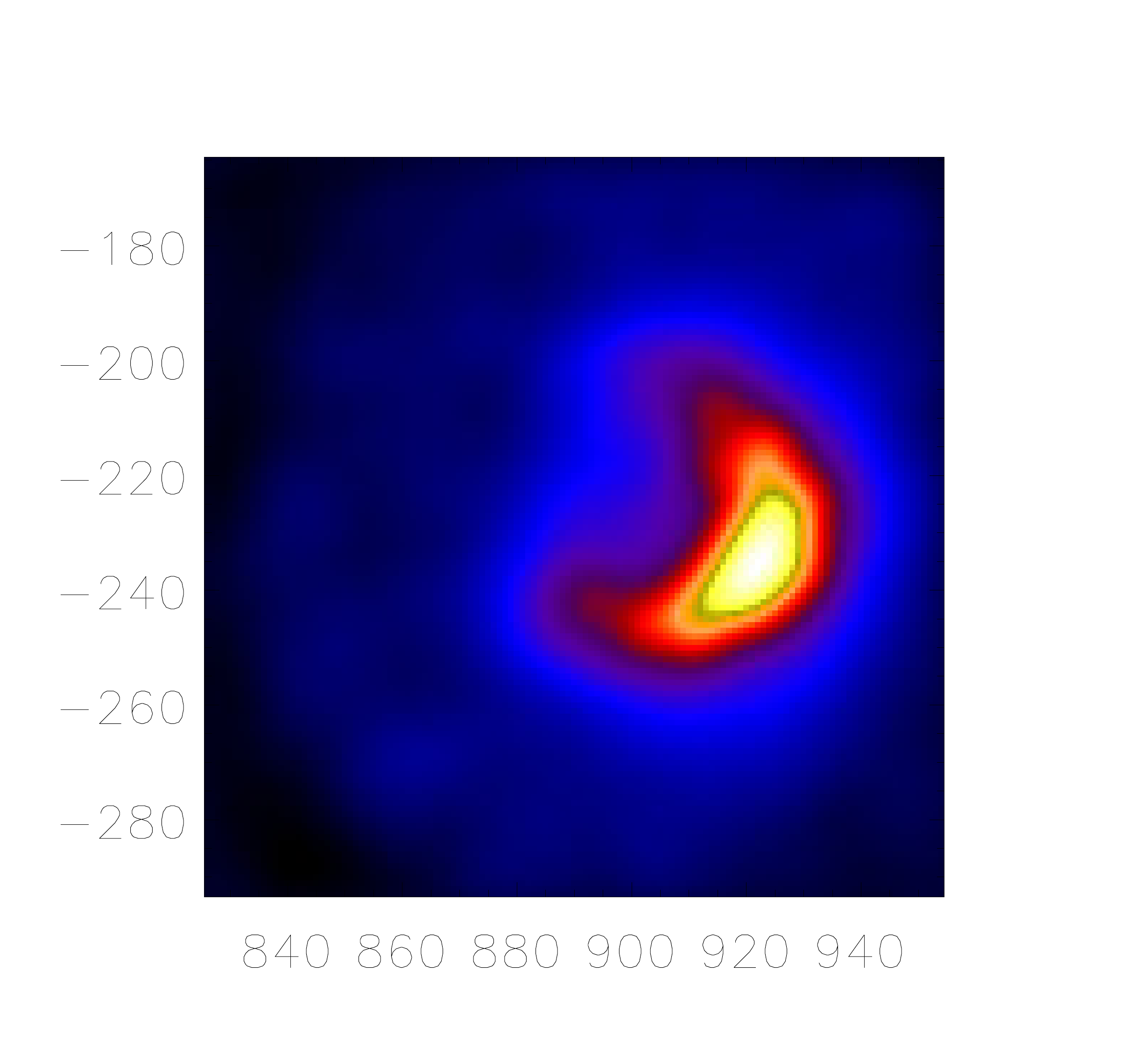}\hspace{-1cm}
    \includegraphics[height=3.7cm]{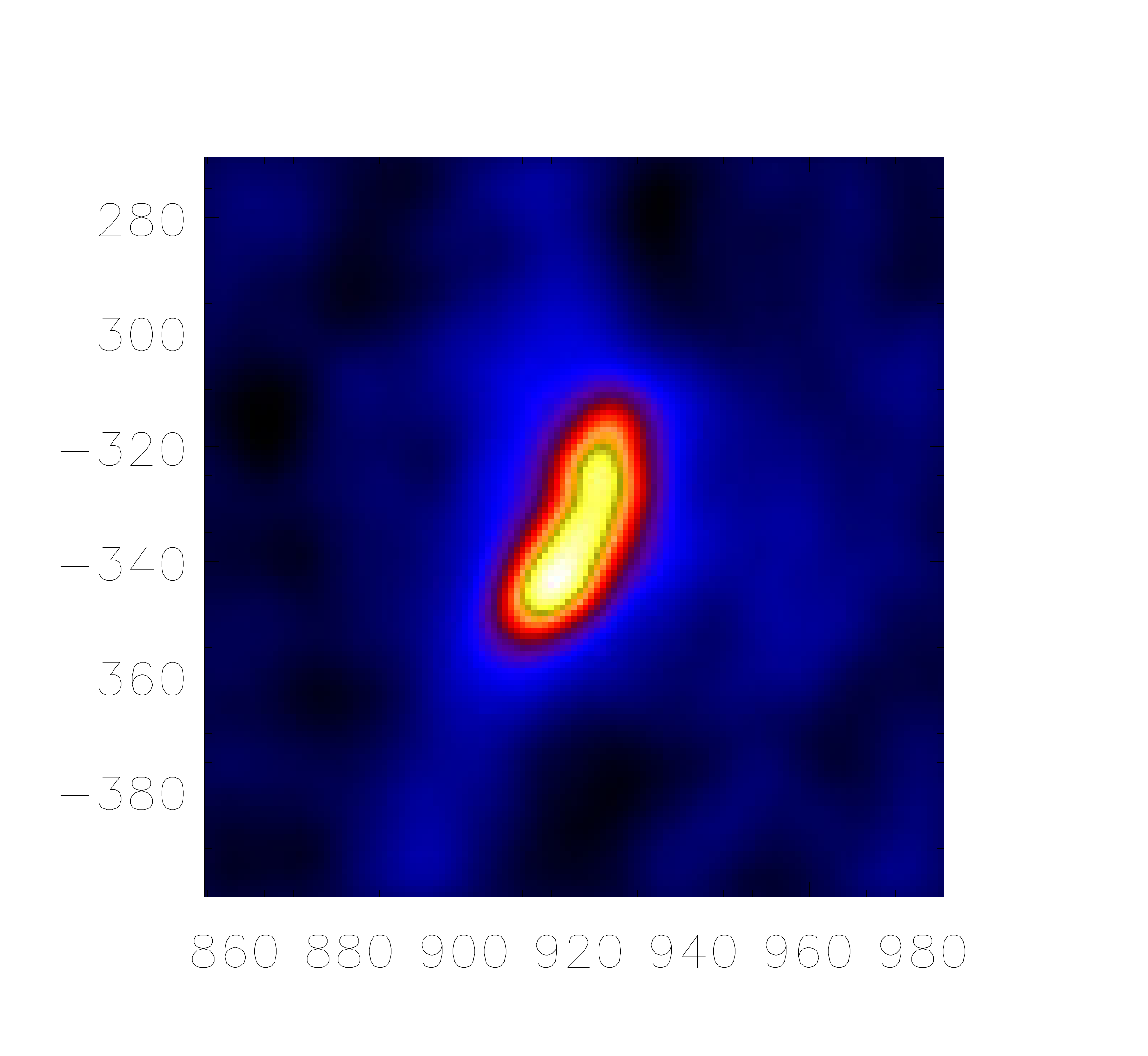}\hspace{-1cm}
    \includegraphics[height=3.7cm]{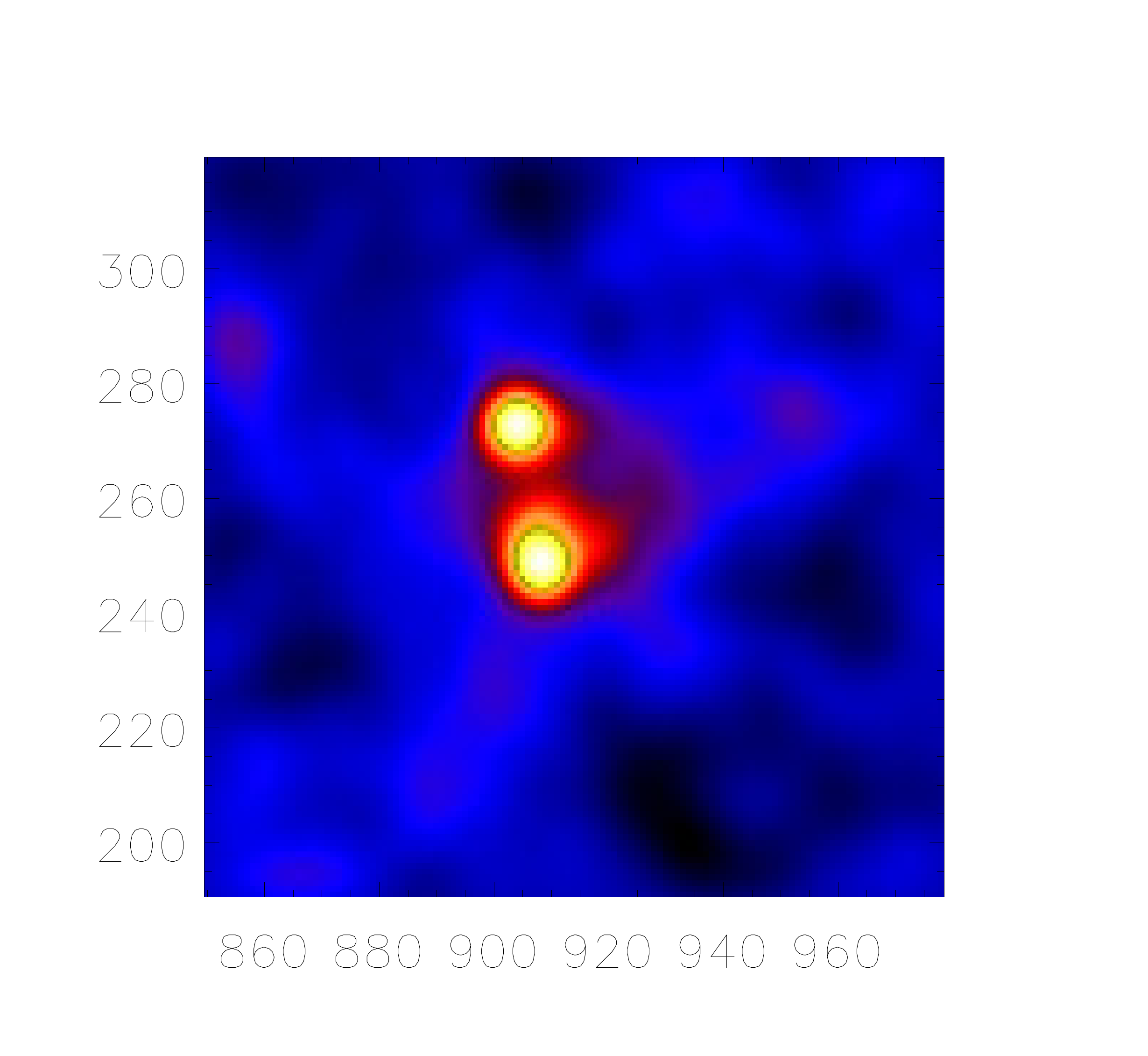}\hspace{-1cm}
    \includegraphics[height=3.7cm]{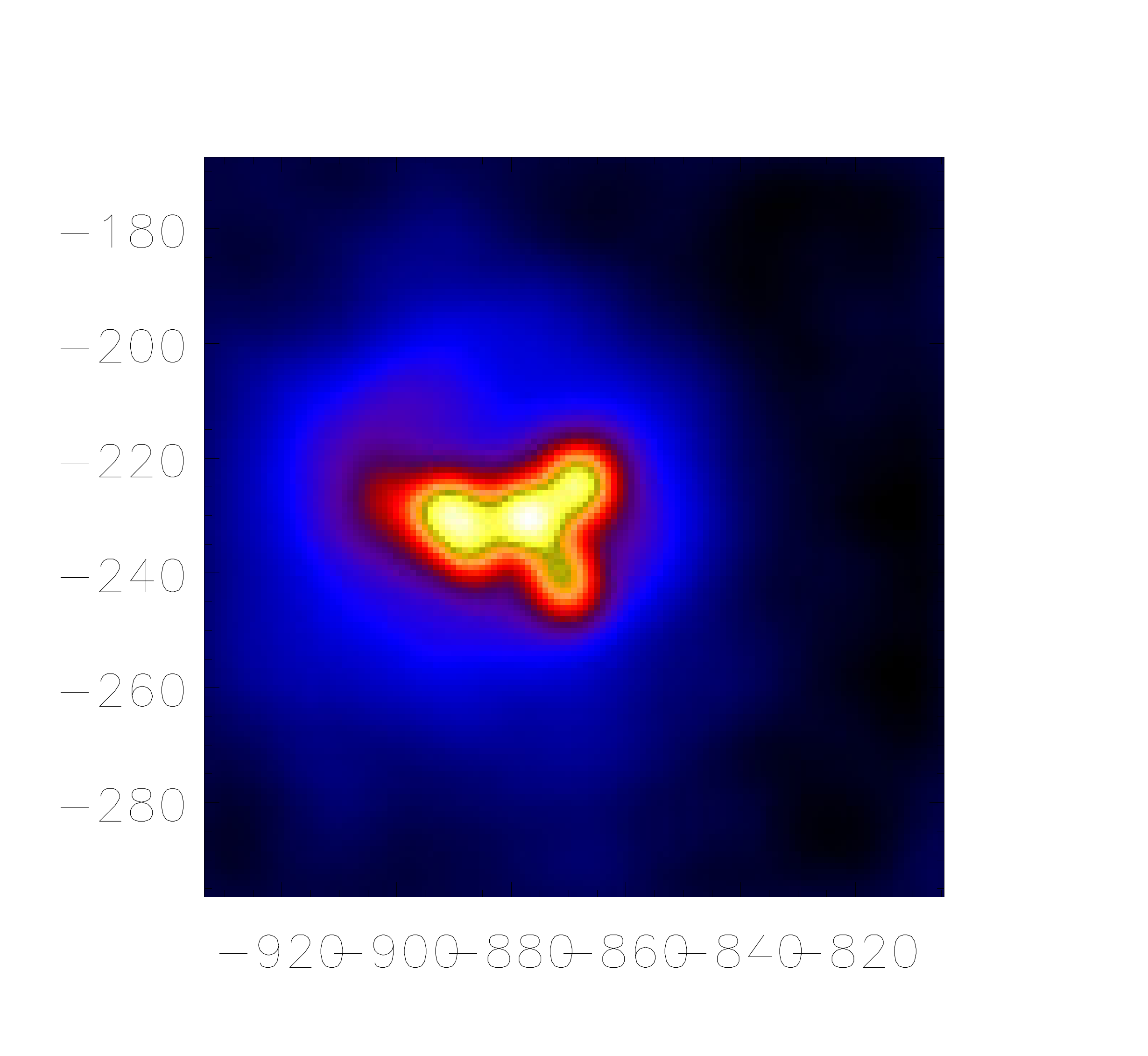}\\ \vspace{-1cm}
    \end{minipage}
            \begin{minipage}{.11\textwidth}
\centering
VIS\texttt{\_} FWDFIT
\end{minipage}
\begin{minipage}{.88\textwidth}
 \includegraphics[height=3.7cm]{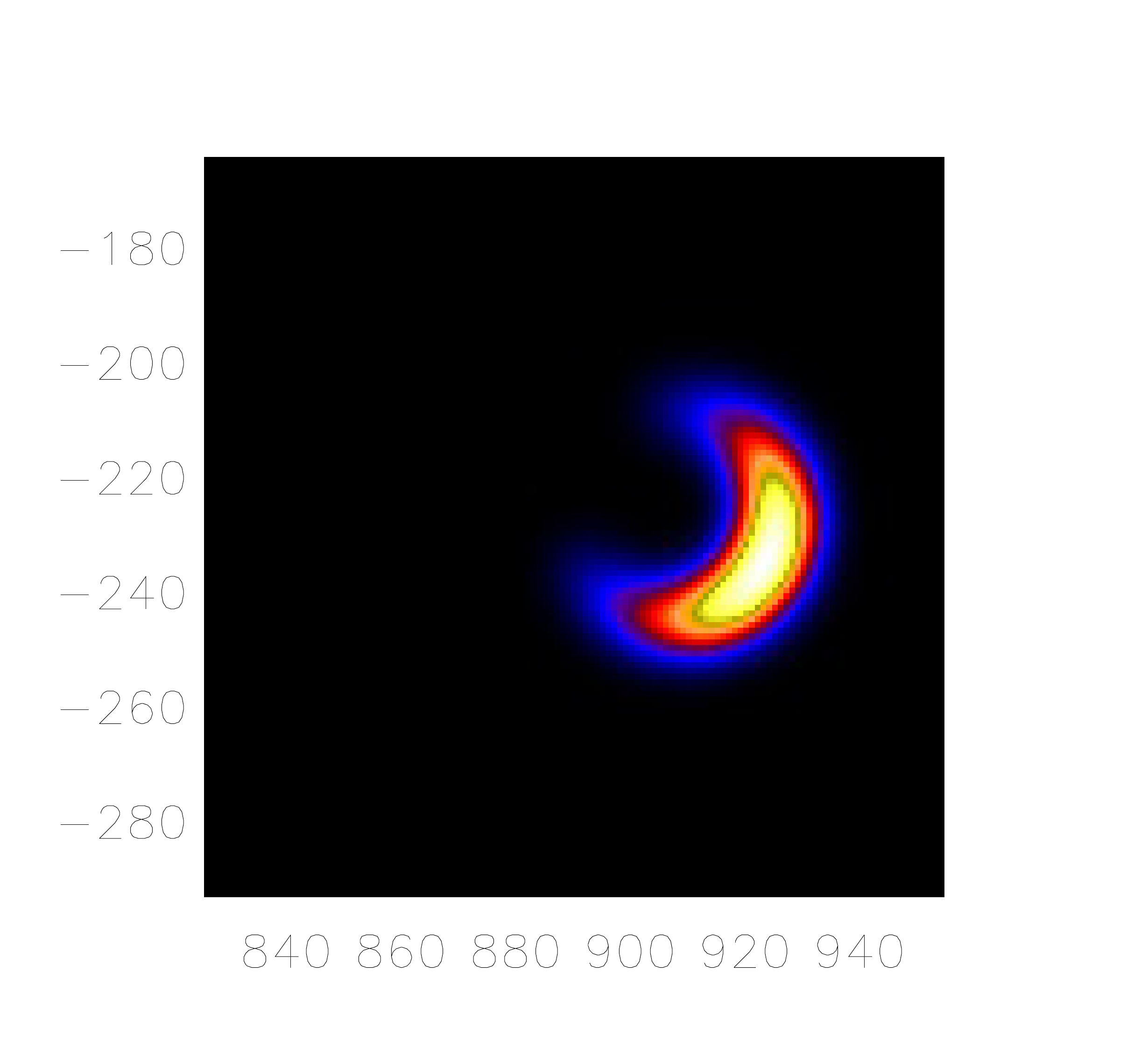}\hspace{-1cm}
    \includegraphics[height=3.7cm]{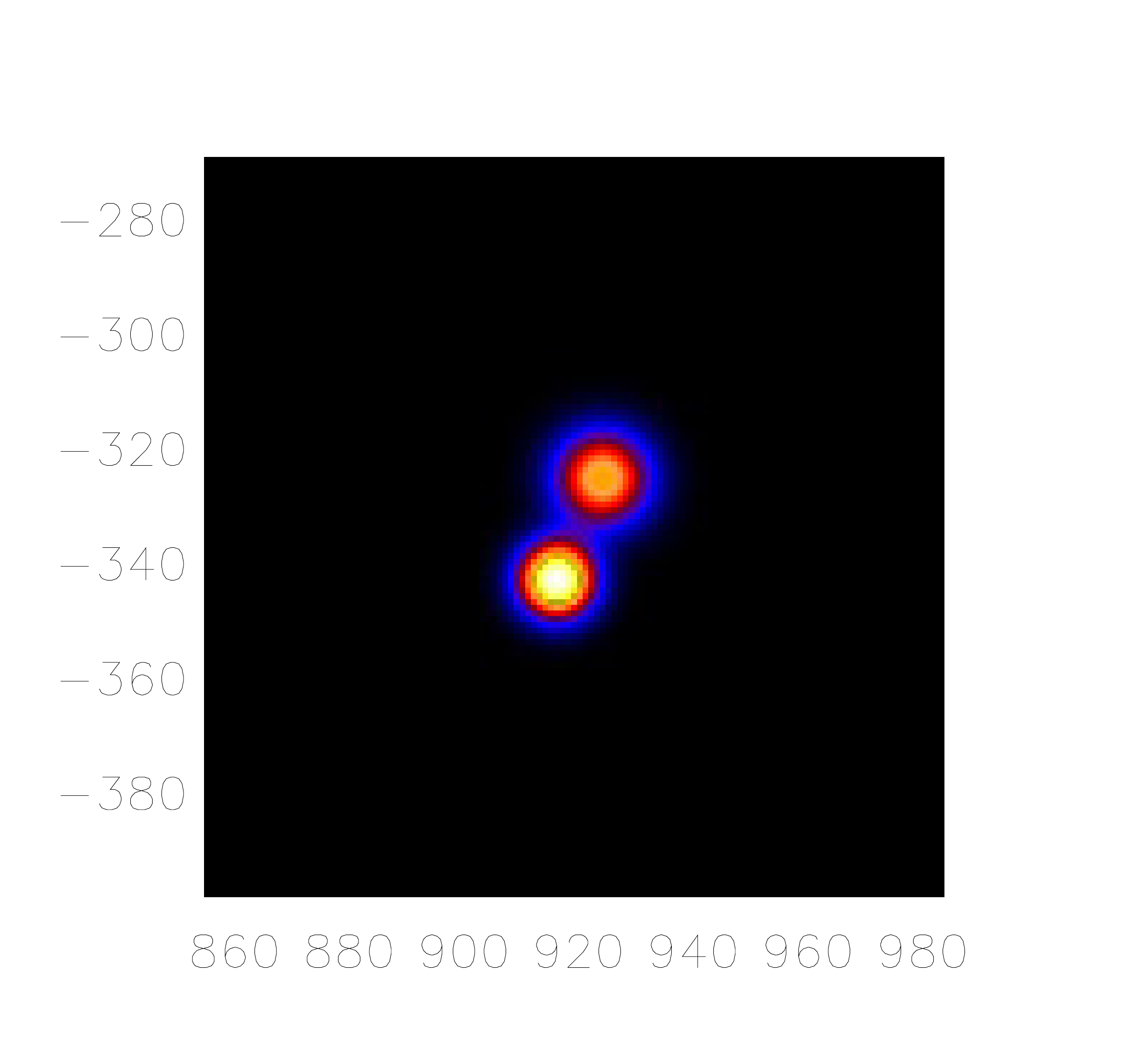}\hspace{-1cm}
    \includegraphics[height=3.7cm]{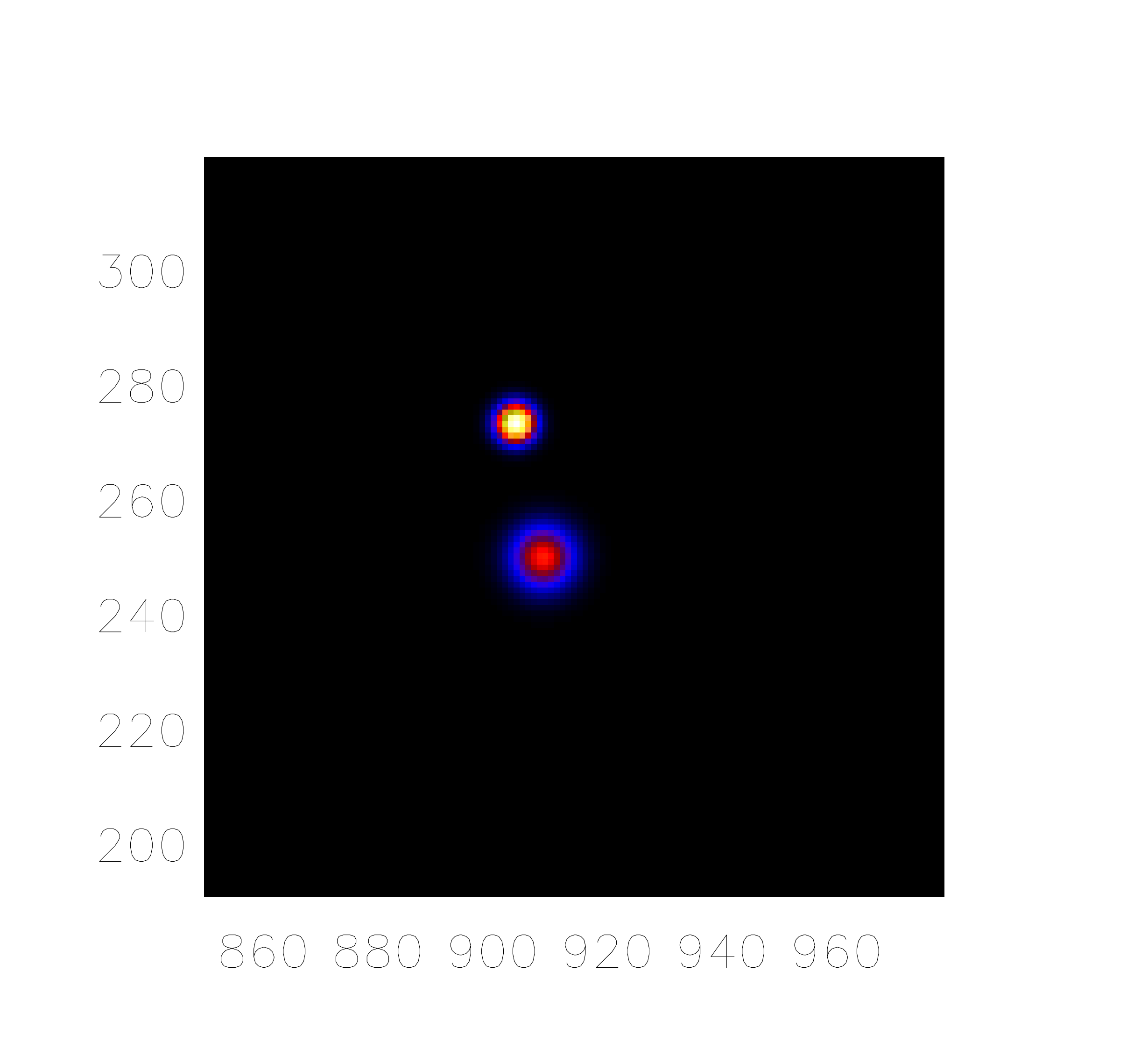}\hspace{-1cm}
    \includegraphics[height=3.7cm]{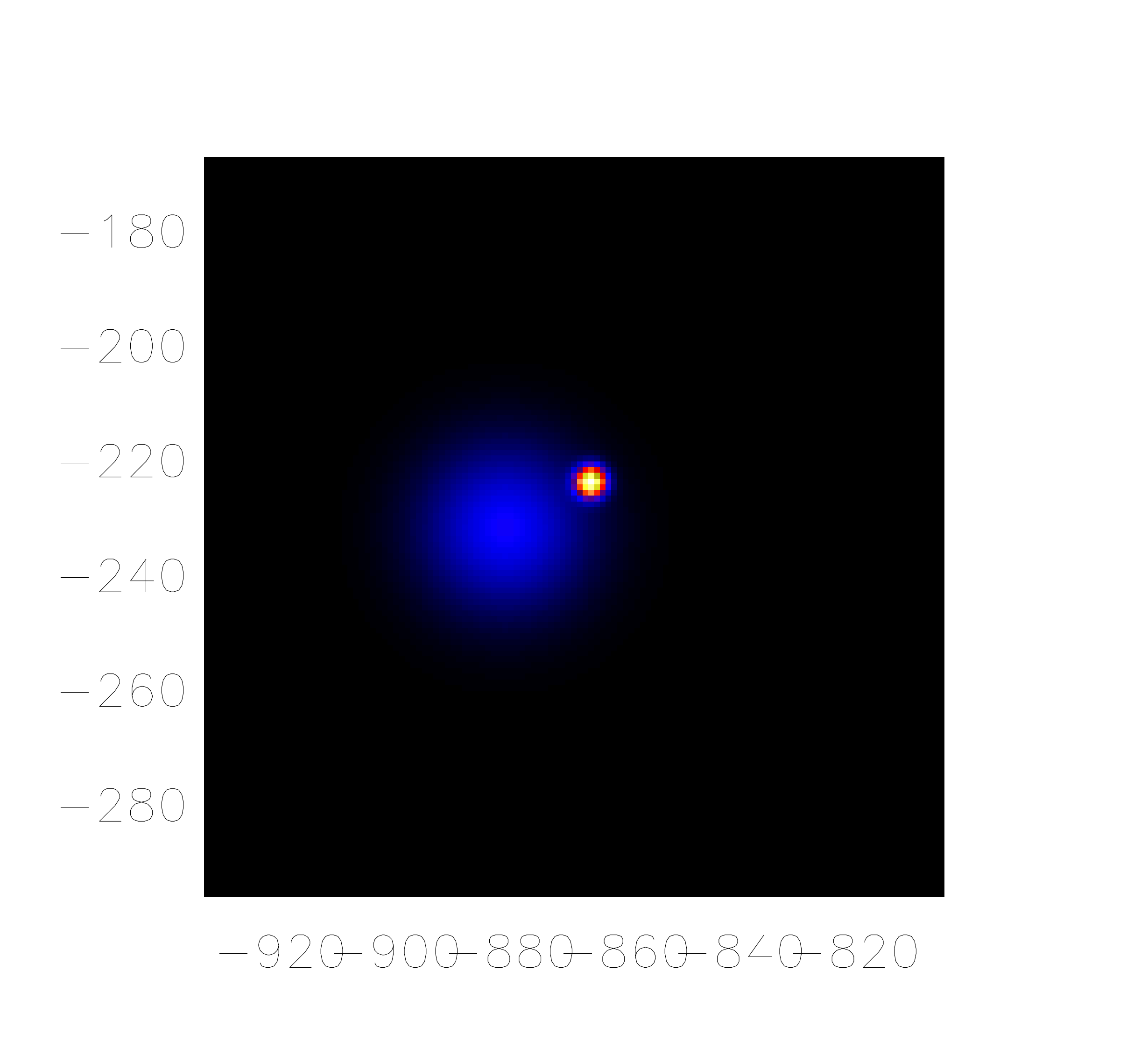}\\ \vspace{-1cm}
    \end{minipage} 
        \begin{minipage}{.11\textwidth}
\centering
VIS\texttt{\_}CS
\end{minipage}
\begin{minipage}{.88\textwidth}
 \includegraphics[height=3.7cm]{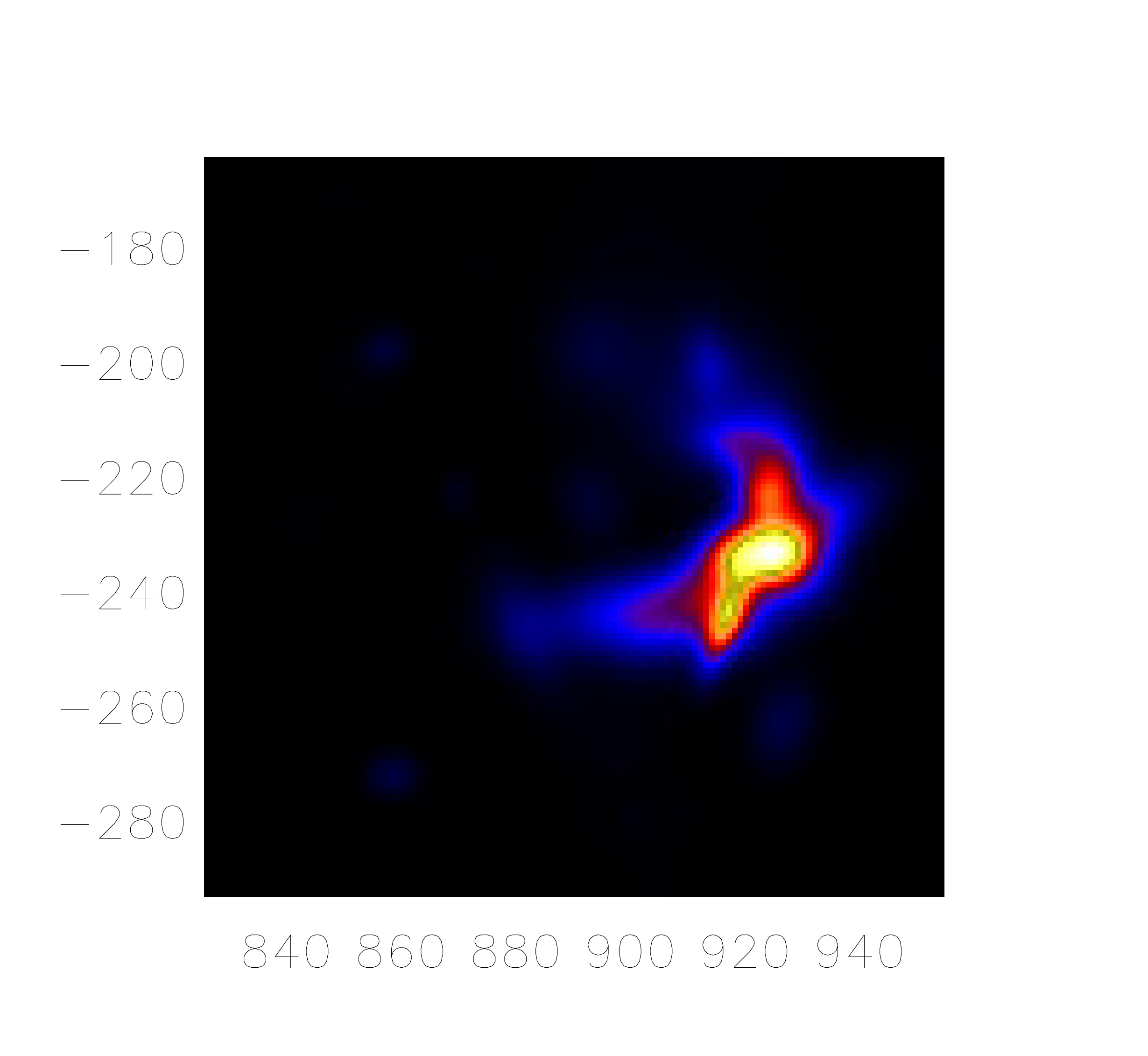}\hspace{-1cm}
    \includegraphics[height=3.7cm]{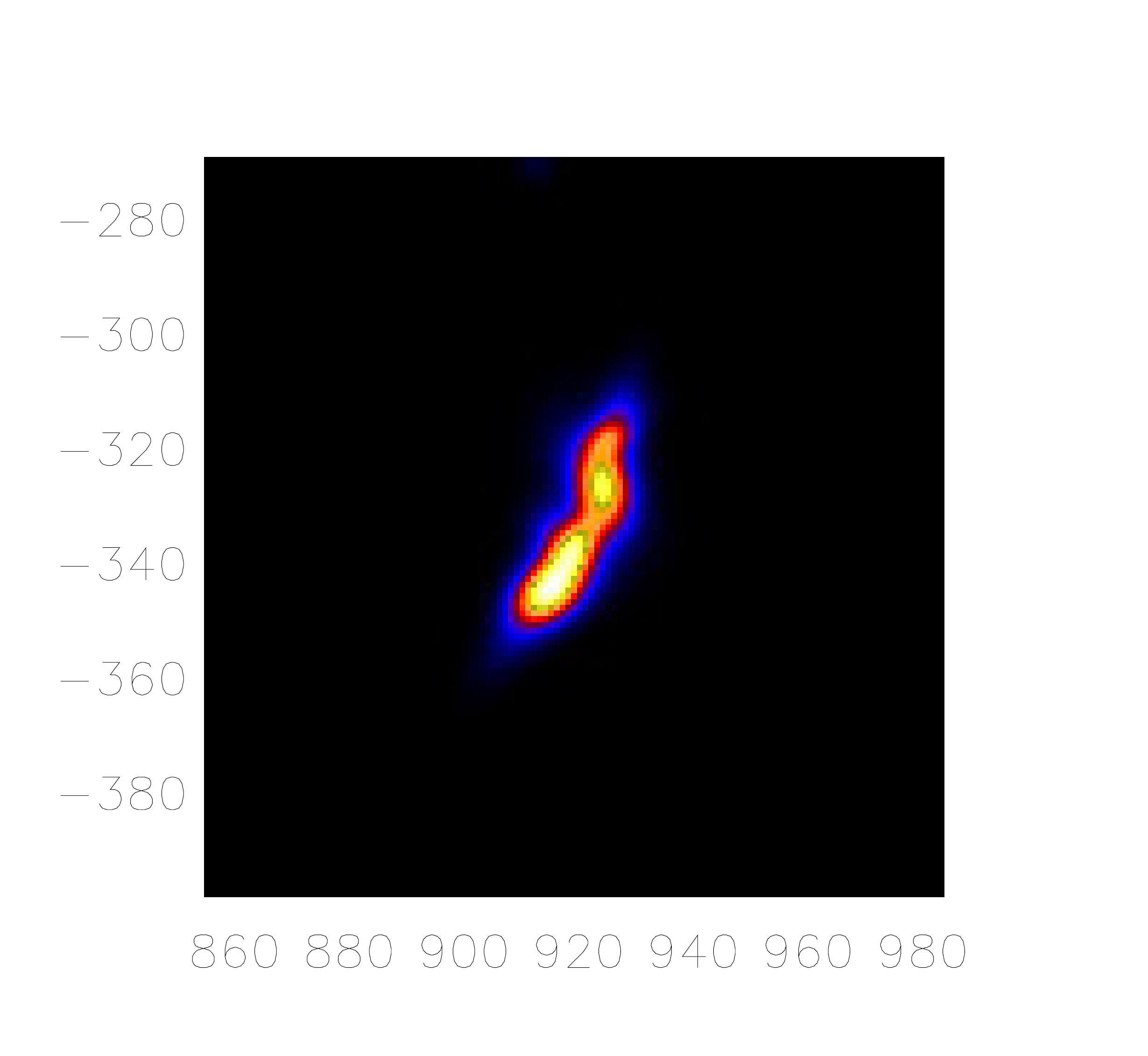}\hspace{-1cm}
    \includegraphics[height=3.7cm]{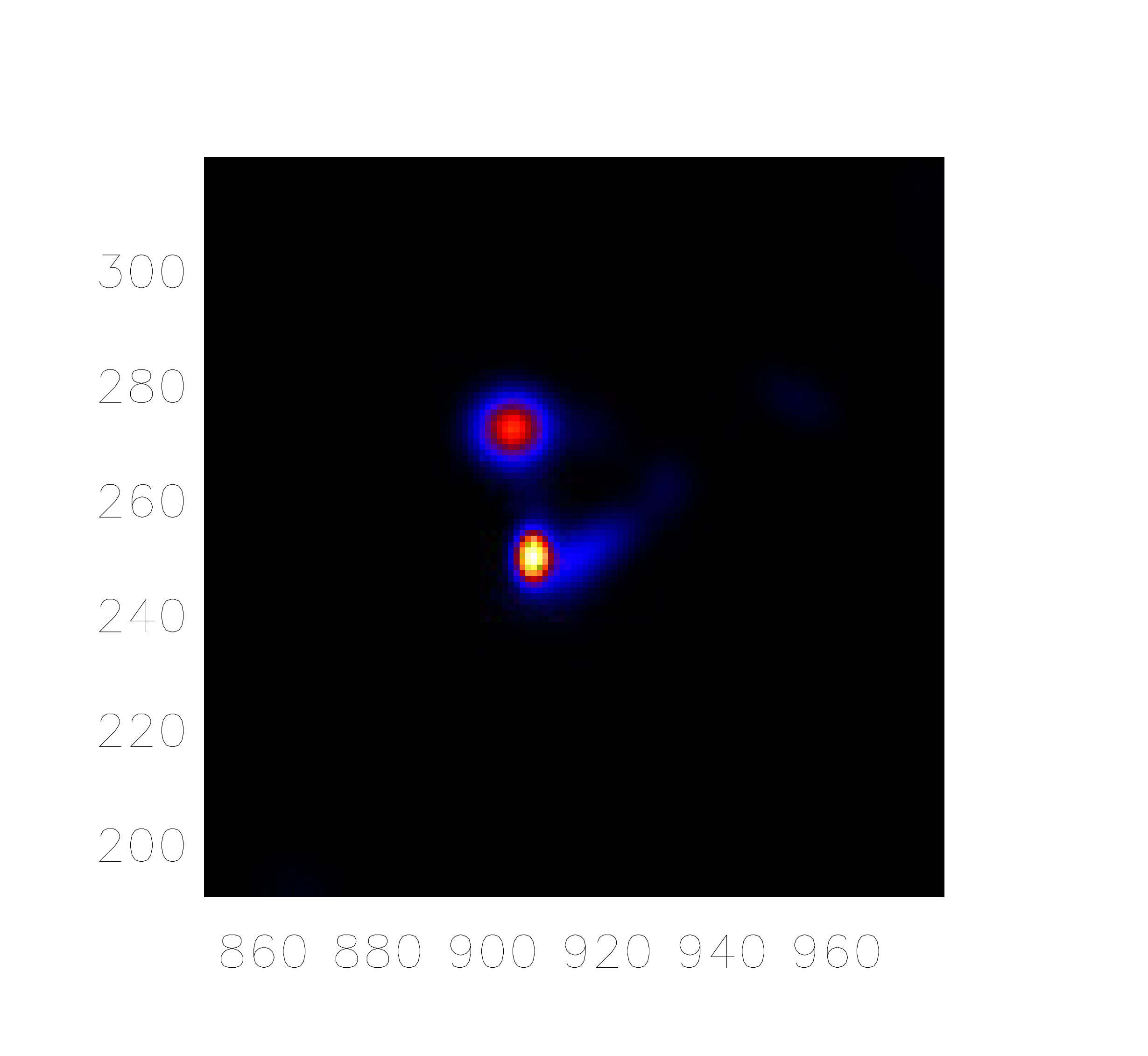}\hspace{-1cm}
    \includegraphics[height=3.7cm]{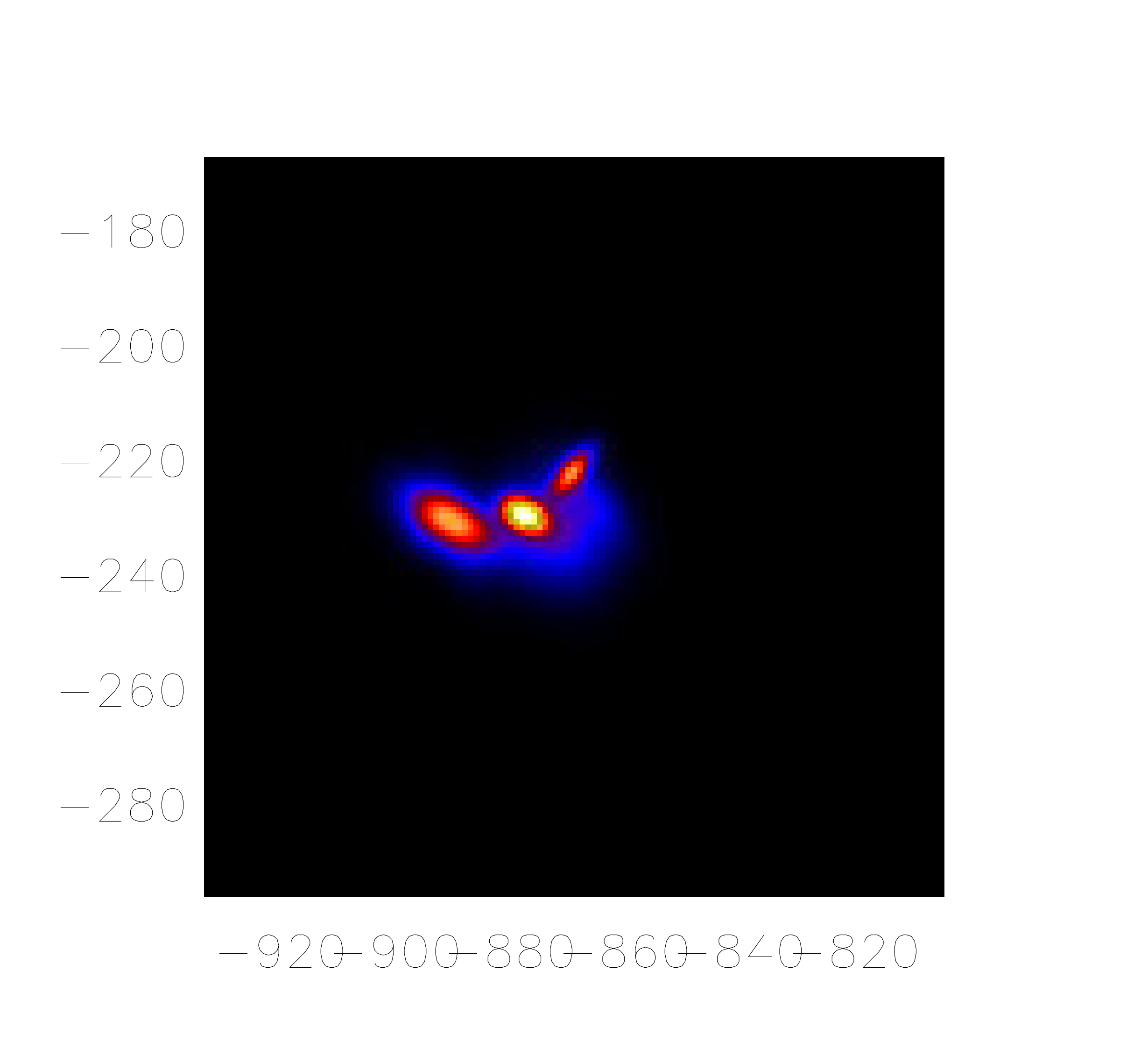}\\ \vspace{-1cm}
    \end{minipage}       
        \begin{minipage}{.11\textwidth}
\centering
VIS\texttt{\_}WV 
\end{minipage}
\begin{minipage}{.88\textwidth}
 \includegraphics[height=3.7cm]{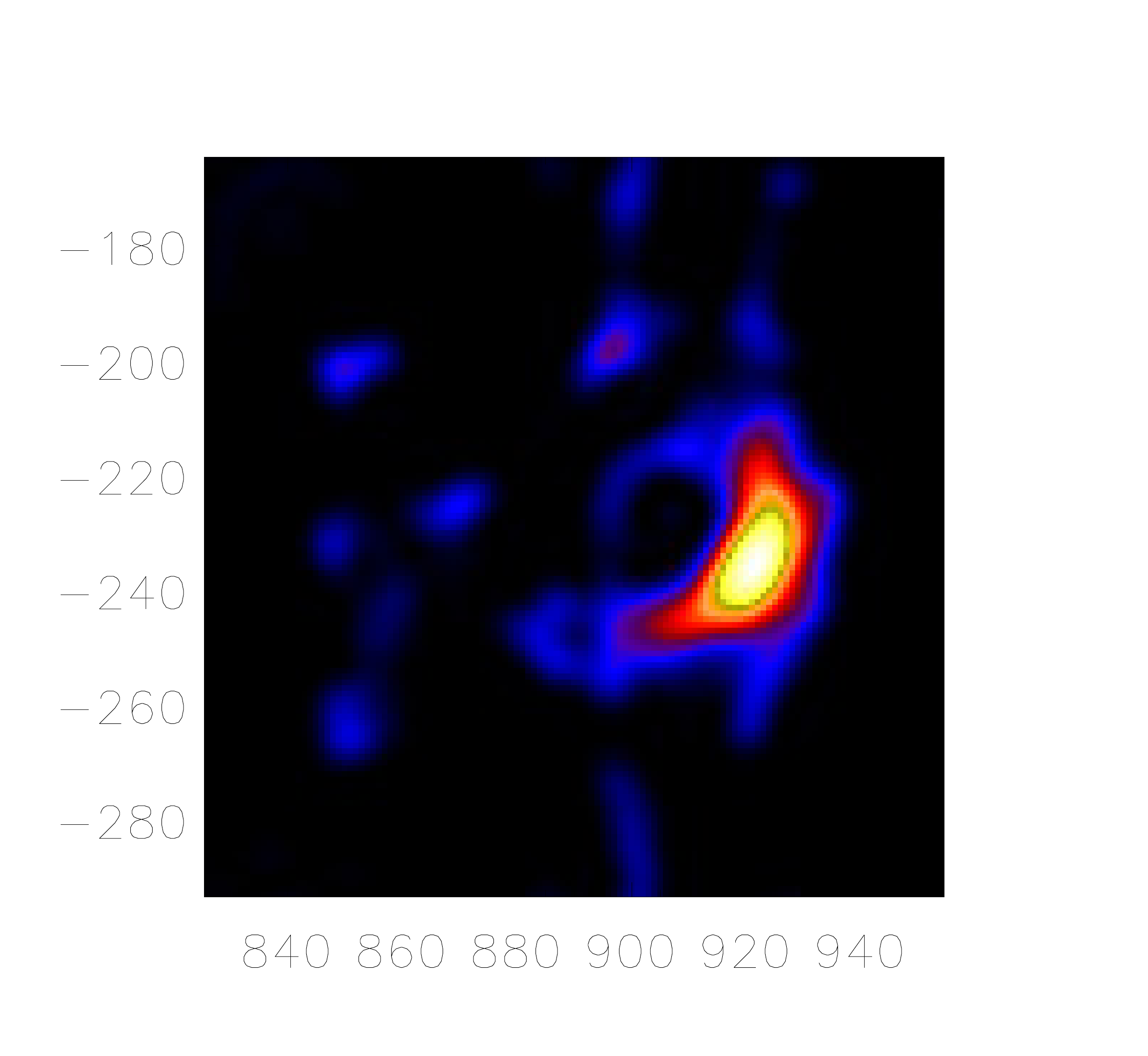}\hspace{-1cm}
    \includegraphics[height=3.7cm]{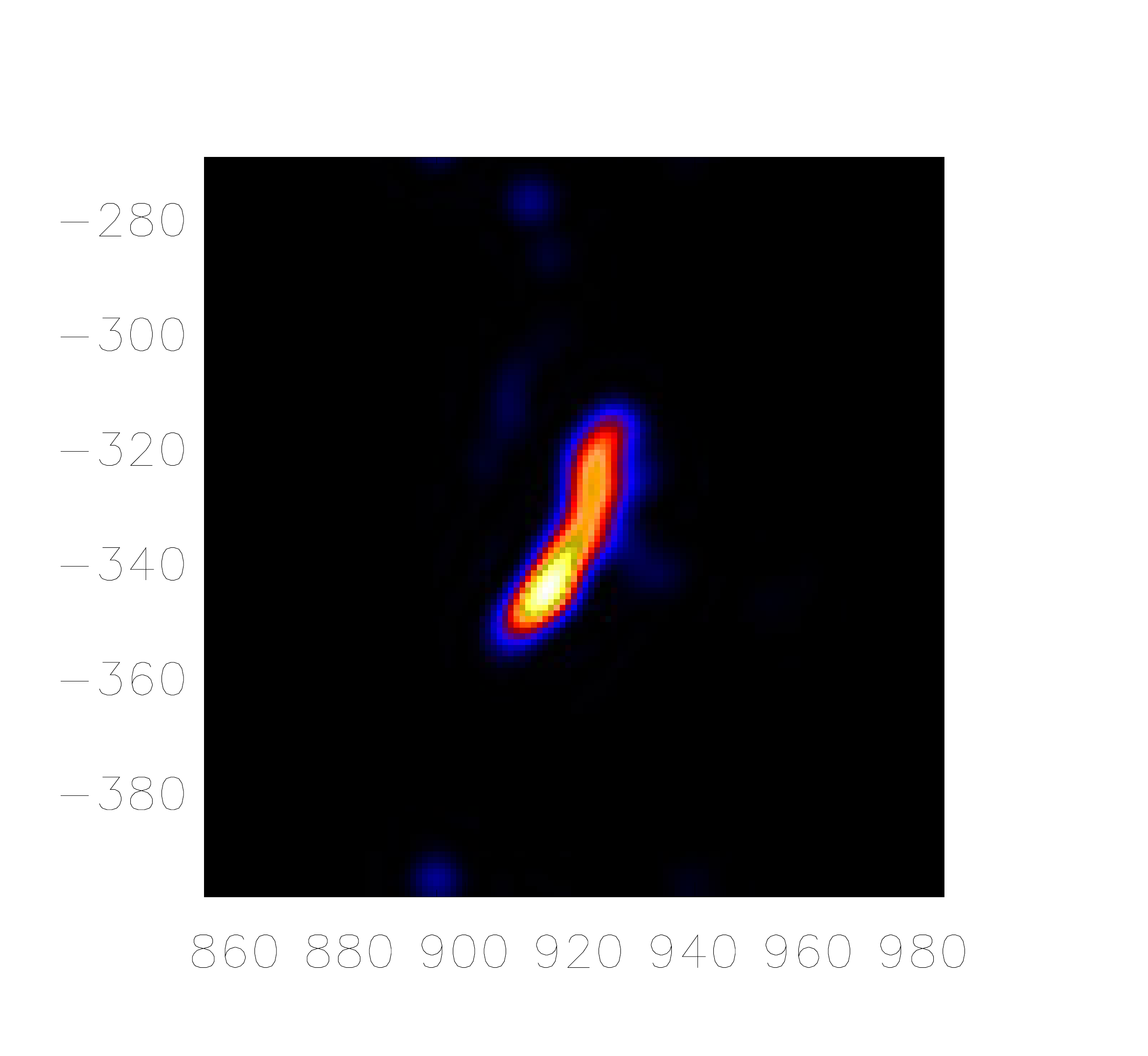}\hspace{-1cm}
    \includegraphics[height=3.7cm]{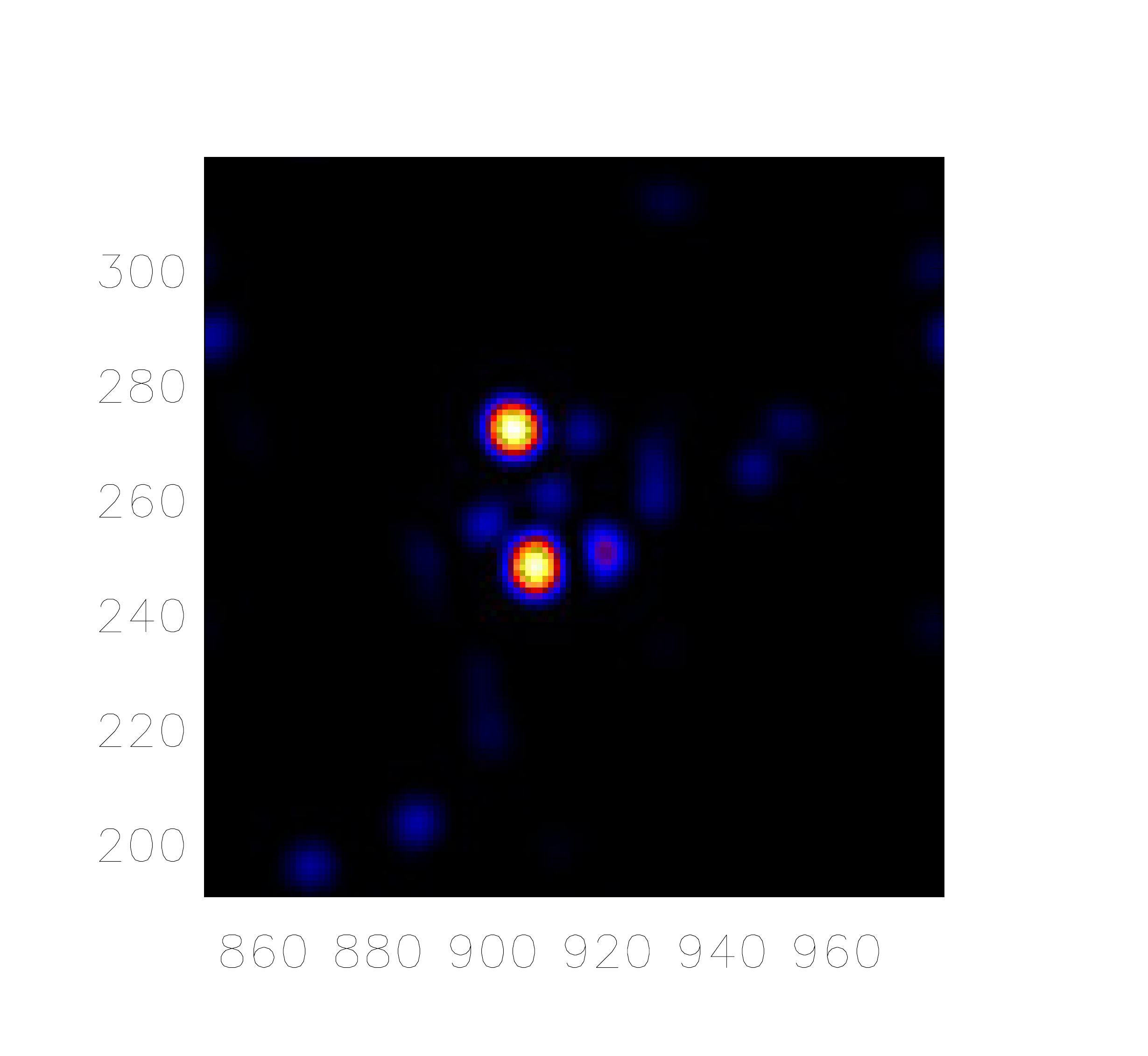}\hspace{-1cm}
    \includegraphics[height=3.7cm]{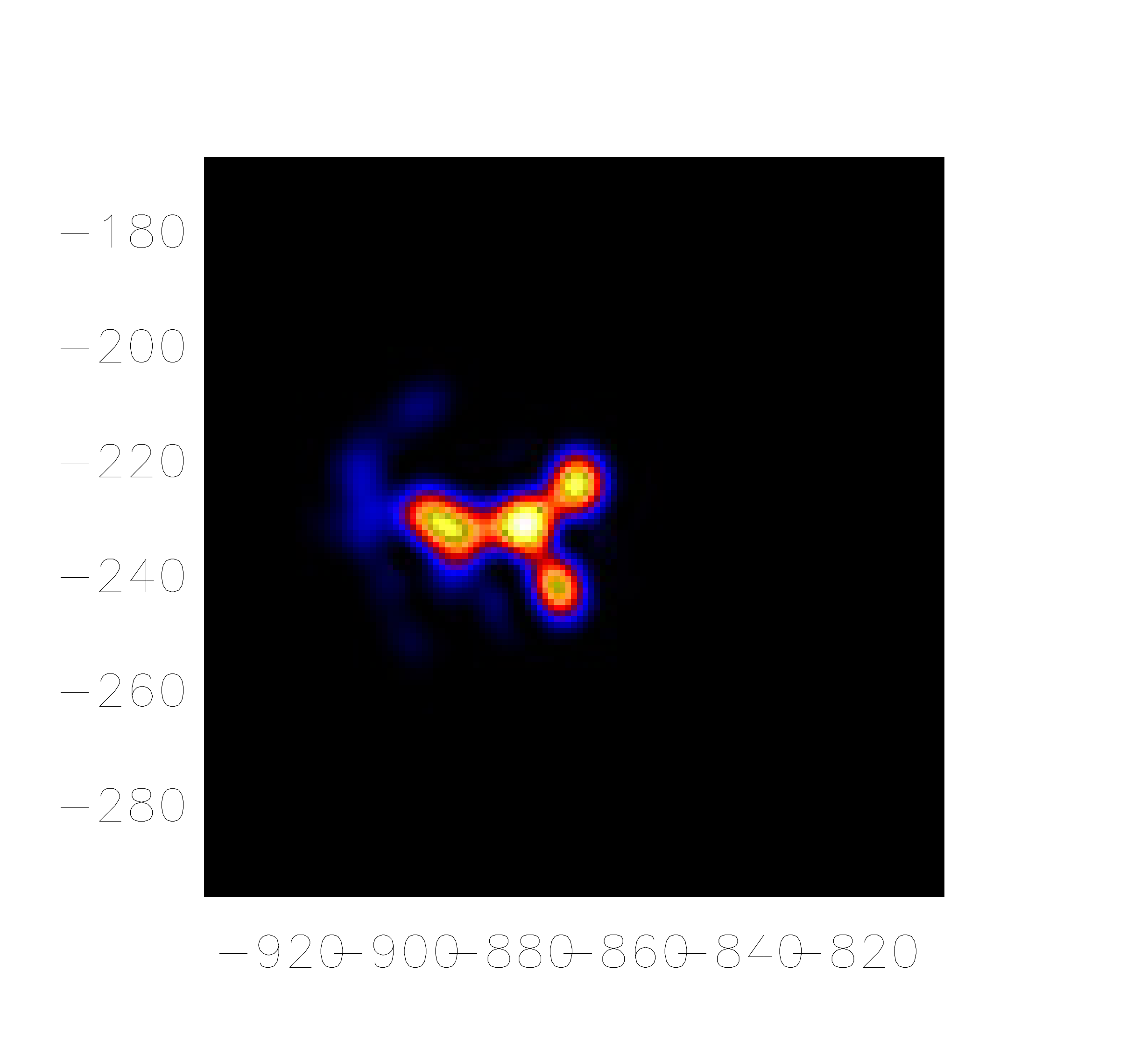}\\ \vspace{-1cm}
    \end{minipage}
        \begin{minipage}{.11\textwidth}
\centering
ASMC
\end{minipage}
\begin{minipage}{.88\textwidth}
 \includegraphics[height=3.7cm]{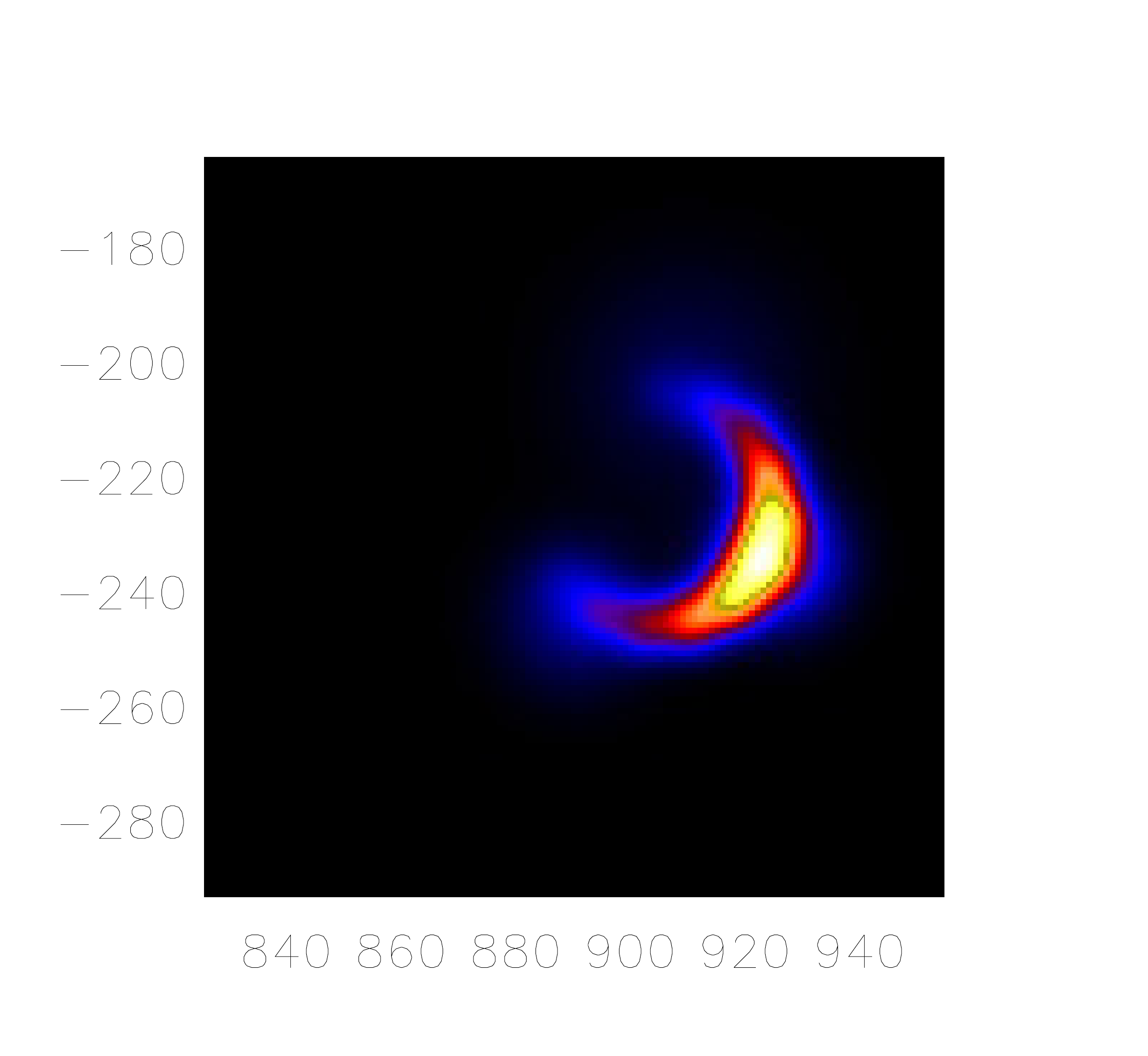}\hspace{-1cm}
    \includegraphics[height=3.7cm]{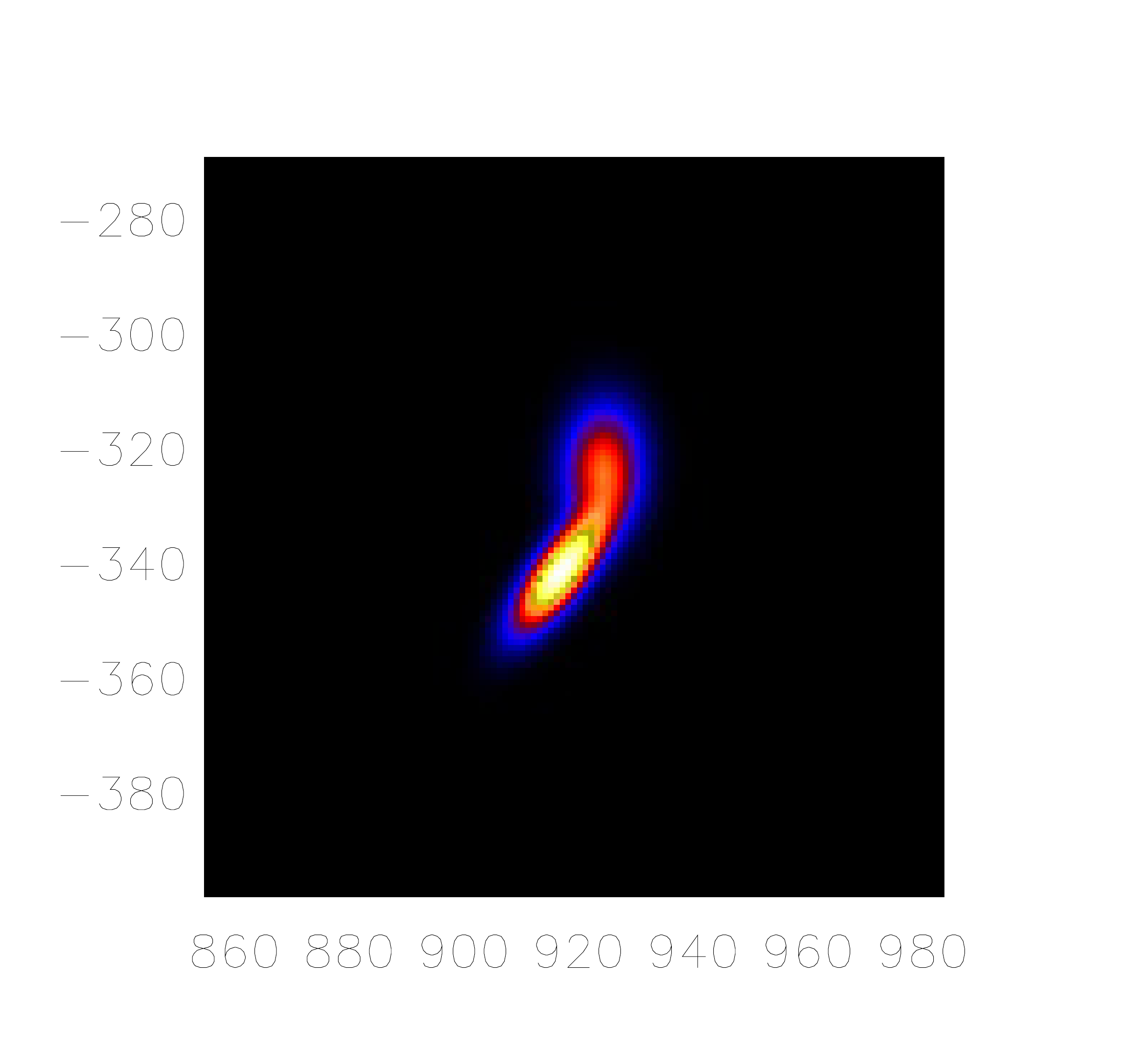}\hspace{-1cm}
    \includegraphics[height=3.7cm]{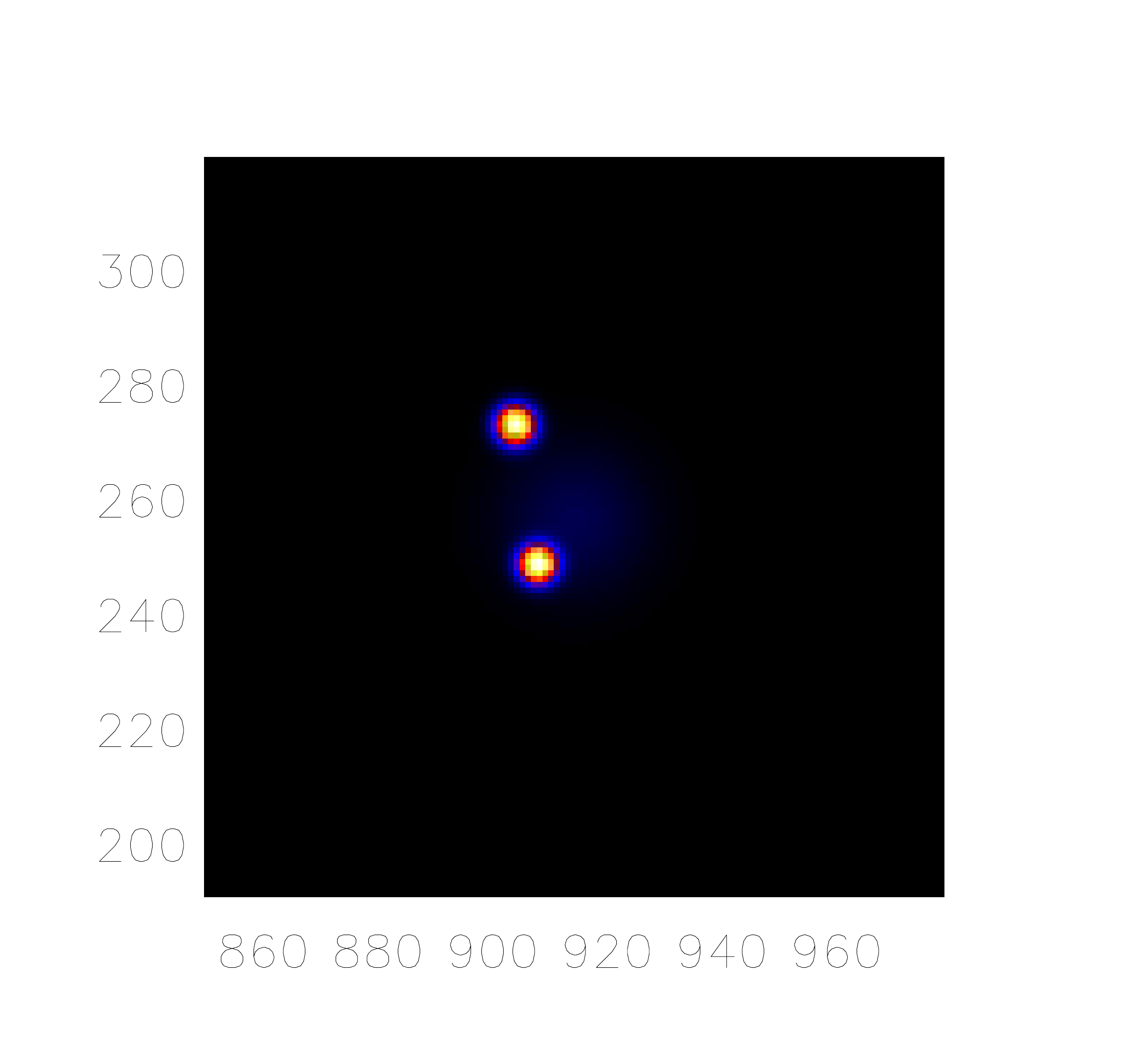}\hspace{-1cm}
    \includegraphics[height=3.7cm]{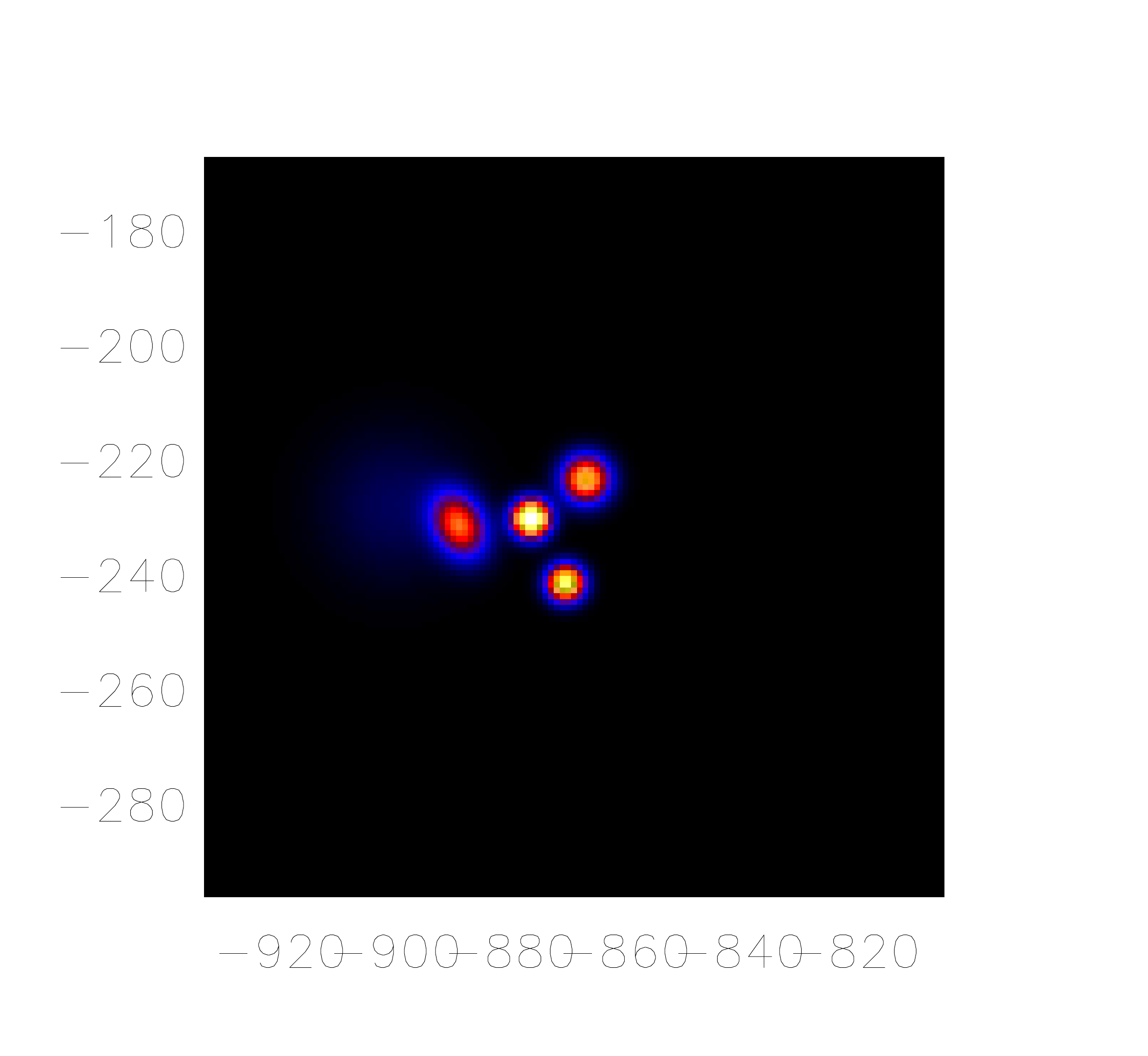}\\ \vspace{-1cm}
    \end{minipage}
  \caption{From left to right: comparison of the images obtained for the 23 August 2005, the 02 December 2003 , the 20 February  2002, and the 23 July   2002 events, by using VIS{\_}CLEAN (first row), VIS{\_}FWDFIT (second row), VIS{\_}CS (third row),  VIS\texttt{\_}WV (fourth row), and ASMC (fifth row). The number of sources estimated by the ASMC method is four, two, three, and five, respectively. }
  \label{fig:real_data}
\end{figure}

In \cref{fig:real_data}, we show the images of the four flaring events obtained by employing VIS{\_}CLEAN, VIS{\_}FWDFIT, VIS{\_}CS, VIS{\_}WV and the Bayesian method proposed here. For all the events, we observe that VIS{\_}CLEAN tends to produce overly smooth images, with little separation between different sources, even when these sources are relatively far from each other (third column). In addition, it can be difficult to tell the difference between weaker sources and artifacts, such as in the case of the bottom right source in the fourth column. As before, the images provided by VIS{\_}FWDFIT suffer from some limitations, see for instance the December 2003 and July 2002 events, where only two circles have been recovered. The reconstructions obtained with  VIS\texttt{\_}WV are better, inasmuch as they provide more detailed images and allow to better isolate different sources; however, they suffer from the presence of a relatively high number of small spots which should not be present (similar behaviour as in the synthetic case). VIS{\_}CS recovers accurate images, but some events are not perfectly recovered, we refer for instance to the fourth weak source in the July 2002 event; this problem can probably be overcome by tuning the parameters. The Bayesian approach confirms its capability of recovering tidy images even from experimental data. Despite the simplicity of the underlying geometrical model, the reconstructed images do not appear substantially simpler than those obtained with the other methods, while being less affected by spurious sources. Locations and intensities of the estimated sources are in accordance with those obtained with the other methods; the size of the sources estimated by our approach is, at times, smaller than that estimated by the other methods; this fact needs to be investigated further. 

We also notice that the number of sources estimated from these experimental data (reported in the caption of \cref{fig:real_data}) is, at times, larger than the number of objects in the image. This is mainly due to the more complex shape of real sources, which present asymmetries that cannot be explained by our simple geometric shapes. This phenomenon is particularly manifest for the August 2005 flare, in which four sources are recontructed to explain a single loop: in addition to the wide loop-source, three small and low intensity circles are estimated here.

\section{Conclusions}
\label{sec:conclusions}

We described a Bayesian approach to RHESSI imaging that is based on the assumption that the image can be represented by a small set of parametric objects, the number of objects being a priori unknown. This setting can be considered as an alternative to classical regularization methods with sparse penalty; here, the parameters in the prior distribution and in the likelihood function (noise standard deviation) play the role of regularization parameters, and the prior on the number of sources is the sparsity--promoting element. We set up a Monte Carlo algorithm that approximates the posterior distribution of our Bayesian model, by constructing an artificial sequence of distributions that starts from the prior distribution and evolves smoothly towards the desired posterior. Interestingly, each distribution of the sequence can be interpreted as a different posterior distribution corresponding to a different, decreasing value of the noise standard deviation; as a consequence, the algorithm explores models of increasing complexity as iterations proceed.

We applied the proposed method to a set of synthetic data, already used to assess the reliability of RHESSI imaging algorithms, and to experimental data. In both cases, the Bayesian approach proved to perform similarly to or better than well known and state--of--the--art methods. Importantly, the Bayesian solutions appeared to have a relatively weak dependence on the parameters in the prior distribution, which implies that little or no tuning is needed to run the algorithm, while the other RHESSI reconstruction methods require an in-depth knowledge of the algorithms to fine-tune the parameters, which can be impractical for certain users and applications. In addition, the proposed algorithm provides posterior probabilities for the estimated parameters, thus allowing to quantify the uncertainty of the estimated values; such additional information can be important in astrophysics studies (see \cite{sciacchitano2018identification}).

The main drawback of the method is perhaps the relatively high computational cost which is inherent in the sampling procedure. However, taking into account that almost no parameter tuning is needed, the proposed method can be considered to provide a relatively good balance between computational efficiency and user interaction, when compared to the other well known methods. In addition, the algorithm can be quite strongly parallelized in order to reduce the computing time, if necessary.

The positive results of this work encourage further investigation and methodological development. For instance, one could exploit the linear dependence of the data on the source fluxes, and marginalize the posterior distribution using a technique known as Rao--Blackwellization \cite{caro96}, that entails sampling less parameters and therefore having smaller Monte Carlo variance. On a different level, RHESSI images are always obtained for a specific energy interval; nearby energy intervals typically produce similar, but different images; such ``dynamic'' characteristic of RHESSI data could be exploited by setting up a Monte Carlo filtering procedure \cite{doucet2009tutorial}, in which the posterior distribution at a given energy is used as a prior distribution for the next energy interval.

Finally we remark that, even  though in this work we considered only application to RHESSI imaging, the proposed method can be generalized/adapted to any sparse imaging problem, provided that the image can be represented by a relatively small number of parametric objects. In particular, early tests on STIX simulated data show that the proposed method can be used on the upcoming STIX instrument.

\section*{Acknowledgments}
FS kindly acknowledges the Gruppo Nazionale per il Calcolo Scientifico (GNCS) for financial support.
FS and AS have been supported by the H2020 grant Flare Likelihood And Region Eruption foreCASTing (FLARECAST), project number 640216.
Further, the authors wish to thank  A. M. Massone and M. Piana for useful discussions on the numerical results,  R. A. Schwartz for providing the simulated dataset as well as the anonymous referees for  their insightful and constructive comments.
\bibliographystyle{siamplain}
\bibliography{references}
\end{document}